\pdfoutput=1
\documentclass[twoside, openright, 12pt]{report}

\usepackage{thestyle}
\usepackage{thevars}

\begin{document}

  \newgeometry{top=1.5cm, bottom=1.5cm, inner=2cm, outer=2cm}

  \begin{titlepage}

    \centering
    \fontfamily{helvetica}
    \includegraphics[width=0.4\textwidth]{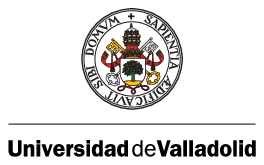}\par\vspace{1cm}

    {\scshape\Huge \theschool \par}
    \vspace{0.5cm}

    {\scshape\Large \thesubject\par}
    \vspace{1.25cm}

    {\scshape\Large \thegrade\par}
    \vspace{4cm}

    {\Huge\textbf{\thetitle}\par}
    \vspace{4cm}

    {\Large\hfill Autor: \par}
    {\Large\hfill\bfseries \theauthor\par}
    \vfill

  \end{titlepage}

  \clearpage\mbox{}\thispagestyle{empty}\clearpage

  \begin{titlepage}

    \centering
    \fontfamily{helvetica}
    \includegraphics[width=0.4\textwidth]{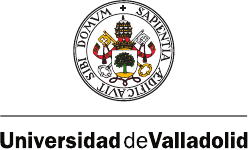}\par\vspace{1cm}

    {\scshape\Huge \theschool \par}
    \vspace{0.5cm}

    {\scshape\Large \thesubject\par}
    \vspace{1.25cm}

    {\scshape\Large \thegrade\par}
    \vspace{4cm}

    {\Huge\textbf{\thetitle}\par}
    \vspace{4cm}

    {\Large\hfill Autor: \par}
    {\Large\hfill\bfseries \theauthor\par}
    \vspace{0.5cm}

    {\Large\hfill Tutor: \par}
    {\Large\hfill\bfseries \thesupervisor\par}
    \vfill

  \end{titlepage}

  \clearpage\mbox{}\thispagestyle{empty}\clearpage

  \restoregeometry
  \setcounter{page}{1}

  \begin{abstractpage}
    \addcontentsline{toc}{chapter}{\protect\numberline{}Resumen}
    \begin{abstract-lang}{english}
      This work consists of a study of a set of techniques and strategies related with algorithm's design, whose purpose is the resolution of problems on massive data sets, in an efficient way. This field is known as \emph{Algorithms for Big Data}. In particular, this work has studied the \emph{Streaming Algorithms}, which represents the basis of the data structures of sublinear order $o(n)$ in space, known as \emph{Sketches}. In addition, it has deepened in the study of problems applied to \emph{Graphs} on the \emph{Semi-Streaming} model. Next, the \emph{PageRank} algorithm was analyzed as a concrete case study. Finally, the development of a library for the resolution of graph problems, implemented on the top of the intensive mathematical computation platform known as \emph{TensorFlow} has been started.
    \end{abstract-lang}
    \begin{abstract-lang}{spanish}
      En este trabajo se ha realizado un estudio acerca de las diferentes técnicas y estrategias de diseño de algoritmos, pensadas para la resolución de problemas sobre conjuntos de datos de tamaño masivo, de manera eficiente. Este campo es conocido conocido como \emph{Algoritmos para Big Data}. En concreto, en este trabajo se ha profundizado en el estudio de los \emph{Algoritmos para Streaming}, que representan la base de las estructuras de datos de orden sublineal $o(n)$ en espacio, conocidas como \emph{Sketches}. Además, se ha profundizado en el estudio de problemas aplicados a \emph{Grafos} sobre el modelo en \emph{Semi-Streaming}. Seguidamente, se ha analizado el algoritmo \emph{PageRank} como caso concreto de estudio. Por último, se ha comenzado el desarrollo de una biblioteca para la resolución de problemas de grafos, implementada sobre la plataforma de cómputo matemático intensivo \emph{TensorFlow}.
    \end{abstract-lang}

    \centering
    Este trabajo ha sido publicado en: \url{https://github.com/garciparedes/tf_G}

  \end{abstractpage}

  \chapter*{Agradecimientos}
  \label{sec:acknowledgements}
  \addcontentsline{toc}{chapter}{\protect\numberline{}Agradecimientos}

    \paragraph{}
    Me gustaría agradecer a todas aquellas personas y entidades que han estado cerca de mí durante estos cuatro años de carrera.

    \paragraph{}
    Entre ellos, me gustaría remarcar la labor realizada por la \emph{Universidad de Valladolid} en general, y de la \emph{Escuela de Ingeniería Informática de Valladolid} en concreto, por su colaboración siempre que se ha requerido. También me gustaría agradecer a la empresa \emph{Brooktec S.L.} su acogida durante mi periodo de prácticas.

    \paragraph{}
    Muchas gracias a todos los profesores que siempre han estado ahí para recibirme en su despacho cuando he necesitado su ayuda. En especial, muchas gracias a \emph{Manuel Barrio Solórzano} por supervisar este trabajo.

    \paragraph{}
    También me gustaría agradecer a mi familia su apoyo incondicional, tanto a nivel económico como emocional. En especial, a mi madre por todos esos días de estrés en que aguantó siempre con una sonrisa mi mal humor y preocupaciones.

    \paragraph{}
    Por último, tengo que resaltar el apoyo de todas aquellas personas que me han estado apoyando estos años, tanto a los compañeros de clase, con los que sin su compañía todo esto habría sido mucho más complicado. A los amigos de siempre, por no dejar nunca de estar ahí y a todas esas personas que han estado aguantando mis dramas este último año, ellas saben quienes son.

  \chapter*{Prefacio}
  \addcontentsline{toc}{chapter}{\protect\numberline{}Prefacio}

    \paragraph{}
    Para entender el contenido de este documento así como la metodología seguida para su elaboración, se han de tener en cuenta diversos factores, entre los que se encuentran el contexto académico en que ha sido redactado, así como el tecnológico y social. Es por ello que a continuación se expone una breve descripción acerca de los mismo, para tratar de facilitar la compresión sobre el alcance de este texto.

    \paragraph{}
    Lo primero que se debe tener en cuenta es el contexto académico en que se ha llevado a cabo. Este documento se ha redactado para la asignatura de \textbf{Trabajo de Fin de Grado (mención en Computación)} para el \emph{Grado de Ingeniería Informática}, impartido en la \emph{E.T.S de Ingeniería Informática} de la \emph{Universidad de Valladolid}. Dicha asignatura se caracteriza por ser necesaria la superación del resto de las asignaturas que componen los estudios del grado para su evaluacion. Su carga de trabajo es de \textbf{12 créditos ECTS}, cuyo equivalente temporal es de \emph{300 horas} de trabajo del alumno, que se han llevado a cabo en un periodo de 4 meses.

    \paragraph{}
    La temática escogida para realizar dicho trabajo es \textbf{Algoritmos para Big Data}. El Big Data es la disciplina que se encarga de \say{todas las actividades relacionadas con los sistemas que manipulan grandes conjuntos de datos. Las dificultades más habituales vinculadas a la gestión de estas cantidades de datos se centran en la recolección y el almacenamiento, búsqueda, compartición, análisis, y visualización. La tendencia a manipular enormes cantidades de datos se debe a la necesidad en muchos casos de incluir dicha información para la creación de informes estadísticos y modelos predictivos utilizados en diversas materias.}\cite{wiki:big_data}

    \paragraph{}
    Uno de los puntos más importantes para entender la motivación por la cual se ha escogido dicha temática es el contexto tecnológico en que nos encontramos. Debido a la importante evolución que están sufriendo otras disciplinas dentro del mundo de la informática y las nuevas tecnologías, cada vez es más sencillo y económico recoger gran cantidad de información de cualquier proceso que se dé en la vida real. Esto se debe a una gran cantidad de factores, entre los que se destacan los siguientes:

    \begin{itemize}

      \item \textbf{Reducción de costes derivados de la recolección de información}: Debido a la constante evolución tecnológica cada vez es más barato disponer de mecanismos (tanto a nivel de hardware como de software), a partir de los cuales se puede recabar datos sobre un determinado suceso.

      \item \textbf{Mayor capacidad de cómputo y almacenamiento}: La recolección y manipulación de grandes cantidades de datos que se recogen a partir de sensores u otros métodos requieren por tanto del apoyo de altas capacidades de cómputo y almacenamiento. Las tendencias actuales se están apoyando en técnicas de virtualización que permiten gestionar sistemas de gran tamaño ubicados en distintas zonas geográficas como una unidad, lo cual proporciona grandes ventajas en cuanto a reducción de complejidad algorítmica a nivel de aplicación.

      \item \textbf{Mejora de las telecomunicaciones}: Uno de los factores que ha permitido una gran disminución de la problemática relacionada con la virtualización y su capacidad de respuesta ha sido el gran avance a nivel global que han sufrido las telecomunicaciones en los últimos años, permitiendo disminuir las barreras geográficas entre sistemas tecnológicos dispersos.

    \end{itemize}

    \paragraph{}
    Gracias a este conjunto de mejoras se ha llegado al punto en que existe la oporturnidad de poder utilizar una gran cantidad de conocimiento, que individualmente o sin un apropiado procesamiento, carece de valor a nivel de información.

    \paragraph{}
    El tercer factor que es necesario tener en cuenta es la tendencia social actual, que cada vez más, está concienciada con el valor que tiene la información. Esto se ve reflejado en un amplio abanico de aspectos relacionados con el comportamiento de la población:

    \begin{itemize}

      \item \textbf{Monitorización de procesos laborales}: Muchas empresas están teniendo en cuenta la mejora de la productividad de sus empleados y máquinas. Por tanto, buscan nuevas técnicas que les permitan llevar a cabo dicha tarea. En los últimos años se ha dedicado mucho esfuerzo en implementar sistemas de monitorización que permitan obtener información para después procesarla y obtener resultados valiosos para dichas organizaciones.

      \item \textbf{Crecimiento exponencial de las plataformas de redes sociales}: La inherente naturaleza social del ser humano hace necesaria la expresión pública de sus sentimientos y acciones, lo cual, en el mundo de la tecnología se ha visto reflejado en un gran crecimiento de las plataformas de compartición de información así como de las de comunicación.

      \item \textbf{Iniciativas de datos abiertos por parte de las administraciones públicas}: Muchas insitituciones públicas están dedicando grandes esfuerzos en hacer visible la información que poseen, lo que conlleva una mejora social aumentando el grado de transparencia de las mismas, así como el nivel de conocimiento colectivo, que puede ser beneficioso tampo para ciudadanos como para empresas.

    \end{itemize}

    \paragraph{}
    Como consecuencia de este cambio social, posiblemente propiciado por el avance tecnológico anteriormente citado, la población tiene un mayor grado de curiosidad por aspectos que antes no tenia la capacidad de entender, debido al nivel de complejidad derivado del tamaño de los conjuntos de muestra necesarios para obtener resultados fiables.

    \paragraph{}
    En este documento no se pretenden abordar temas relacionados con las técnicas utilizadas para recabar nuevos datos a partir de los ya existentes. A pesar de ello se realizará una breve introducción sobre dicho conjunto de estrategias, entre las que se encuentran:  \emph{Heurísticas}, \emph{Regresión Lineal}, \emph{Árboles de decisión}, \emph{Máquinas de Vector Soporte (SVM)} o \emph{Redes Neuronales Artificiales}.

    \paragraph{}
    Por contra, se pretende realizar un análisis acerca de los diferentes algoritmos necesarios para manejar dichas cantidades ingentes de información, en especial de su manipulación a nivel de operaciones básicas, como operaciones aritméticas, búsqueda o tratamiento de campos ausentes. Para ello, se tratará de acometer dicha problemática teniendo en cuenta estrategias de paralelización, que permitan aprovechar en mayor medida las capacidades de cómputo existentes en la actualidad.

    \paragraph{}
    Otro de los aspectos importantes en que se quiere orientar este trabajo es el factor dinámico necesario para entender la información, lo cual conlleva la búsqueda de nuevas estrategias algorítmicas de procesamiento en tiempo real. Por lo tanto, se pretende ilustrar un análisis acerca de las soluciones existentes en cada caso con respecto a la solución estática indicando las ventajas e inconvenientes de la versión dinámica según corresponda.

  \tableofcontents

  \chapter{Introducción}
  \label{chap:intro}

    \section{Contexto}
    \label{sec:introduction_context}

      \paragraph{}
      El \emph{Trabajo de Fin de Grado} representa la última fase para la obtención de una titulación de \emph{Grado} en el modelo educativo Español. Para poder que el alumno pueda presentar su trabajo final ante un tribunal, es necesario que este haya completado el resto de créditos de la titulación. Por tanto, el \emph{Trabajo de Fin de Grado} representa la última barrera antes de convertirse en graduado. En este trabajo, se espera que el alumno demuestre las capacidades y conocimientos adquiridos a lo largo de su formación universitaria desde una perspectiva práctica y más cercana a lo que se espera que realice una vez comience su andadura por el mundo laboral.

      \paragraph{}
      Estas ideas son de carácter general y no dependen de la titulación que se esté realizando. Sin embargo, el trabajo de fin de grado depende fuertemente de la titulación a la cual se refiera. Es trivial entender que un alumno que haya cursado estudios de \emph{Filología Hispánica} no tenga nada que ver con el de un alumno que cuyos estudios estén referidos a otros ámbitos del conocimiento como la \emph{Física}, puesto que sus competencias son muy diferentes. Todos ellos tendrán una base común, realizando una introducción previa al tema que pretenden desarrollar, posiblemente describiendo el contexto histórico, seguidamente desarrollando el tema principal para finalmente llegar a unas conclusiones específicas.

      \paragraph{}
      En este caso, este \emph{Trabajo de Fin de Grado} se refiere a la titulación del \emph{Grado} en \emph{Ingeniería Informática} impartido en la \emph{Escuela Técnica Superior de Ingeniería Informática} de \emph{Valladolid}, dependiente de la \emph{Universidad de Valladolid}. Por esta razón, el trabajo será referido completamente al ámbito de la informática. Sin embargo, en este caso sucede una característica similar al descrito en el párrafo anterior. En esta titulación existen 3 menciones (o especialidades) que tratan de segregar las competencias que se enseñan, de tal manera que los alumnos puedan llegar a un mayor grado de conocimiento en la disciplina que más prefieran.

      \paragraph{}
      La razón de dicha separación durante el segundo ciclo de la titulación de grado se debe al amplísimo crecimiento que se está llevando a cabo en los últimos años, de tal manera que a pesar de haber una serie de conocimientos comunes que todo \emph{Ingeniero Informático} debe conocer, llega un punto en que la diversificación de áreas dificultan la tarea de adquicisión de todos aquellos conceptos en profundidad. Por tanto, parece apropiado dividir dichas disciplinas en ramas separadas. En el caso de la titulación para la cual se realiza este trabajo, existen 3 menciones: \emph{Tecnologías de Información}, \emph{Ingeniería de Software} y \emph{Computación}.

      \paragraph{}
      En este documento no se describirán cada una de ellas, ni se realizará una diferenciación de las mismas, puesto que esto ya puede ser consultado a través de la página web de la \emph{Escuela Técnica Superior de Ingeniería Informática} de \emph{Valladolid} a través de \url{https://www.inf.uva.es/}. Sin embargo, si que se indicará que este trabajo ha sido realizado tras haber seguido la mención en \emph{Computación}, la cual se refiere a los aspectos más teóricos, matemáticos y abstractos de la informática, tratando de dejar de lado el contexto del problema para centrarse en la búsqueda eficiente de la solución.

      \paragraph{}
      La razón por la cual se indica dicha explicación acerca de las distintas menciones sobre las que completar la titulación de \emph{Grado} en \emph{Ingeniería Informática}, así como el ejemplo inicial acerca de la diferenciación a nivel de contenido entre distintos trabajos de \emph{Fin de Grado} dependiendo de la titulación se debe a lo siguiente:

      \paragraph{}
      Este trabajo se ha pretendido focalizar en el estudio de \emph{Algoritmos para Big Data, Grafos y PageRank} desde una perspectiva mayoritariamente teórica, dejando de lado aspectos y cuestiones referidas a la implementación de los mismos. A pesar de que se ha realizado una implementación en el trabajo, esta ha sido de carácter ilustrativo, teniendo que requerir de trabajo adicional si pretender convertirse en una implementación adecuada para ser usada en entornos de producción.

      \paragraph{}
      Esto se contrapone con los temas que se desarrollan comúnmente para los estudios en cuestión, que generalmente basan un mayor esfuerzo en la parte de la  implementación para llegar en muchas ocasiones a un producto o servicio final. Esto se debe a las competencias desarrolladas, que se centran en ese tipo de actividades. Sin embargo, esto se contrapone con las competencias adquiridas durante el desarrollo de la mención en \emph{Computación}, la cual, tal y como se ha indicado anteriormente, se centra mayoritariamente en el apartado teórico y matemático de la resolución de problemas de manera eficiente.

      \paragraph{}
      Una vez realizada dicha distinción, ya se está en condiciones de comenzar a tratar el tema sobre el cual trata este \emph{Trabajo de Fin de Grado}. A continuación se hablará acerca de las motivaciones tanto personales como académicas que han propiciado la selección de dicho tema en la sección \ref{sec:introduction_motivation}. Para ello, se ha creído conveniente realizar una descripción acerca de las ideas iniciales que se tenían antes de comenzar el trabajo, las cuales son drásticamente diferentes de las que se tiene una vez se ha finalizado el mismo. Esto se realiza en la sección \ref{sec:introduction_initial_ideas}. Posteriormente, en las secciones \ref{sec:introduction_big_data} y \ref{sec:introduction_graphs} se realiza una descripción superficial acerca del \emph{Big Data} y la modelización de \emph{Grafos} respectivamente, puesto que son los temas principales de dicho trabajo. Por último, en la sección \ref{sec:introduction_goals} se indican una serie de objetivos que se han pretendido conseguir mediante la realización de este trabajo.

    \section{Motivación}
    \label{sec:introduction_motivation}

      \paragraph{}
      La razón original por la cual se decidió escoger la temática del \emph{Big Data} para la realización de este trabajo está motivada por la consulta con el profesor y doctor \emph{Manuel Barrio-Solórzano} (\texttt{mbarrio@infor.uva.es}), que creyó apropiado un proyecto cercano a la investigación de los algoritmos subyacentes que permiten resolver problemas de tamaño masivo como \emph{Trabajo de Fin de Grado}.

      \paragraph{}
      Una vez indicada la explicación acerca del modo en que el trabajo fue propuesto, a continuación se indican distintas razones por las cuales se cree que dicho tema es interesante y ferviente en la actualidad: Durante los últimos años se han producido grandes avances a nivel tecnológico que han permitido la construcción de sistemas computacionales cuyo coste es mucho más reducido y su capacidad de cálculo mucho más elevada.

      \paragraph{}
      Estos avances son claramente apreciables, tanto directamente por los usuarios a través de los dispositivos móviles, las televisiones inteligentes o los ordenadores personales, que ofrecen capacidades de cómputo inimaginables en décadas posteriores. Esto también se puede apreciar internamente en la industria de la informática, con la construcción de supercomputadores como el \emph{Sunway TaihuLight} que duplica las capacidades de su predecesor. La necesidad por agilizar los cálculos matemáticoss intensivos ha llevado a empresas como \emph{Google} a diseñar sus propios chips específicamente para dicha función, que denominan \emph{Tensor Processor Units} \cite{jouppi2017datacenter}. Además, distintas técnicas basadas en virtualización de equipos y paralelización han permitido un mejor aprovechamiento de las capacidades computacionales existentes.

      \paragraph{}
      Todo ello ha propiciado una explosión a nivel de información, de tal manera que año a año la cantidad de datos generados por los usuarios está creciendo en un orden asombrante. Por tanto, esto ha generado nuevos retos, tanto a nivel de almacenamiento y recuperación, como de procesamiento y obtención de nuevas conclusiones a partir de su análisis. Sin embargo, debido a distintos factores, entre los que destacan el elevado tamaño del mismo, así como la componente dinámica que muchas veces se da, generando una ventana temporal subyacente a partir de la cual se restringe el periodo de validez de estos, en los últimos años se han realizado distintos trabajos centrados en la investigación de distintas técnicas que tratan de agilizar dicho procesamiento y análisis.

      \paragraph{}
      Muchos de los procesos que se llevan a cabo en la vida cotidiana se basan en la interrelación de objetos entre sí, lo cual genera una red de interrelaciones de manera subyacente lo cual puede ser modelizado matemáticamente a través del concepto de grafo. Este tipo de sucesos también están siendo estudiados y analizados por lo cual, también forman parte del mundo del \emph{Big Data}. No es difícil darse cuenta de que muchas de las empresas más populares del sector tecnológico basan su actividad en procesos de este tipo. Algunos ejemplos son la búsqueda de Sitios Web de \emph{Google} y las interconexiones que estos tienen a través de enlaces, las redes de amigos que se forman en \emph{Facebook}, las relaciones de similitud de contenido en servicios como \emph{Netflix} o \emph{Spotify}, los sistemas de planificación de rutas de empresas como \emph{Tesla} para sus sistemas de conducción autónoma, etc. Por tanto, se cree interesante estudiar las distintas técnicas y conceptos que permiten agilizar la obtención de dichos resultados.

      \paragraph{}
      En cuanto al algoritmo \emph{PageRank}, se ha creído adecuado como punto de combinación entre las técnicas de \emph{Big Data} y procesamiento masivo de información, con el modelo matemático de \emph{Grafos}. Además, se cree que los conceptos matemáticos en que se basa son de gran utilidad para entender tanto su comportamiento como servir de introducción en el área de modelos gráficos probabilísticos (\emph{Cadenas de Markov}). Otra de las razones por las cuales se ha creído conveniente el estudio de dicho algoritmo es la importante relevancia que ha tenido en el mundo de la informática, permitiendo mejorar drásticamente los resultados de búsquedas que se obtenían hasta el momento de su publicación y convirtiendo al motor de búsquedas \emph{Google} en la alternativa más utilizada respecto de la competencia.

      \paragraph{}
      Tras indicar los motivos por los cuales se ha creído interesante el estudio de \emph{Algoritmos para Big Data}, centrándose especialmente en el caso de los problemas de \emph{Grafos} y discutiendo en profundidad el \emph{Algoritmo PageRank}, se ha creído conveniente añadir una sección en la cual se indique la visión que se tenía sobre dichas estrategias al comienzo del trabajo. Esto se realizará en la siguiente sección.

    \section{Ideas Iniciales}
    \label{sec:introduction_initial_ideas}

      \paragraph{}
      Se ha creído oportuno añadir un apartado en el documento, en el cual se explicara la visión y conocimientos previos que se tenían sobre la temática del trabajo antes de comenzar el proceso de investigación y estudio. Por tanto, en esta sección se pretende realizar una descripción acerca de lo que se conocía previamente por \emph{Big Data}, las intuiciones acerca de las técnicas utilizadas para hacer frente a la resolución de problemas sobre grafos de tamaño masivo y el algoritmo \emph{PageRank}. Habría sido más apropiado redactar esta sección al comienzo del proyecto, de tal manera que la apreciación del contraste entre los conocimientos iniciales y los adquiridos durante el proyecto sería mucho más visible y realista. Sin embargo, esta tarea se ha realizado al final del proyecto, por lo que se tratará de ser lo más riguroso posible con respecto de la visión previa.

      \paragraph{}
      Hasta el comiendo del estudio en profundidad y al seguimiento de cursos como el de \emph{Algorithms for Big Data} \cite{bigdata2015jelani} impartido por \emph{Jelani Nelson}, la visión que se tenía de estos era muy reducida. Se entendía que se trataban de técnicas para extraer más rápidamente información de conjuntos de datos contenidos en bases de datos, o incluso en tiempo real a través de \emph{streams} de datos. Se intuía que para el procesamiento de tal magnitud de información, las implementaciones se apoyaban en \emph{Algoritmos Probabilistas}, que agilizaran los cálculos a coste de una determinada tasa de error. Sin embargo, no se tenía ningún conocimiento sobre las estrategias en que se basaban las soluciones.

      \paragraph{}
      Algo a remarcar es la necesidad de clarificar el concepto de \emph{Big Data}, que al comienzo del trabajo se concebía únicamente como la obtención de métricas a partir de conjuntos de datos de tamaño elevado. Sin embargo, tras la realización del trabajo se ha ampliado el entendimiento de dicho concepto, para llegar a la conclusión de que \say{\emph{Big Data} consiste en todas aquellas soluciones diseñadas para tratar de resolver problemas para los cuales el cálculo de su solución no es asumible utilizando únicamente la memoria del sistema}. En la sección \ref{sec:introduction_big_data}, destinada a la descripción superficial acerca del \emph{Big Data} se indican las distintas alternativas propuestas para tratar de hacer frente a dicha problemática.

      \paragraph{}
      En cuanto a la modelización de \emph{Grafos}, entendida como la representación matemática de estructuras relacionales de tal manera que estas puedan ser vistas como una red de interconexiones entre puntos, desde el comienzo de este trabajo se poseía una cierta base en la materia. Dichos conocimientos fueron adquiridos a partir del conjunto de asignaturas impartidas en la titulación, destacando la de \emph{Matemática Discreta} \cite{matematicaDiscreta2016notes}, en la cual se estudia el formalismo matemático, además de un amplio conjunto de definiciones básicas relacionadas con propiedades de \emph{Grafos} o sus vértices. En dicha asignatura se describen algunos algoritmos como el de \emph{Prim} o \emph{Kruskal} para encontrar la solución al problema del árbol recubridor mínimo.

      \paragraph{}
      Sin embargo, debido al carácter general e introductorio de esta asignatura, en ella se describen estos algoritmos sin tener en cuenta el coste computacional de los mismos, por tanto, no se tiene en cuenta la escalabilidad de dichas estrategias sobre grafos formados por trillones de aristas \cite{ching2015one}. Tal y como se ha comprendido a lo largo del desarrollo del trabajo, existen diferentes estrategias para hacer frente al elevado tamaño de los grafos, dependiendo del problema a tratar. La idea que se tenía previamente era el tratamiento de sub-grafos seleccionados de manera acertada para resolver problemas que después fueran extrapolables al grafo general. Sin embargo, no se tenía conocimiento acerca de qué propiedades se pretendía mantener ni cómo se llevaban a cabo dichas técnicas.

      \paragraph{}
      En cuanto al algoritmo \emph{PageRank} estudiado en detalle en el trabajo, al igual que en los casos anteriores, se tenía una vaga intuición acerca de su funcionamiento, así como la información que proporciona. Se sabía que inicialmente se diseñó para la obtención del grado de importancia de un determinado vértice de un grafo, al igual que sucede en el grafo formado por los sitios web y los enlaces que los relacionan (\emph{Web Graph}). Sin embargo, tan solo se conocía que este basaba su puntuación en la propagación de influencias entre vértices, de tal manera que relacionarse con un número reducido de vértices importantes genera más relevancia que relacionarse con un mayor número de vértices menos importantes. A pesar de tener esta visión del algoritmo, no se tenían ideas claras acerca de la forma en que este puede ser calculado, ni de las capacidades de personalización o su relación con el concepto de \emph{Cadenas de Markov}.

      \paragraph{}
      Tras tratar de realizar una breve explicación acerca de la base de conocimientos relacionados con el tema al comienzo del trabajo, el siguiente paso que se realiza es exponer los objetivos que se pretenden conseguir tras la realización de este trabajo en la siguiente sección.

    \section{Objetivos}
    \label{sec:introduction_goals}

      \paragraph{}
      Para la realización de este trabajo, se han fijado una serie de objetivos, lo cual ha servido como guía para la realización del mismo. Sin embargo, dicha tarea no ha sido simple por la naturaleza exploratoria del proyecto. Esta razón, tal y como se ha tratado de exponer en el apéndice \ref{chap:methodology}, ha permitido que a pesar de que el trabajo tuviera una temática fijada \emph{a-priori}, la especificación del mismo en un ámbito concreto y estudiable haya sido guiada por el proceso de investigación. Por tanto, a continuación se indican los objetivos generales que se pretendía conseguir al comienzo del trabajo, además de la indicación acerca de los temas escogidos de manera concreta.

      \begin{itemize}

        \item Obtención de una visión panorámica acerca de las distintas técnicas y estrategias utilizadas para resolver problemas sobre conjuntos de datos de tamaño masivo (\emph{Big Data}).

        \item Selección de un tema concreto para ser estudiado con mayor profundidad, cuya relación con las estrategias y algoritmos estudiados desde el punto de vista del \emph{Big Data} sea elevada. Para esta tarea se ha decidido escoger el ámbito de los \emph{Grafos} de tamaño masivo y las distintas técnicas de reducción de su tamaño manteniendo la estructura de relaciones semejante.

        \item Implementación y estudio de un algoritmo concreto ampliamente relacionado con el resto del trabajo realizado, que permita poner en práctica el conjunto de conceptos estudiados a lo largo del proyecto. En este caso, el algoritmo escogido ha sido \emph{PageRank}, por su relación conceptual con el modelo de \emph{Grafos}, su base conceptual ampliamente relacionada con la \emph{Estadística} y la fuerte necesidad de ser implementado de manera eficiente para hacer frente al elevado tamaño de los problemas en que es aplicado \emph{Big Data}.

      \end{itemize}

      \paragraph{}
      Dichos objetivos principales no son los únicos que se han pretendido conseguir durante la realización del trabajo, sino que existe una cantidad más amplia de sub-objetivos necesarios para poder llegar al cumplimiento de estos. En este grupo se encuentra la realización de tareas de carácter investigatorio, requiriendo la necesidad de mantener un determinado índice de curiosidad que facilite la búsqueda de nuevas definiciones. Esto conlleva la lectura de distintos artículos de carácter científico, junto con la correspondiente dificultad propiciada por el tono extremadamente formal de los mismo. Además, esto requiere rigurosidad, tanto desde el punto de vista de la comprensión y citación, como del mantenimiento de unos objetivos claros que no conlleven un proceso de divagación entre un conjunto de temas muy dispersos entre sí.

      \paragraph{}
      También se pueden incluir dentro de estos sub-objetivos la necesidad de mantener un nivel personal de disciplina apropiado para la realización del trabajo, puesto que tanto la envergadura como la cantidad de tiempo para llevarlo a cabo son de gran tamaño. Sin el apropiado orden esto puede generar problemas derivados de dejar el trabajo para última hora, por tanto, se ha creído conveniente incluir dicho orden como sub-objetivo.

      \paragraph{}
      En cuanto a la implementación a realizar, también se ha creído conveniente el cumplimiento de una serie de objetivos a nivel de calidad del software. El primero de ellos es el apropiado funcionamiento de la implementación, que debido a su importancia debería incluso presuponerse. Además, se ha creído conveniente el diseño de la implementación como un módulo auto-contenido que tan solo requiera un conjunto reducido de dependencias para facilitar su instalación y distribución. En cuanto al código, se ha creído conveniente prestar especial atención a la parte de claridad del código fuente, de tal manera que la legibilidad del mismo sea sencilla. También se ha fijado como objetivo la generación de un conjunto de pruebas que permitan validar el funcionamiento del mismo, así como la inclusión de un sistema de auto-documentación que permita a otros usuarios utilizar la implementación siguiendo las indicaciones, sin necesidad de tener que comprender el código fuente subyacente.

      \paragraph{}
      En las secciones posteriores se realiza una visión superficial acerca de las diferentes disciplinas que abarca el ámbito del conocimiento del \emph{Big Data} así como los grafos de tamaño masivo.

    \section{Big Data}
    \label{sec:introduction_big_data}

      \paragraph{}
      El procesamiento de cantidades masivas de información presenta un gran reto a nivel computacional, debido a un elevado coste originado por el gran tamaño en la entrada. Para solventar dicha problemática, se prefieren algoritmos que posean un orden de complejidad sub-lineal ($o(N)$) sobre todo en espacio. Dichas técnicas se llevan a cabo sobre paradigmas de computación paralela, lo que permite aprovechar en mayor medida las restricciones a nivel de hardware.

      \subsection{Algoritmos para Streaming}

        \paragraph{}
        Los \emph{Algoritmos para Streaming} se caracterizan por procesar las instancias del conjunto de datos secuencialmente e imponen como restricción que el orden de dicha operación sea irrelevante para el resultado final. La ventaja que presentan respecto de otras alternativas en tiempo real, como los \emph{Algoritmos Online}, es la utilización de propiedades estadísticas (se enmarcan por tanto, dentro de los \emph{Algoritmos Probabilísticos}) para reducir su coste, lo que por contra, añade una determinada tasa de error. El descubrimiento de métodos altamente eficientes para estimar los \emph{Momentos de Frecuencia} ha marcado un gran hito dentro de esta categoría algorítmica.

      \subsection{Estrategias de Sumarización}

        \paragraph{}
        Para reducir el coste derivado de la obtención de resultados valiosos sobre conjuntos masivos de datos, es necesario apoyarse en diferentes estrategias que los sinteticen, de manera que el coste de procesamiento a partir de estas estructuras se convierta en una tarea mucho más asequible. Se utilizan sobre conjuntos de datos de distinta índole, como \emph{streamings en tiempo real}, \emph{bases de datos estáticas} o \emph{grafos}. Existen distintas técnicas como \emph{Sampling}, \emph{Histogram}, \emph{Wavelets} o \emph{Sketch}. A continuación se realiza una breve descripción acerca de esta última técnica.

        \subsection{Sketch}

          \paragraph{}
          Son estructuras de datos que se basan en la idea de realizar sobre cada una de las instancias del conjunto de datos la misma operación (lo que permite su uso en entornos tanto estáticos como dinámicos) para recolectar distintas características. Destacan los \emph{Sketches lineales}, que permiten su procesamiento de manera distribuida. Para mantener estas estructuras se utilizan \emph{Algoritmos para Streaming}, puesto que se encajan perfectamente en el contexto descrito. Los \emph{Sketches} permiten realizar distintas preguntas sobre propiedades estadísticas referentes al conjunto de datos. Los ejemplos más destacados son: \emph{Count-Sketch}, \emph{CountMin-Sketch}, \emph{AMS Sketch}, \emph{HyperLogLog}, etc.

      \subsection{Redución de la Dimensionalidad}

        \paragraph{}
        Los algoritmos que utilizan técnicas de reducción de dimensionalidad se basan en la intuición originada a partir del lema de \emph{Johnson–Lindenstrauss}, que demuestra la existencia de funciones para la redución de la dimensión espacial con un ratio de distorsión acotado. Estas técnicas son utilizadas en algoritmos para la \emph{búsqueda de los vecinos más cercanos}, la \emph{multiplicación aproximada de matrices} o el aprendizaje mediante \emph{Manifold Leaning}.

      \subsection{Paralelización a gran Escala}

        \paragraph{}
        El paradigma de alto nivel sobre el que se lleva a cabo el procesamiento de conjuntos de datos de gran escala se apoya fuertemente en técnicas de paralelización. La razón se debe al elevado tamaño de la entrada, que no permite su almacenamiento en la memoria de un único sistema.

        \subsection{Modelo MapReduce}

          \paragraph{}
          El modelo \emph{MapReduce} ha sufrido un crecimiento exponencial en los últimos años debido a su alto grado de abstracción, que oculta casi por completo cuestiones relacionadas con la implementación de bajo nivel al desarrollador, y su capacidad para ajustarse a un gran número de problemas de manera eficiente.

      \subsection{Técnicas de Minería de Datos}

        \paragraph{}
        Una de las razones por las cuales es necesaria la investigación de nuevos algoritmos de carácter sub-lineal es la necesidad de obtención de información valiosa a partir de conjuntos masivos de datos. A este fenómeno se le denomina \emph{Minería de Datos}. Existen dos grandes categorías denominadas: \emph{Clasificación} (determinar una clase de pertenencia) y \emph{Regresión} (determinar un valor continuo). Para ello, se utilizan distintas técnicas como: \emph{Árboles de Decisión}, \emph{Métodos Bayesianos}, \emph{Redes Neuronales}, \emph{Máquinas de Vector Soporte}, \emph{Manifold Leaning}, etc.

    \section{Grafos}
    \label{sec:introduction_graphs}

      \paragraph{}
      Los grafos representan un método de representación para la resolución de problemas desde una perspectiva matemática mediante la modelización de una red de objetos que se relacionan a través de interconexiones. Esta abstracción, que deja de lado el contexto de aplicación para basarse únicamente en las relaciones y la estructura de las mismas, permite diseñar algoritmos de manera más simple al tener en cuenta únicamente la información necesaria para resolver el problema.

      \paragraph{}
      Los problemas referidos a grafos han sido ampliamente en la literatura desde hace mucho tiempo. Sin embargo, en los últimos años se ha producido un elevado crecimiento de distintas técnicas que permiten resolver estos, de tal manera que el coste sea más reducido. Esto genera una reducción de tiempo y espacio en su resolución a costa de la inclusión de una determinada tasa de error.

      \paragraph{}
      Una propuesta interesante es la generación de un sub-grafo de menor tamaño que mantenga las propiedades a nivel de estructura lo más semejantes posibles respecto del grafo sobre el cual se pretende resolver el problema en cuestión. Existen distintas técnicas para esta labor, conocidas como \emph{Spanners} y \emph{Sparsifiers}. Los últimos trabajos de investigación relacionados con el tema pretenden diseñar algoritmos que apliquen dichas técnicas siguiendo las mismas ideas que los \emph{Sketches} para el caso de valores numéricos.

      \paragraph{}
      Un algoritmo basado en conceptos de estadística y aplicado a grafos de tamaño masivo es el algoritmo \emph{PageRank}, el cual genera un ranking de importancia entre los puntos de la red, basándose únicamente en la estructura de interconexiones de la misma. Este ranking está íntimamente relacionado con conceptos de probabilidad como las \emph{Cadenas de Markov}.

    \paragraph{}
    Debido a la ingente cantidad de tiempo necesaria para realizar un trabajo de investigación que contuviera descripciones acerca de todos los conceptos relacionados con el \emph{Big Data} que se han resumido en las secciones posteriores, se han seleccionado un sub-conjunto de ellas. Por tanto, a continuación se indica cómo se organiza el resto del documento: en el capítulo \ref{chap:streaming} se realiza una descripción acerca de los \emph{Algoritmos para Streaming}. Seguidamente, en el capítulo \ref{chap:summaries} se indican distintas \emph{Estrategias de Sumarización}. A continuación, se cambia de perspectiva para hablar de \emph{Grafos} en el capítulo \ref{chap:graphs}. Después, se describe el algoritmo \emph{PageRank} en detalle en el capítulo \ref{chap:pagerank}. Por último, se describen distintos detalles de implementación, así como de los resultados obtenidos y se realiza una conclusión acerca del trabajo realizado en el capítulo \ref{chap:implementation}.

    \paragraph{}
    De manera adicional, también se han incluido distintos anexos: En el anexo \ref{chap:methodology} se indica la metodología de trabajo seguida durante el desarrollo del proyecto. En el anexo \ref{chap:how_it_was_build} se indica cómo ha sido construido este documento mediante la herramienta \LaTeX. Por último, se ha incluido una guía de usuario de la implementación en el anexo \ref{chap:user_guide}.

  \chapter{Algoritmos para Streaming}
  \label{chap:streaming}

    \section{Introducción}
    \label{sec:streaming_intro}

      \paragraph{}
      Los \emph{algoritmos para streaming} son una estrategia de diseño algorítmica basada en el procesamiento secuencial de la entrada, lo cual encaja en marcos en los cuales los datos tienen una componente dinámica. Además, este contexto se amolda perfectamente en los casos en que el tamaño de los mismos es tan elevado que no es posible mantenerlos de manera completa en la memoria del sistema. Dicha problemática es precisamente la que surge con el denominado \emph{Big Data}, que al trabajar con conjuntos masivos de datos en el orden de gigabytes, terabytes o incluso petabytes, no se pueden procesar utilizando estrategias clásicas que presuponen que se dispone de todos los datos de manera directa e inmediata.

      \paragraph{}
      Por tanto, en dicho contexto se presupone un espacio de almacenamiento en disco de tamaño infinito, mientras que se restringe el espacio de trabajo o memoria a un tamaño limitado y mucho menor que el del conjunto de datos con el que se trabaja. Mediante estas presuposiciones fijadas a priori cobra especial importancia el diseño de algoritmos en el modelo en streaming, que tratan de reducir el número de peticiones al espacio de almacenamiento o disco, lo cual genera una gran reducción en el tiempo de procesamiento.

      \paragraph{}
      Además, bajo este modelo es trivial realizar una extensión de los algoritmos y técnicas para su uso en entornos dinámicos en los cuales el conjunto de datos varía con respecto del tiempo, añadiendo y eliminando nuevos datos. Debido a estas características la investigación en el campo de los \emph{algoritmos para streaming} a cobrado una gran importancia. En este capítulo se pretende realizar una introducción conceptual acerca de los mismos, además de realizar una exposición acerca de los algoritmos más relevantes en dicha área.

      \subsection{Computación en Tiempo Real}
      \label{sec:realtime_computing}
        \paragraph{}
        El primer concepto del que hablaremos es \textbf{Computación en Tiempo Real}, que tal y cómo describen Shin y Ramanathan \cite{259423} se carácteriza por tres términos que se exponen a continuación:

        \begin{itemize}

          \item \textbf{Tiempo}\emph{(time)}: En la disciplina de \emph{Computación en Tiempo Real} el tiempo de ejecución de una determinada tarea es especialmente crucial para garantizar el correcto desarrollo del cómputo, debido a que se asume un plazo de ejecución permitido, a partir del cual la solución del problema deja de tener validez. Shin y Ramanathan\cite{259423} diferencian entre tres categorías dentro de dicha restricción, a las cuales denominan \emph{hard}, \emph{firm} y \emph{soft}, dependiendo del grado de relajación de la misma.

          \item \textbf{Confiabilidad}\emph{(correctness)}: Otro de los puntos cruciales en un sistema de \emph{Computación en Tiempo Real} es la determinación de una unidad de medida o indicador acerca de las garantías de una determinada solución algorítmica para cumplir lo que promete de manera correcta en el tiempo esperado.

          \item \textbf{Entorno}\emph{(environment)}: El último factor que indican Shin y Ramanathan\cite{259423} para describir un sistema de \emph{Computación en Tiempo Real} es el entorno del mismo, debido a que este condiciona el conjunto de tareas y la periodicidad en que se deben llevar a cabo. Por esta razón, realizan una diferenciación entre:
          \begin{enumerate*}[label=\itshape\alph*\upshape)]
  					\item tareas periódicas \emph{periodic tasks} las cuales se realizan secuencialmente a partir de la finalización de una ventana de tiempo, y
  					\item tareas no periódicas \emph{periodic tasks} que se llevan a cabo debido al suceso de un determinado evento externo.
  				\end{enumerate*}

        \end{itemize}

      \subsection{Problemas Dinámicos}
      \label{sec:dynamic_problems}

        \paragraph{}
        Una vez completada la descripción acerca de lo que se puede definir como \emph{Computación en Tiempo Real}, conviene realizar una descripción desde el punto de vista de la \emph{teoría de complejidad computacional}. Para definir este tipo de problemas, se utiliza el término \emph{problemas dinámicos}, los cuales consisten en aquellos en los cuales es necesario recalcular su solución conforme el tiempo avanza debido a variaciones en los parámetros de entrada del problema (Nótese que dicho término no debe confundirse con la estrategia de \emph{programación dinámica} para el diseño de algoritmos).

        \paragraph{}
        Existen distintas vertientes dependiendo del punto de vista desde el que se estudien, tanto de la naturaleza del problema (soluciones dependientes temporalmente unas de otras o soluciones aisladas) como de los parámetros de entrada (entrada completa en cada nueva ejecución o variación respecto de la anterior). Los \emph{Algoritmos para Streaming} están diseñados para resolver \emph{problemas dinámicos}, por lo que en la sección \ref{sec:streaming_model}, se describe en profundidad el modelo en que se enmarcan.

        \paragraph{}
        A continuación se indican los principales indicadores utilizados para describir la complejidad de una determinada solución algorítmica destinada a resolver un problema de dicha naturaleza:

        \begin{itemize}
          \item Espacio: Cantidad de espacio utilizado en memoria durante la ejecución del algoritmo.
          \item Inicialización: Tiempo necesario para la inicialización del algoritmo.
          \item Procesado: Tiempo necesario para procesar una determinada entrada.
          \item Consulta: Tiempo necesario para procesar la solución a partir de los datos de entrada procesados hasta el momento.
        \end{itemize}

      \subsection{Algoritmos Online vs Algoritmos Offline}

        \paragraph{}
        Una vez descrita la problemática de \emph{Computación en Tiempo Real} en la sección \ref{sec:realtime_computing} y la categoría de \emph{Problemas Dinámicos} en la sección \ref{sec:dynamic_problems}, en esta sección se pretende ilustrar la diferencia entre los \emph{Algoritmos Online} y los \emph{Algoritmos Offline}. Para ello, se ha seguido la diferenciación propuesta por Karp \cite{Karp:1992:OAV:645569.659725}, en la cual se plantea el problema de la siguiente manera (Se utilizará la misma notación que sigue Muthukrishnan\cite{Muthukrishnan:2005:DSA:1166409.1166410} para tratar mantener la consistencia durante todo el documento): Sea $A$ el conjunto de datos o eventos de entrada, siendo cada $A[i]$ el elemento \emph{$i$-ésimo} del conjunto, y que en el caso de los \emph{Algoritmos Online} supondremos que es el elemento recibido en el instante \emph{i}. A continuación se muestran las características de cada subgrupo:

        \begin{itemize}
          \item \textbf{Algoritmos Offline}: Esta categoría contiene todos los algoritmos que realizan el cómputo suponiendo el acceso a cualquier elemento del conjunto de datos $A$ durante cualquier momento de su ejecución. Además, en esta categoría se impone la restricción de que el $A$ debe ser invariante respecto del tiempo, lo que conlleva que para la adaptación del resultado a cambios en la entrada, este tenga que realizar una nueva ejecución desde su estado inicial. Nótese por tanto, que dentro de este grupo se engloba la mayoría de algoritmos utilizados comúnmente.

          \item \textbf{Algoritmos Online}: Son aquellos que calculan el resultado a partir de una secuencia de sucesos $A[i]$, los cuales generan un resultado dependiente de la entrada actual, y posiblemente de las anteriores. A partir de dicha estrategia, se añade una componente dinámica, la cual permite que el tamaño del conjunto de datos de entrada $A$ no tenga impuesta una restricción acerca de su longitud \emph{a-priori}. Por contra, en este modelo no se permite conocer el suceso $A[i+1]$ en el momento $i$.  Esto encaja perfectamente en el modelo que se describirá en la sección \ref{sec:streaming_model}.

        \end{itemize}

        \paragraph{}
        Según la diferenciación que se acaba de indicar, estas dos estrategias de diseño de algoritmos encajan en disciplinas distintas, teniendo una gran ventaja a nivel de eficiencia en el caso estático los \emph{Algoritmos Offline}, pero quedando prácticamente in-utilizables cuando la computación es en tiempo real, donde es mucho más apropiado el uso de estrategias de diseño de \emph{Algoritmos Online}.

        \paragraph{}
        Como medida de eficiencia para los \emph{Algoritmos Online}, Karp \cite{Karp:1992:OAV:645569.659725} propone el \textbf{Ratio Competitivo}, el cual se define como la cota inferior del coste de cualquier nueva entrada con respecto de la que tiene menor coste. Sin embargo, dicha medida de eficiencia no es comúnmente utilizada en el caso de los \emph{Algoritmos para Streaming} por la componente estocástica de los mismos, para los cuales son más apropiadas medidas probabilistas. A continuación se describen las ventajas de estos respecto de su vertiente determinista.

      \subsection{Algoritmos Probabilistas}

        \paragraph{}
        Los \emph{Algoritmos Probabilistas} son una estrategia de diseño que emplea en un cierto grado de aleatoriedad en alguna parte de su lógica. Estos utilizan distribuciones uniformes de probabilidad para tratar de conseguir un incremento del rendimiento en su caso promedio. A continuación se describen los dos tipos de algoritmos probabilísticos según la clasificación realizada por Babai \cite{Babai79monte-carloalgorithms}:

        \begin{itemize}

          \item \textbf{Algoritmos Las Vegas}: Devuelven un resultado incorrecto con una determinada probabilidad, pero avisan del resultado incorrecto cuando esto sucede. Para contrarrestar este suceso basta con llevar a cabo una nueva ejecución del algoritmo, lo cual tras un número indeterminado de ejecuciones produce un resultado válido.

          \item \textbf{Algoritmos Monte Carlo}: Fallan con un cierto grado de probabilidad, pero en este caso no avisan del resultado incorrecto. Por lo tanto, lo único que se puede obtener al respecto es una indicador de la estimación del resultado correcto hacia la que converge tras varias ejecuciones. Además, se asegura una determinada cota del error $\epsilon$, que se cumple con probabilidad $\delta$.

        \end{itemize}

        \paragraph{}
        La razón anecdótica por la cual Babai \cite{Babai79monte-carloalgorithms} decidió denominar dichas categorías de algoritmos de esta manera se debe a lo siguiente (teniendo en cuenta el contexto de lengua inglesa): cuando se va a un casino en \emph{Las Vegas} y se realiza una apuesta el \emph{croupier} puede decir si se ha ganado o perdido porque habla el mismo idioma. Sin embargo, si sucede la misma situación en \emph{Monte Carlo}, tan solo se puede conocer una medida de probabilidad debido a que en este caso el \emph{croupier} no puede comunicarlo por la diferencia dialéctica.

      \subsection{Algoritmos Online Probabilistas vs Deterministas}

        \paragraph{}
        La idea subyacente acerca del diseño de los \emph{Algoritmos Online} es la mejora de eficiencia con respecto de sus homónimos estáticos cuando el conjunto de valores de entrada es dependiente de los resultados anteriores. Sin embargo, existen casos en que la frecuencia de ejecución del algoritmo, debido a una alta tasa de llegada de valores en la entrada, las soluciones deterministas se convierten en alternativas poco escalables.

        \paragraph{}
        Dicha problemática se ha incrementado de manera exponencial debido al avance tecnológico y la gran cantidad de información que se genera en la actualidad, que sigue creciendo a un ritmo desorbitado. Este fenómeno ha convertido en algo necesario el diseño de estrategias basadas en técnicas probabilísticas que reduzcan en gran medida el coste computacional que como consecuencia eliminan el determinismo de la solución.

    \section{Modelo en Streaming}
    \label{sec:streaming_model}

      \paragraph{}
      En esta sección se describen los aspectos formales del \emph{Modelo en Streaming}. Para ello se ha seguido la representación definida por Muthukrishnan \cite{Muthukrishnan:2005:DSA:1166409.1166410}. Lo primero por tanto, es definir un flujo de datos o \emph{Data Stream} como una \say{secuencia de señales digitalmente codificadas utilizadas para representar una transmisión de información} \cite{ITS-def-data-stream}. Muthukrishnan \cite{Muthukrishnan:2005:DSA:1166409.1166410} hace una aclaración sobre dicha definición y añade la objeción de que los datos de entrada deben tener un ritmo elevado de llegada. Debido a esta razón existe complejidad a tres niveles:

      \begin{itemize}

        \item \textbf{Transmisión}: Ya que debido a la alta tasa de llegada es necesario diseñar un sistema de interconexiones que permita que no se produzcan congestiones debido a la obtención de los datos de entrada.

        \item \textbf{Computación}: Puesto que la tarea de procesar la gran cantidad de información que llega por unidad de tiempo produce cuellos de botella en el cálculo de la solución por lo que es necesario implementar técnicas algorítmicas con un reducido nivel de complejidad computacional para contrarrestar dicha problemática.

        \item \textbf{Almacenamiento}: Debido a la gran cantidad de datos que se presentan en la entrada, deben existir técnicas que permitan almacenar dicha información de manera eficiente. Esto puede ser visto desde dos puntos de vista diferences: \begin{enumerate*}[label=\itshape\alph*\upshape)]
          \item tanto desde el punto de vista del espacio, tratando de minimizar el tamaño de los datos almacenados, maximizando la cantidad de información que se puede recuperar de ellos,
          \item como desde el punto de vista del tiempo necesario para realizar operaciones de búsqueda, adicción, eliminación o edición.
        \end{enumerate*}. Además, se debe prestar especial atención en la información que se almacena, tratando de reducirla al máximo prescindiendo de datos redundantes o irrelevantes.

      \end{itemize}

      \subsection{Formalismo para Streaming}
      \label{sec:streaming_formalism}

        \paragraph{}
        Una vez descritos los niveles de complejidad a los que es necesario hacer frente para abordar problemas en el \emph{Modelo en Streaming}, se realiza una descripción de los distintos modelos que propone Muthukrishnan \cite{Muthukrishnan:2005:DSA:1166409.1166410} en los apartados \ref{sec:streaming_time_series}, \ref{sec:streaming_cash_register} y \ref{sec:streaming_turnstile}. La especificación descrita en dichos apartados será seguida durante el resto del capítulo. Para ello nos basaremos en el siguiente formalismo:

        \paragraph{}
        Sea $a_1 ,a_2 ,... ,a_t,... $ un flujo de datos de entrada (\emph{Input Stream}), de tal manera que cada elemento debe presentar un orden de llegada secuencial respecto de $t \in \mathbb{M}$. Esto también se puede ver de la siguiente manera: el elemento siguiente a la llegada de $a_{t-1}$ debe ser $a_{t}$ y, por inducción, el próximo será $a_{t+1}$. Es necesario aclarar que $t$ no se refiere a unidades propiamente temporales, sino a la posición en la entrada.

        \begin{equation}
        \label{eq:streaming_A_function}
          \boldsymbol{A}_t:[1...N] \rightarrow \mathbb{R}^2
        \end{equation}

        \paragraph{}
        El siguiente paso para describir el formalismo es añadir la función $\boldsymbol{A}_t$, cuyo dominio e imagen se muestran en la ecuación \eqref{eq:streaming_A_function}. Esta función tiene distintas interpretaciones dependientes del \emph{Modelo en Streaming} bajo el cual se esté trabajando en cada caso, pero la idea subyacente puede resumirse asumiendo que la primera componente almacena el valor, mientras que la segunda almacena el número de ocurrencias de dicho valor. Algo común a todos ellos es la variable $t$, que se corresponde con resultado de la función en el instante de tiempo $t$. Por motivos de claridad, en los casos en que nos estemos refiriendo un único momento, dicha variable será obviada en la notación.

      \subsection{Modelo de Serie Temporal}
      \label{sec:streaming_time_series}

        \paragraph{}
        El \emph{Modelo de Serie Temporal} o \emph{Time Series Model} se refiere, tal y como indica su nombre, a una serie temporal, es decir, modeliza los valores que toma la variable $i$ respecto de $t$, codificados en el modelo como $a_t = (i,1)$. Nótese que se utiliza el valor $1$ en la segunda componente de $a_t$, la razón de ello se debe a la definición de la imagen de $\boldsymbol{A}$ en la ecuación \eqref{eq:streaming_A_function}. A pesar de ello, dicho campo es irrelevante en este modelo, por lo que se podría haber escogido cualquier otro arbitrariamente. La razón por la cual se ha utilizado el valor $1$ ha sido el refuerzo de la idea de que en este caso, el valor que toma $a_t$ en un determinado momento, no volverá a variar su valor, pero quedará obsoleto con la llegada de $a_{t+1}$.

        \paragraph{}
        El modelo se describe de manera matemática mediante la función $\boldsymbol{A}$, tal y como se ilustra en la ecuación \eqref{eq:streaming_time_series}. Textualmente, esto puede traducirse diciendo que la función $\boldsymbol{A}$ representa una estructura de datos que almacena el valor de todos los elementos recibidos en la entrada hasta el instante de tiempo $t$, es decir, actúa como un historial. Un ejemplo de este modelo son los valores en bolsa que toma una determinada empresa a lo largo del tiempo.

        \begin{equation}
  			\label{eq:streaming_time_series}
  				\boldsymbol{A}(t) = a_t
  			\end{equation}

      \subsection{Modelo de Caja Registradora}
      \label{sec:streaming_cash_register}

        \paragraph{}
        El \emph{Modelo de Caja Registradora} o \emph{Cash Register Model} consiste en la recepción de incrementos de un determinado valor $i$. El nombre del modelo hace referencia al funcionamiento de una caja registradora (suponiendo que el pago se realiza de manera exacta), que recibe billetes o monedas de tipos diferentes de manera secuencial.

        \paragraph{}
        Para describir dicho modelo, previamente hay que realizar una aclaración acerca del contenido del elemento $a_t = (i, I_t)$, de manera que $i$ representa el valor recibido, mientras que $I_t \geq 0$ indica el incremento en el instante $t$. Una vez aclarada esta definición, la función $\boldsymbol{A}_{t}$, se construye tal y como se indica en la ecuación \eqref{eq:streaming_cash_register}.

        \begin{equation}
  			\label{eq:streaming_cash_register}
  				\boldsymbol{A}_{t}(i) = {A}_{t-1}(i) + I_{t}
  			\end{equation}

        \paragraph{}
        El \emph{Modelo de Caja Registradora} es ampliamente utilizado en la formalización de problemas reales debido a que muchos fenómenos siguen esta estructura. Un ejemplo de ello es el conteo de accesos a un determinado sitio web, los cuales se corresponden con incrementos $I_t$, en este caso de carácter unitario realizados por un determinado usuario $i$ en el momento $t$.

      \subsection{Modelo de Molinete}
      \label{sec:streaming_turnstile}

        \paragraph{}
        El \emph{Modelo de Molinete} o \emph{Turnstile Model} se corresponde con el caso más general, en el cual no solo se permiten incrementos, sino que también se pueden realizar decrementos en la cuenta. El nombre que se le puso a este modelo se debe al funcionamiento de los molinetes que hay en las estaciones de metro para permitir el paso a los usuarios, que en la entrada incrementan la cuenta del número de personas, mientras que en la salida los decrementan. La relajación originada por la capacidad de decremento ofrece una mayor versatilidad, que permite la contextualización de un gran número de problemas en este modelo. Por contra, añade un numerosas complicaciones a nivel computacional, tal y como se verá a lo largo del capítulo.

        \paragraph{}
        Al igual que ocurre en el caso anterior, para describir este modelo, lo primero es pensar en la estructura de los elementos en la entrada, que están formados por $a_t = (i, U_t)$, algo muy semejante a lo descrito en el \emph{Modelo de Caja Registradora}. Sin embargo, en este caso $U_t$ no tiene restricciones en su imagen, sino que puede tomar cualquier valor tanto positivo como negativo, lo cual añade el concepto de decremento. La construcción de la función $\boldsymbol{A}_{t}$ se describe en la ecuación \eqref{eq:streaming_turnstile}.

        \begin{equation}
        \label{eq:streaming_turnstile}
          \boldsymbol{A}_{t}(i) = {A}_{t-1}(i) + U_{t}
        \end{equation}

        \paragraph{}
        Muthukrishnan \cite{Muthukrishnan:2005:DSA:1166409.1166410} hace una diferenciación dentro de este modelo dependiendo del nivel de exigencia que se le pide al modelo, se dice que es un \emph{Modelo de Molinete estricto} cuando se añade la restricción $\forall i, \forall t \ \boldsymbol{A}_{t}(i) \geq 0$, mientras que se dice que es un \emph{Modelo de Molinete relajado} cuando dicha restricción no se tiene en cuenta.

        \paragraph{}
        Un ejemplo de este modelo es el conteo del número de usuarios que están visitando un determinado sitio web, tomando $U_t$ el valor $1$ en el caso de una nueva conexión y $-1$ en el caso de de una desconexión. En este ejemplo el valor $i$ representa una determinada página dentro del sitio web.

    \section{Estructura básica}
    \label{sec:streaming_structure}

      \paragraph{}
      Puesto que la naturaleza intríseca de los \emph{Algoritmos para Streaming} hace que procesen los elementos de entrada según van llegando, esto presenta peculiaridades con respecto de otras categorías algorítmicas más asentadas y utilizadas en la actualidad. Por tanto, primero se describirá la estructura básica que siguen los algoritmos más comúnmente utilizados para después mostrar la estrategia seguida en el caso de Streaming.

      \paragraph{}
      Los algoritmos clásicamente estudiados para resolver la mayoría de problemas se basan la idea de funciones matemáticas. Es decir, se les presenta un conjunto de valores en la entrada, y a partir de ellos, realizan una determinada transformación sobre los datos, que genera como resultado una salida. Nótese que esta idea no impone ninguna restricción acerca de lo que puede suceder en dicho proceso, es decir, no se restringe el uso de estructuras de datos auxiliares o técnicas similares.

      \paragraph{}
      Esta visión no se enmarca correctamente en el contexto de los \emph{Algoritmos para Streaming}. La razón se debe a que la entrada no es propiamente un conjunto de datos, sino que se refiere a un flujo en sí mismo. Esta característica tiene como consecuencia que en un gran número de ocasiones ya no sea necesario obtener los resultados tras cada llamada al algoritmo, ya que estos podrían carecer de interés o requerir un sobrecoste innecesario. Por lo tanto, el concepto de función matemática pierde el sentido en este caso, ya que estas exigen la existencia de un valor como resultado.

      \paragraph{}
      Un concepto más acertado para modelizar un \emph{Algoritmo para Streaming} podría ser lo que en los lenguajes de programación derivados de \emph{Fortran} se denomina subrutina, es decir, una secuencia de instrucciones que realizan una tarea encapsulada como una unidad. Sin embargo, para poder describir correctamente la estructura de un \emph{Algoritmo para Streaming} hace falta algo más. La razón de ello es que a partir de dicho modelo de diseño no sería posible realizar peticiones sobre el resultado calculado, es decir, sería una estrategia puramente procedural. Para corregir dicha problemática surge el concepto de consulta o \emph{query}. A través de dicha idea se pretende representar la manera de obtener un resultado a partir del cómputo realizado hasta el momento actual.

      \paragraph{}
      En resumen, con dicha estrategia de diseño se consigue separar la parte de procesado de la entrada de la parte de consulta del resultado, lo cual proporciona amplias ventajas para el modelo seguido por los \emph{Algoritmos para Streaming}. Sin embargo, dicha estrategia produce un sobrecoste espacial con respecto del modelo de algoritmo clásico. Este se debe a la necesidad de mantener una estructura de datos en la cual se almacenen los resultados parciales referentes al flujo de entrada.

      \paragraph{}
      Los algoritmos \emph{Algoritmos para Streaming} se componen por tanto de algoritmo de procesamiento del flujo de datos, una estructura de datos que almacena dichos resultados, y por último, un algoritmo de procesamiento de la consulta o \emph{query} necesaria para obtener los resultados requeridos. a continuación se dividen las fases para el funcionamiento de un algoritmo de dichas características.

      \begin{itemize}

        \item \textbf{Inicialización}: En esta fase se llevan a cabo el conjunto de tareas necesarias para inicializar la estructura de datos que actuará como almacen de información durante el procesamiento del flujo de datos de entrada. Generalmente esto consite en el proceso de reservar memoria, inicializar a un valor por defecto la estructura de datos, etc. Sin embargo, existen técnicas más sofisticadas que requieren de una mayor carga computacional en esta fase.

        \item \textbf{Procesado}: Se corresponde con el procesamiento del flujo de datos de manera secuencial. La idea subyacente en esta fase es la de realizar una determinada operación sobre la estructura de datos y el elemento de entrada actual, de manera que se lleve a cabo una actualización sobre la misma. Nótese en que la manera en que se manipula dicha estructura de datos condiciona en gran medida el conjunto de peticiones que se podrán realizar sobre ella.

        \item \textbf{Consulta}: La fase de consulta se diferencia con respecto de la anterior por ser de carácter consultivo. Con esto nos estamos refiriendo a que dicha tarea no modifica el estado actual de la estructura de datos, sino que recoge información de la misma, que posiblemente transforma mediante alguna operación, para después obtener un valor como resultado de dicha petición.

      \end{itemize}

    \section{Medidas de Análisis y Conceptos Matemáticos}
    \label{sec:streaming_analysis}

      \paragraph{}
      Los \emph{Algoritmos para Streaming} se caracterizan por utilizar propiedades estadísticas en alguna parte (generalmente en el procesado) de su cómputo para obtener la solución con un menor coste computacional. En este caso, el coste que se pretende minimizar es el referido al espacio necesario para almacenar la estructura de datos auxiliar. Tal y como se ha dicho anteriormente, la razón de ello es debida a que se presupone un conjunto masivo de datos en la entrada, por lo que se pretende que el orden de complejidad espacial respecto de la misma sea de carácter sub-lineal ($o(N)$).

      \paragraph{}
      El objetivo es encontrar soluciones con un intervalo de error acotado que permitan llegar a la solución en un orden espacial de complejidad logarítmica ($O(log(N))$). Sin embargo, existen ocasiones en que no es posible llegar a una solución en dicho orden de complejidad, como es el caso de \emph{Algoritmos para Streaming} aplicados a problemas de \emph{Grafos}, en los cuales se relaja dicha restricción a un orden de complejidad \emph{poli-logarítmico} ($O(polylog(N))$).

      \paragraph{}
      El orden de complejidad \emph{poli-logarítmico} engloba el conjunto de funciones cuyo orden de complejidad presenta un crecimiento acorde a una función polinomial formada logaritmos. Matemáticamente esto se modeliza a través de la ecuación \eqref{eq:polylog-complexity}

      \begin{equation}
      \label{eq:polylog-complexity}
        a_{k}\log ^{k}(N)+\cdots +a_{1}\log(N)+a_{0} = O(polylog(N)) \in o(N).
      \end{equation}

      \paragraph{}
      En esta sección se muestran distintas estrategias para poder llevar a cabo la demostración de pertenencia a un determinado orden de complejidad de un \emph{Algoritmo para Streaming}. Debido a la elevada base estadística que requieren dichas demostraciones, a continuación se definen algunos conceptos básicos relacionadas con estimadores estadísticos, para después realizar una breve demostración acerca de distintas cotas de concentración de valores en una distribución en las sub-secciones \ref{sec:markov_inequality}, \ref{sec:chebyshev_inequality} y \ref{sec:chernoff_inequality}. Las definiciones que se exponen a continuación han sido extraídas de los apuntes del curso sobre \emph{Randomized Algorithms} \cite{aspnes2014notes} impartido por Aspnes en la \emph{Universidad de Yale} así como las de la asignatura de \emph{Estadística}\cite{estadistica2016notes} impartida en el Grado de Ingeniería Informática de la \emph{Universidad de Valladolid}.

      \subsection{Conceptos básicos de Estadística}
      \label{sec:basic_statistics}

        \paragraph{}
        Denotaremos como $x_1$ a una observación cualquiera contenida en el espacio de todas los posibles. Al conjunto de todas las observaciones posibles lo denotaremos como $\Omega$ y lo denominaremos espacio muestral, por lo tanto, $x \in \Omega$. Este concepto se puede entender de manera más sencilla mediante el siguiente ejemplo. Supongamos el lanzamiento de una moneda, que como resultado puede tomar los valores cara o cruz. Definiremos entonces $x_1 = \text{cara}$ y $x_2 = \text{cruz}$ como los sucesos posibles de lanzar una moneda. Por tanto el espacio $\Omega$ se define como $\Omega = \{x_1, x_2\} = \{\text{cara}, \text{cruz}\}$.

        \paragraph{}
        El siguiente paso es definir el concepto de \textbf{Variable Aleatoria}, que representa una función que mapea la realización de un determinado suceso sobre el espacio $\Omega$. Dicha función se denota con letras mayúsculas y puesto que sus parámetros de entrada son desconocidos, estos se ignoran en la notación. Por tanto denotaremos las variables aleatorias como ($\boldsymbol{E}, \boldsymbol{X}, \boldsymbol{Y}, \text{etc.}$). Para la variable aleatoria $\boldsymbol{X}$, sean $x_1, x_2, ..., x_i,...$ cada una de las observaciones posibles. Siguiendo el ejemplo anterior, se puede modelizar el lanzamiento de una moneda como $X$. Nótese por tanto, que una variable aleatoria puede definirse de manera textual como la modelización del resultado de un suceso \emph{a-priori} desconocido.

        \paragraph{}
        Definiremos probabilidad como la medida de certidumbre asociada a un suceso o evento futuro, expresada como un valor contenido en el intervalo $[0,1]$, tomando el valor $0$ un suceso imposible y $1$ un suceso seguro. La notación seguida para representar esto será $Pr[\boldsymbol{X} = x_i]$. Suponiendo la equi-probabilidad en el ejemplo de la moneda, podemos definir sus valores de probabilidad como $Pr[\boldsymbol{X} = \text{cara}] = \tfrac{1}{2}$ y $Pr[\boldsymbol{X} = \text{cruz}] = \tfrac{1}{2}$

        \paragraph{}
        Una vez descritos estos conceptos simples, a continuación hablaremos sobre distintos conceptos estadísticos utilizados en el análisis de algoritmos probabilísticos tales como \emph{Esperanza}, \emph{Varianza}, \emph{Variables Independientes} y \emph{Probabilidad Condicionada}.

        \paragraph{}
        Denominaremos \textbf{Esperanza Matemática} al valor medio o más probable que se espera que tome una determinada variable aleatoria. La modelización matemática de dicho concepto se muestra en la ecuación \eqref{eq:expectation}. Además, la esperanza matemática es de carácter lineal, por lo que se cumplen las ecuaciones \eqref{eq:expectation_l1} y \eqref{eq:expectation_l2}

        \begin{equation}
        \label{eq:expectation}
          \mathbb{E}[\boldsymbol{X}] = \sum_{i=1}^\infty x_i \cdot Pr[\boldsymbol{X} = x_i]
        \end{equation}

        \begin{equation}
        \label{eq:expectation_l1}
          \mathbb{E}[c \boldsymbol{X}] = c \mathbb{E}[\boldsymbol{X}]
        \end{equation}

        \begin{equation}
        \label{eq:expectation_l2}
          \mathbb{E}[\boldsymbol{X} + \boldsymbol{Y}] = \mathbb{E}[\boldsymbol{X}] + \mathbb{E}[\boldsymbol{Y}]
        \end{equation}

        \paragraph{}
        La \textbf{Varianza} se define como una medida de dispersión de una variable aleatoria. Dicho estimador representa el error cuadrático respecto de la esperanza. Su modelización matemática se muestra en la ecuación \eqref{eq:variance}. Aplicando propiedades algebraicas se puede demostrar la veracidad de las propiedades descritas en las ecuaciones \eqref{eq:variance_p1} y \eqref{eq:variance_p2}.

        \begin{equation}
        \label{eq:variance}
          Var[\boldsymbol{X}] = \mathbb{E}[(\boldsymbol{X} - \mathbb{E}[\boldsymbol{X}])^2]
        \end{equation}

        \begin{equation}
        \label{eq:variance_p1}
          Var[\boldsymbol{X}] = \mathbb{E}[\boldsymbol{X}^2] - \mathbb{E}^2[\boldsymbol{X}]
        \end{equation}

        \begin{equation}
        \label{eq:variance_p2}
          Var[c \boldsymbol{X}] = c^2 Var[\boldsymbol{X}]
        \end{equation}

        \paragraph{}
        A continuación se describe el concepto de \textbf{Independencia} entre dos variables aleatorias $\boldsymbol{X}, \boldsymbol{Y}$. Se dice que dos variables son independientes cuando los sucesos de cada una de ellas no están condicionados por los de otras. Esto puede verse a como el cumplimiento de la igualdad de la ecuación \eqref{eq:independence}.

        \begin{equation}
        \label{eq:independence}
          Pr[\boldsymbol{X} = x \cap \boldsymbol{Y} = y] = Pr[\boldsymbol{X} = x] \cdot Pr[\boldsymbol{Y} = y]
        \end{equation}

        \paragraph{}
        Cuando nos referimos al concepto de independencia referido a un conjunto $n$ variables aleatorias $\boldsymbol{X_1}, \boldsymbol{X_2},..., \boldsymbol{X_n}$ lo denominaremos \textbf{Independencia Mutua}, que impone la restricción descrita en la ecuación \eqref{eq:mutual_independence}.

        \begin{equation}
        \label{eq:mutual_independence}
          Pr \bigg[ \bigcap_{i=1}^n \boldsymbol{X_i} = x_i \bigg] = \prod_{i=1}^n Pr[\boldsymbol{X_i} = x_i]
        \end{equation}

        \paragraph{}
        También es de especial interés en el campo de los algoritmos probabilísticos el caso de la \textbf{k-independencia} sobre un conjunto de $n$ variables aleatorias $\boldsymbol{X_1}, \boldsymbol{X_2},..., \boldsymbol{X_n}$. Dicha idea se puede resumir como la independencia de todas las variables en grupos de $k$ variables. Este concepto tiene mucha importancia en el ámbito de los \emph{Sketches}, tal y como se verá en la sección \ref{sec:sketch}. El caso más simple es para $k = 2$, el cual se denomina \textbf{independencia pareada}, cuya modelización matemática se muestra en la ecuación \eqref{eq:pairwise_independence}.

        \begin{equation}
        \label{eq:pairwise_independence}
          \forall i, \forall j \ Pr[\boldsymbol{X_i} = x_i \cap \boldsymbol{X_j} = x_j] = Pr[\boldsymbol{X_i} = x_i] \cdot Pr[\boldsymbol{X_j} = x_j]
        \end{equation}

        \paragraph{}
        Desde el punto de vista de conjuntos de $n$ variables aleatorias  $\boldsymbol{X_1}, \boldsymbol{X_2},..., \boldsymbol{X_n}$, existen distintas propiedades de linealidad que se cumplen entre ellas a nivel del cálculo de la \emph{Esperanza} y la \emph{Varianza}. En el caso de la \emph{Esperanza}, la linealidad respecto de la suma (ecuación \eqref{eq:expectation_linearity}) se cumple para variables dependientes e independientes. Sin embargo, en el caso de la \emph{Varianza}, la linealidad respecto de la suma (ecuación \eqref{eq:variance_linearity}) se cumple tan solo para variables \textbf{independientes pareadas}.

        \begin{equation}
        \label{eq:expectation_linearity}
          \mathbb{E}\bigg[\sum_{i=1}^n \boldsymbol{X_i}\bigg] = \sum_{i=1}^n \mathbb{E}[\boldsymbol{X_i}]
        \end{equation}

        \begin{equation}
        \label{eq:variance_linearity}
          Var\bigg[\sum_{i=1}^n \boldsymbol{X_i}\bigg] = \sum_{i=1}^n Var[\boldsymbol{X_i}]
        \end{equation}

        \paragraph{}
        La \textbf{Probabilidad Condicionada} entre dos variables aleatorias $\boldsymbol{E_1}$ y $\boldsymbol{E_2}$ se puede definir como la medida de verosimilitud de la ocurrencia del suceso $\boldsymbol{E_1}$ sabiendo que ya ha ocurrido $\boldsymbol{E_2}$. Esto se puede modelizar matemáticamente tal y como se muestra en la ecuación \eqref{eq:conditional_probability}.

        \begin{equation}
        \label{eq:conditional_probability}
          Pr[\boldsymbol{E_1} \rvert \boldsymbol{E_2}] = \frac{Pr[\boldsymbol{E_1} \cap \boldsymbol{E_2}]}{Pr[\boldsymbol{E_1}]}
        \end{equation}

        \paragraph{}
        En el caso de la \textbf{Probabilidad Condicionada} sobre variables independientes, surge la propiedad descrita en la ecuación \eqref{eq:conditional_probability_independence}. Es fácil entender la razón, que se apoya en la idea de que si dos variables aleatorias no guardan relación, entonces la ocurrencia de una de ellas, no condicionará el resultado de la otra.

        \begin{equation}
        \label{eq:conditional_probability_independence}
          Pr[\boldsymbol{X_1} = x_1 \rvert \boldsymbol{X_2} = x_2] =
          \frac{Pr[\boldsymbol{X_1} = x_1 \cap  \boldsymbol{X_2} = x_2]}{Pr[\boldsymbol{X_1} = x_1]} =
          \frac{Pr[\boldsymbol{X_1} = x_1 \cdot \boldsymbol{X_2} = x_2]}{Pr[\boldsymbol{X_1} = x_1]} =
          Pr[\boldsymbol{X_2} = x_2]
        \end{equation}

        \paragraph{}
        Una vez descritos los conceptos estadísticos básicos para el análisis de algoritmos probabilísticos, lo siguiente es realizar una exposición acerca de las distintas cotas de concentración de valores, lo cual permite obtener resultados aproximados acerca de los resultados esperados por dichos algoritmos, así como sus niveles de complejidad. Primero se describirá \emph{Desigualdad de Boole}, para después tratar las desigualdades de \emph{Markov}(\ref{sec:markov_inequality}), \emph{Chebyshev}(\ref{sec:chebyshev_inequality}) y \emph{Chernoff}(\ref{sec:chernoff_inequality})

        \paragraph{}
        La \textbf{Desigualdad de Boole} consiste en una propiedad básica que indica que la probabilidad de que se cumpla la ocurrencia de un suceso es menor o igual que ocurrencia de la suma de todas ellas. Esto se modeliza matemáticamente en la ecuación \eqref{eq:boole_inequality}.

        \begin{equation}
        \label{eq:boole_inequality}
          Pr\bigg[\bigcup_{i=1}^n \boldsymbol{E_i}\bigg] \leq \sum_{i=1}^n Pr[\boldsymbol{E_i}]
        \end{equation}

      \subsection{Desigualdad de Markov}
      \label{sec:markov_inequality}

        \paragraph{}
        La \emph{Desigualdad de Markov} es la técnica base que utilizan otras desigualdades más sofisticadas para tratar de acotar la \emph{Esperanza} de una determinada \emph{Variable Aleatoria}. Proporciona una cota superior de probabilidad respecto de la \emph{Esperanza} tal y como se muestra en la ecuación \eqref{eq:markov_inequality}. Tal y como se puede intuir, dicha cota es muy poco ajustada, sin embargo, presenta una propiedad muy interesante como estructura base. Sea $f: \boldsymbol{X} \rightarrow \mathbb{R}^+$ una función positiva, entonces también se cumple la desigualdad de la ecuación \eqref{eq:markov_inequality_function}. El punto interesante surge cuando se escoge la función $f$ de tal manera que sea estrictamente creciente, entonces se cumple la propiedad descrita ecuación \eqref{eq:markov_inequality_function_positive}, a través de la cual podemos obtener cotas mucho más ajustadas. Dicha idea se apoya en la \emph{Desigualdad de Jensen}.

        \begin{equation}
        \label{eq:markov_inequality}
          \forall \lambda \geq 0, \ Pr[\boldsymbol{X} \geq \lambda ] \leq \frac{\mathbb{E}[\boldsymbol{X}]}{\lambda}
        \end{equation}

        \begin{equation}
        \label{eq:markov_inequality_function}
          \forall \lambda \geq 0, \ Pr[f(\boldsymbol{X}) \geq f(\lambda) ] \leq \frac{\mathbb{E}[f(\boldsymbol{X})]}{f(\lambda)}
        \end{equation}

        \begin{equation}
        \label{eq:markov_inequality_function_positive}
          \forall \lambda \geq 0, \ Pr[\boldsymbol{X} \geq \lambda ] = Pr[f(\boldsymbol{X}) \geq f(\lambda) ] \leq \frac{\mathbb{E}[f(\boldsymbol{X})]}{f(\lambda)}
        \end{equation}

      \subsection{Desigualdad de Chebyshev}
      \label{sec:chebyshev_inequality}

        \paragraph{}
        La \emph{Desigualdad de Chebyshev} utiliza la técnica descrita en la sub-sección anterior apoyándose en al idea de la función $f$ para obtener una cota de concentración mucho más ajustada basándose en la \emph{Varianza}. Dicha propiedad se muestra en la ecuación \eqref{eq:chebyshev_inequality}. En este caso se utiliza $f(\boldsymbol{X}) = \boldsymbol{X}^2$, que es estrictamente creciente en el dominio de aplicación de $\boldsymbol{X}$. Además, se selecciona como variable aleatoria $|\boldsymbol{X} - \mathbb{E}[\boldsymbol{X}]$, es decir, el error absoluto de una  $\boldsymbol{X}$ respecto de su valor esperado. La demostración de esta idea se muestra en la ecuación \eqref{eq:chebyshev_inequality_demo}.

        \begin{equation}
        \label{eq:chebyshev_inequality}
          \forall \lambda \geq 0, \ Pr[|\boldsymbol{X} - \mathbb{E}[\boldsymbol{X}]| \geq \lambda]  \leq \frac{Var[\boldsymbol{X}]}{\lambda^2}
        \end{equation}

        \begin{equation}
        \label{eq:chebyshev_inequality_demo}
          \forall \lambda \geq 0, \
          Pr[|\boldsymbol{X} - \mathbb{E}[\boldsymbol{X}]| \geq \lambda] =
          Pr[(\boldsymbol{X} - \mathbb{E}[\boldsymbol{X}])^2 \geq \lambda^2] \leq
          \frac{\mathbb{E}[(\boldsymbol{X} - \mathbb{E}[\boldsymbol{X}])^2]}{\lambda^2} =
          \frac{Var[\boldsymbol{X}]}{\lambda^2}
        \end{equation}

      \subsection{Desigualdad de Chernoff}
      \label{sec:chernoff_inequality}

        \paragraph{}
        En este apartado se realiza una descripción acerca de la \emph{Desigualdad de Chernoff}. Dicha descripción ha sido extraída de los apuntes de la asignatura de Algoritmos Probabilísticos (\emph{Randomized Algorithms}) \cite{chawla2004chernoff} impartida por Shuchi Chawla en la \emph{Carnegie Mellon University} de Pennsylvania.

        \paragraph{}
        La \emph{Desigualdad de Chernoff} proporciona cotas mucho más ajustadas que por contra, exigen unas presunciones más restrictivas para poder ser utilizada. La variable aleatoria en este caso debe ser de la forma $\boldsymbol{S} = \sum_{i=1}^n \boldsymbol{X_i}$ donde cada $\boldsymbol{X_i}$ es una variable aleatoria uniformemente distribuida e independiente del resto. También describiremos la esperanza de cada una de las variables $\boldsymbol{X_i}$ como $\mathbb{E}[\boldsymbol{X}] = p_i$.

        \paragraph{}
        Denotaremos como $\mu$ a la esperanza de $\boldsymbol{S}$, tal y como se describe en la ecuación \eqref{eq:chernoff_inequality_expectation}. También se define la función $f$ como $f(\boldsymbol{S}) = e^{t\boldsymbol{S}}$.

        \begin{equation}
        \label{eq:chernoff_inequality_expectation}
          \mu = \mathbb{E}\bigg[\sum_{i=1}^n \boldsymbol{X_i}\bigg] = \sum_{i=1}^n  \mathbb{E}[\boldsymbol{X_i}] = \sum_{i=1}^n p_i
        \end{equation}

        \paragraph{}
        El siguiente paso es utilizar la ecuación \eqref{eq:markov_inequality_function_positive} de la \emph{Desigualdad de Markov} con la función $f$, que en este caso es posible puesto que es estrictamente creciente. En este caso en lugar de utilizar $\lambda$ como constante, se prefiere $\delta \in [0,1]$, que se relacionada con la anterior de la siguiente manera: $\lambda = (1 + \delta)$. Entonces la ecuación \eqref{eq:chernoff_inequality_raw} muestra el paso inicial para llegar a la \emph{Desigualdad de Chernoff}.

        \begin{equation}
        \label{eq:chernoff_inequality_raw}
          Pr[ \boldsymbol{X} > (1+\delta) \mu] =
          Pr[ e^{t \mathbb{X}} > e^{(1+\delta)t\mu}] \leq
          \frac{\mathbb{E}[ e^{t \boldsymbol{X}}] }{e^{(1 + \delta) t \mu}}
        \end{equation}

        \paragraph{}
        Aplicando operaciones aritméticas y otras propiedades estadísticas, se puede demostrar la veracidad de las ecuaciones \eqref{eq:chernoff_inequality_upper} y \eqref{eq:chernoff_inequality_lower}, que proporcionan cotas mucho muy ajustadas de concentración de la distribución de una \emph{variable aleatoria} formada por la suma de $n$ variables aleatorias independientes uniformemente distribuidas.

        \begin{equation}
        \label{eq:chernoff_inequality_upper}
          \forall \delta \geq 0, \ Pr[\boldsymbol{X} \geq (1 + \delta)\mu]  \leq e^\frac{-\lambda^2\mu}{2 + \lambda}
        \end{equation}

        \begin{equation}
        \label{eq:chernoff_inequality_lower}
          \forall \delta \geq 0, \ Pr[\boldsymbol{X} \leq (1 - \delta)\mu]  \leq e^\frac{-\lambda^2\mu}{2 + \lambda}
        \end{equation}

      \subsection{Funciones Hash}
      \label{sec:hash_functions}

        \paragraph{}
        Las funciones hash son transformaciones matemáticas basadas en la idea de trasladar un valor en un espacio discreto con $n$ posibles valores a otro de $m$ valores de tamaño menor tratando de evitar que se produzcan colisiones en la imagen. Por tanto son funciones que tratan tratan de ser inyectivas en un sub-espacio de destino menor que el de partida. Sin embargo, tal y como se puede comprender de manera intuitiva es una propiedad imposible de cumplir debido a los tamaños del espacio de partida y de destino.

        \paragraph{}
        Las familias de funciones hash universales se refieren a distintas categorías en las cuales se pueden agrupar las funciones hash según el nivel de propiedades que cumplen. La categorías en las que se agrupan se denominan \emph{funciones hash k-universales}, de tal manera que el valor $k$ indican la dificultad de aparición de colisiones. Las funciones \emph{2-universales} se refieren a aquellas en las cuales se cumple que $Pr[h(x) = h(y)] \leq 1/m$ siendo $h$ la función hash, $x$ e $y$ dos posibles entradas tales que $x \neq y$ y $m$ el cardinal del conjunto de todas las posibles claves. Se dice que una función hash es \emph{fuertemente 2-universal} (\emph{strongly 2-universal}) si cumple que $Pr[h(x) = h(y)] \leq 1/m^2$ y genéricamente se dice que una función es \emph{k-universal} si cumple que $Pr[h(x) = h(y)] \leq 1/m^-k$. A continuación se describen dos estrategias básicas de diseño de funciones hash. Posteriormente se realiza una descripción acerca de las funciones hash sensibles a la localización.

        \subsubsection{Hash basado en Congruencias}
        \label{sec:hash_congruential}

          \paragraph{}
          Las funciones hash basadas en congruencias poseen la propiedad de ser \emph{2-universales}. Se basan en la idea de utilizar como espacio de destino aquel formado por $\mathbb{Z}_p$, es decir, todos los enteros que son congruentes con $p$ siendo $p \geq m$ un número primo. Además, se utilizan los enteros $a,b \in \mathbb{Z}_p$ con $a \neq 0$. La función hash entonces se describe tal y como se indica en la ecuación \eqref{eq:hash_congruential}.

          \begin{equation}
          \label{eq:hash_congruential}
            h_{ab}(x) = (ax + b) mod p
          \end{equation}

        \subsubsection{Hash basado en Tabulación}
        \label{sec:hasch_tabulation}

          \paragraph{}
          El hash basado en tabulación consiste en un método de generación de valores hash que restringe su dominio de entrada a cadenas de texto de tamaño fijo (u otras entradas que puedan codificarse de dicha forma). Denominaremos $c$ al tamaño fijo y  $T_i, i \in [1,c]$ a vectores que mapean el carácter $i$-ésimo de manera aleatoria. Entonces la función hash realiza una operación \emph{or-exclusiva} sobre los $T_i[x_i]$ valores tal y como se indica en la ecuación \eqref{eq:hash_tabulation}. Las funciones hash que se construyen siguiendo esta estrategia poseen la propiedad pertenecer a la categoría de las \emph{3-universales}.

          \begin{equation}
          \label{eq:hash_tabulation}
            h(x) = T_1[x_1] \oplus T_2[x_2] \oplus ... \oplus T_c[x_c]
          \end{equation}

      \subsubsection{Funciones Hash sensibles a la localización}
      \label{sec:hash_lsh}

        \paragraph{}
        Una categoría a destacar son las \emph{funciones hash sensibles a la localización}. Estas poseen la propiedad de distribuir los valores cercanos en el espacio de destino tratando mantener las propiedades de cercanía entre los valores. El primer artículo en que se habla de ellas es \emph{Approximate Nearest Neighbor: Towards Removing the Curse of Dimensionality} \cite{indyk1998approximate} de \emph{Indyk} y \emph{Motwani} inicialmente para resolver el problema de la \emph{búsqueda del vecino más cercano (Nearest Neighbor Search)}. Se trata de funciones hash multidimensionales, es decir, en la entrada están compuestas por más de un elemento. Estas funciones han cobrado especial importancia en los últimos años por su uso en problemas de dimensión muy elevada, ya que sirven como estrategia de reducción de la dimensionalidad del conjunto de datos, que como consecuencia reduce el coste computacional del problema.

        \paragraph{}
        \emph{Indyk} indica que para que una función hash sea sensible a la localización o $(r_1,r_2, p_1, p_2)$-sensible, para cualquier par de puntos de entrada $p,q$ y la función de distancia $d$ se cumpla la ecuación \eqref{eq:hash_lsh}. Las funciones hash que cumplen esta propiedad son interesantes cuando $p_1 > p_2$ y $r_1 < r_2$, de tal forma que para puntos cercanos en el espacio, el valor hash obtenido es el mismo, por lo que se asume que dichos puntos se encuentran cercanos en el espacio de partida.

        \begin{align}
        \label{eq:hash_lsh}
          \text{if} \ d(p,q) \leq r_1, \ & \text{then} \ Pr[h(x) = h(y)] \geq p_1 \\
          \text{if} \ d(p,q) \geq r_2, \ & \text{then} \ Pr[h(x) = h(y)] \leq p_2
        \end{align}

      \paragraph{}
      Una vez descritos los conceptos básicos acerca de lo que son los \emph{Algoritmos para Streaming} y las bases estadísticas necesarias para poder entender el funcionamiento de los mismo y sus niveles de complejidad así como de precisión, en las siguientes secciones se realiza una descripción sobre algunos de los algoritmos más relevantes en este área. En especial se explica el algoritmo de \emph{Morris} en la sección \ref{sec:streaming_morris_algorithm}, el de \emph{Flajolet-Martin} en la \ref{sec:streaming_morris_algorithm} y por último se hablará de la \emph{Estimación de Momentos de Frecuencia} en el modelo en streaming en la sección \ref{sec:streaming_frecuency_moment_aproximation}.

    \section{Algoritmo de Morris}
    \label{sec:streaming_morris_algorithm}

      \paragraph{}
      El \emph{Algoritmo de Morris} fue presentado por primera vez en el artículo \emph{Counting Large Numbers of Events in Small Registers} \cite{morris1978counting} redactado por \emph{Robert Morris}. En dicho documento se trata de encontrar una solución al problema de conteo de ocurrencias de un determinado suceso teniendo en cuenta las restriciones de espacio debido a la elevada tasa de ocurrencias  que se da en muchos fenómenos. El problema del conteo (\textbf{Count Problem}) de ocurrencias también se denomina el momento de frecuencia $F_1$ tal y como se verá en la sección \ref{sec:streaming_frecuency_moment_aproximation}.

      \paragraph{}
      Por tanto, \emph{Morris} propone realizar una estimación de dicha tasa para reducir el espacio necesario para almacenar el valor. Intuitivamente, a de partir dicha restricción se consigue un orden de complejidad espacial sub-lineal ($o(N)$) con respecto del número de ocurrencias. Se puede decir que el artículo publicado por \emph{Morris} marcó el punto de comienzo de este área de investigación. El conteo probabilista es algo trivial si se restringe a la condición de incrementar el conteo de ocurrencias siguiendo una distribución de \emph{Bernoulli} con un parámetro $p$ prefijado previamente. Con esto se consigue un error absoluto relativamente pequeño con respecto al valor $p$ escogido. Sin embargo, el error relativo que se obtiene cuando el número de ocurrencias es pequeño es muy elevado, lo cual lo convierte en una solución impracticable.

      \paragraph{}
      Para solucionar dicha problemática y conseguir una cota del error relativo reducida, la solución propuesta por \emph{Morris} se basa en la selección del parámetro $p$ variable con respecto del número de ocurrencias, con lo cual se consigue que la decisión de incrementar el contador sea muy probable en los primeros casos, lo cual elimina el problema del error relativo. \emph{Morris} propone aumentar el contador $X$ con probabilidad $\frac{1}{2^X}$. Tras $n$ ocurrencias, el resultado que devuelve dicho algoritmo es $\widetilde{n} = 2^X -1$. El pseudocódigo se muestra en el algoritmo \ref{code:morris-algorithm}.

      \paragraph{}
      \begin{algorithm}[h]
        \SetAlgoLined
        \KwResult{$\widetilde{n} = 2^X -1$ }
        $X \gets 0$\;
        \For{cada evento}{
          $X \gets X + 1 \ \text{con probabilidad} \frac{1}{2^X}$\;
        }
        \caption{Morris-Algorithm}
        \label{code:morris-algorithm}
      \end{algorithm}

      \paragraph{}
      A continuación se realiza un análisis de la solución. Esta ha sido extraída de los apuntes de la asignatura \emph{Algorithms for Big Data} \cite{bigdata2015jelani} impartida por \emph{Jelani Nelson} en la \emph{Universidad de Harvard}. Denotaremos por $X_n$ el valor del contador $X$ tras $n$ ocurrencias. Entonces se cumplen las igualdades descritas en las ecuaciones \eqref{eq:morris_expectation_1} y \eqref{eq:morris_expectation_2}. Esto se puede demostrar mediante técnicas inductivas sobre $n$.

      \begin{equation}
      \label{eq:morris_expectation_1}
        \mathbb{E}[2^{X_n}] = n + 1
      \end{equation}

      \begin{equation}
      \label{eq:morris_expectation_2}
        \mathbb{E}[2^{2X_n}] = \frac{3}{2}n^2 + \frac{3}{2}n + 1
      \end{equation}

      \paragraph{}
      Por la \emph{Desigualdad de Chebyshev} podemos acotar el error cometido tras $n$ repeticiones, dicha formulación se muestra en la ecuación \eqref{eq:morris_bound}.

      \begin{align}
      \label{eq:morris_bound}
        Pr[|\widetilde{n} - n| > \epsilon n ] < \frac{1}{\epsilon^2n^2}\cdot\mathbb{E}[\widetilde{n} - n]^2
          &= \frac{1}{\epsilon^2n^2}\cdot\mathbb{E}[2^X-1-n]^2
        \\&= \frac{1}{\epsilon^2n^2}\cdot \frac{n^2}{2}
        \\&= \frac{1}{2\epsilon^2}
      \end{align}

      \paragraph{}
      La ventaja de esta estrategia algorítmica con respecto de la trivial es la cota del error relativo producido en cada iteración del algoritmo, lo cual aporta una mayor genericidad debido a que esta se mantiene constante con respecto del número de ocurrencias. Sin embargo, se han propuesto otras soluciones para tratar de reducir en mayor medida dicha cota de error. El algoritmo \emph{Morris+} se basa en el mantenimiento de $s$ copias independientes de \emph{Morris} para después devolver la media del resultado de cada una de ellas. A partir de esta estrategia se consiguen las tasas de error que se indican en la ecuación \eqref{eq:morris+_bound}.

      \begin{align}
      \label{eq:morris+_bound}
        Pr[|\widetilde{n} - n| > \epsilon n ] < \frac{1}{2s\epsilon^2} && s > \frac{3}{2\epsilon^2}
      \end{align}

    \section{Algoritmo de Flajolet-Martin}
    \label{sec:streaming_flajolet_martin_algorithm}

      \paragraph{}
      En esta sección se describe el \emph{Algoritmo de Flajolet-Martin}, cuya descripción aparece en el artículo \emph{Probabilistic Counting Algorithms for Data Base Applications} \cite{flajolet1985probabilistic} redactado por \emph{Philippe Flajolet} y \emph{G. Nigel Martin}. En este caso, la problemática que se pretende resolver no es el número de ocurrencias de un determinado suceso, sino el número de sucesos distintos (\textbf{Count Distinct Problem}) en la entrada. Este problema también se conoce como el cálculo del momento de frecuencia $F_0$ tal y como se verá en la sección \ref{sec:streaming_frecuency_moment_aproximation}. Al igual que en el caso del algoritmo de \emph{Morris}, se apoya en estrategias probabilistas para ajustarse a un orden de complejidad espacial de carácter sub-lineal ($o(N))$) manteniendo una cota de error ajustada.

      \paragraph{}
      La intuición en la cual se basa el \emph{Algoritmo de Flajolet-Martin}, es la transformación de los elementos de entrada sobre una \emph{función Hash} universal binaria con distribución uniforme e independiente de probabilidad. La propiedad de distribución uniforme permite entonces prever que la mitad de los elementos tendrán un $1$ en el bit menos significativo, que una cuarta parte de los elementos tendrán un $1$ en el segundo bit menos significativo y así sucesivamente. Por tanto, a partir de esta idea se puede realizar una aproximación probabilista del número de elementos distintos que han sido presentados en la entrada. Requiere de $L$ bits de espacio para el almacenamiento del número de elementos distintos. Por la notación descrita en anteriores secciones $L = log(n)$, donde $n$ es el número máximo de elementos distintos en la entrada. A continuación se explica esta estrategia, para ello nos apoyaremos en las siguientes funciones:

      \begin{itemize}
        \item $hash(x)$ Es la función hash con distribución uniforme e independiente de probabilidad que mapea una entrada cualquiera a un valor entero en el rango $[0,...,2^L-1]$.
        \item $bit(y, k)$ Esta función devuelve el bit \emph{k-ésimo} de la representación binaria de $y$, de tal manera que se cumple que $y = \sum_{k \geq 0} bit(y,k)2^k$
        \item $\rho(y)$ La función $\rho$ devuelve la posición en la cual se encuentra el bit con valor $1$ empezando a contar a partir del menos significativo. Por convenio, devuelve el valor $L$ si $y$ no contiene ningún $1$ en su representación binaria, es decir, si $y = 0$. Esto se modeliza matemáticamente en la ecuación \eqref{eq:rho_function}.
      \end{itemize}

      \begin{equation}
      \label{eq:rho_function}
        \rho(y) =
          \begin{cases}
            min_{k \geq 0} bit(y, k) \neq 0 & y \geq 0\\
            L & y =0
          \end{cases}
      \end{equation}

      \paragraph{}
      \emph{Flajolet} y \emph{Martin} se apoyan en una estructura de datos indexada a la cual denominan \textbf{BITMAP}, de tamaño $[0...L-1]$ la cual almacena valores binarios $\{ 0, 1\}$ y se inicializa con todos los valores a $0$. Nótese por tanto, que esta estructura de datos puede ser codificada como un string binario de longitud $L$. La idea del algoritmo es marcar con un $1$ la posición \textbf{BITMAP[$\rho(hash(x))$]}. Seguidamente, queda definir el resultado de la consulta sobre cuántos elementos distintos han aparecido en el flujo de datos de entrada. Para ello se calcula $2^{\rho(\textbf{BITMAP})}$. El pseudocódigo se muestra en el algoritmo \ref{code:fm-algorithm}.

      \paragraph{}
      \begin{algorithm}[h]
        \SetAlgoLined
        \KwResult{$2^{\rho(\textbf{BITMAP})}$}
        \For{$i \in [0,...,L-1]$}{
          $\textbf{BITMAP[$i$]} \gets 0$\;
        }
        \For{cada evento}{
          \If{$\textbf{BITMAP[$\rho(hash(x))$]} = 0$}{
            $\textbf{BITMAP[$\rho(hash(x))$]} \gets 1$\;
          }
        }
        \caption{FM-Algorithm}
        \label{code:fm-algorithm}
      \end{algorithm}

      \paragraph{}
      El análisis de esta solución ha sido Extraído de los apuntes del libro \emph{Mining of massive datasets} \cite{leskovec2014mining} de la \emph{Universidad de Cambridge}. En este caso lo representa teniendo el cuenta el número de $0$'s seguidos en la parte menos significativa de la representación binaria de $h(x)$. Nótese que esto es equivalente al valor de la función $\rho(h(y))$, por tanto, adaptaremos dicho análisis a la solución inicial propuesta por \emph{Flajolet} y \emph{Martin}. La probabilidad de que se cumpla $\rho(h(x)) = r$ es $2^{-r}$. Supongamos que el número de elementos distintos en el stream es $m$. Entonces la probabilidad de que ninguno de ellos cumpla $\rho(h(x)) = r$ es al menos $(1- 2^{-r})^m$ lo cual puede ser reescrito como $((1- 2^{-r})^{2^r})^{m2^{-r}}$. Para valores suficientemente grandes $r$ se puede asumir que dicho valor es de la forma $(1-\epsilon)^{1/\epsilon} \approx 1/\epsilon$. Entonces la probabilidad de que no se cumpla que $\rho(h(x)) = r$ cuando han aparecido $m$ elementos distintos en el stream es de $e^{-m2^{-r}}$

      \paragraph{}
      La problemática de este algoritmo deriva de la suposición de la capacidad de generación de claves Hash totalmente aleatorias, lo cual no se ha conseguido en la actualidad. Por lo tanto posteriormente, \emph{Flajolet} ha seguido trabajando el problema de conteo de elementos distintos en \emph{Loglog counting of large cardinalities} \cite{durand2003loglog} y \emph{Hyperloglog: the analysis of a near-optimal cardinality estimation algorithm} \cite{flajolet2007hyperloglog} para tratar de mejorar el grado de precisión de su estrategia de conteo. En el artículo \emph{An optimal algorithm for the distinct elements problem} \cite{kane2010optimal} \emph{Daniel Kane y otros} muestran un algoritmo óptimo para el problema. Los resultados de dichos trabajos se discuten en la sección \ref{sec:hyper_log_log} por su cercana relación con las \emph{estructuras de datos de resumen}.

    \section{Aproximación a los Momentos de Frecuencia}
    \label{sec:streaming_frecuency_moment_aproximation}

      \paragraph{}
      La siguiente idea de la que es interesante hablar para terminar la introducción a los \emph{Algoritmos para Streaming} son los \emph{Momentos de Frecuencia}. Una generalización de los conceptos del número de elementos distintos ($F_0$) y el conteo de elementos ($F_1$) que se puede extender a cualquier $F_k$ para $k \geq 0 $. La definición matemática del momento de frecuencia $k$-ésimo se muestra en la ecuación \eqref{eq:frecuency_moments}. Nótese el caso especial de $F_\infty$ que se muestra en la ecuación \eqref{eq:frecuency_moments_max} y se corresponde con el elemento más veces común en el \emph{Stream}. Estas ideas han sido extraídas del documento \emph{Frequency Moments} \cite{woodruff2009frequency} redactado por \emph{David Woodruff}.

      \begin{equation}
      \label{eq:frecuency_moments}
        F_k = \sum_{i=1}^n m_i^k
      \end{equation}

      \begin{equation}
      \label{eq:frecuency_moments_max}
        F_\infty = max_{1 \leq i \leq n} m_i
      \end{equation}

      \paragraph{}
      El resto de la sección trata sobre la exposición de los algoritmos para el cálculo de los momentos de frecuencia descritos en el articulo \emph{The space complexity of approximating the frequency moments} \cite{alon1996space} redactado por \emph{Noga Alon}, \emph{Yossi Matias} y \emph{Mario Szegedy}, por el cual fueron galardonados con el premio \emph{Gödel} en el año 2005. En dicho trabajo además de presentar \emph{Algoritmos para Streaming} para el cálculo de $F_k$ (cabe destacar su solución para $F_2$), también presentan cotas inferiores para el problema de los \emph{Momentos de Frecuencia}. Posteriormente \emph{Piotr Indyk} y \emph{David Woodruff} encontraron un algoritmo óptimo para el problema de los\emph{Momentos de Frecuencia} tal y como exponen en \emph{Optimal Approximations of the Frequency Moments of Data Streams} \cite{indyk2005optimal}. A continuación se discuten los resultados de dichos trabajos.

      \paragraph{}
      Para el cálculo de $F_k$ para $k \geq 0$ \emph{Alon}, \emph{Matias} y \emph{Szedgedy} proponen un enfoque similar a los propuestos en algoritmos anteriores (la definición de una variable aleatoria $X$ tal que $\mathbb{E}[X] = F_k$). Sin embargo, la novedad en este caso es que su algoritmo no está restringido a un $k$ concreto, sino que en su caso es generalizable para cualquier entero positivo, sin embargo, en este caso la exposición es a nivel teórico.

      \paragraph{}
      Definiremos las constantes $S_1 =O(n^{1-1/k}/\lambda ^{2})$ y $S_2 = O(\log(1/\varepsilon ))$. El algoritmo utiliza $S_2$ variables aleatorias denominadas $Y_1, Y_2, Y_{S_2}$ y devuelve la mediana de estas denominándola $Y$. Cada una de estas variables $Y_i$ está formada por la media de $X_{ij}$ variables aleatorias tales que $1 leq j leq S_1$. La forma en que se actualiza el estado del algoritmo tras cada nueva llegada se apoya en el uso de $S_2$ funciones hash uniformemente distribuidas e independientes entre si que mapean cada símbolo a un determinado indice $j$.

      \paragraph{}
      Para el análisis del algoritmos supondremos que el tamaño $n$ del \emph{Stream} es conocido \emph{a-priori}. La demostración se apoya en una variable aleatoria $X$ construida de la siguiente manera:

      \begin{itemize}
        \item Seleccionaremos de manera aleatoria el elemento $a_{p \in (1,2,...,m)}$ del Stream, siendo $a_p = l \in (1,2,...n)$. Es decir, el elemento procesado en el momento $p$ representa la llegada del símbolo $l$.
        \item Definiremos $r=|\{q:q\geq p,a_{p}=l\}|$ como el número de ocurrencias del símbolo $l$ hasta el momento $p$.
        \item La variable aleatoria $X$ se define como $X=m(r^{k}-(r-1)^{k})$
      \end{itemize}

      \paragraph{}
      Desarrollando la Esperanza Matemática de la variable aleatoria $X$ se puede demostrar que esta tiende al momento de frecuencia $k$ tal y como se muestra en la ecuación \eqref{eq:expectation_frecuency_moments}.

      \begin{equation}
      \label{eq:expectation_frecuency_moments}
        {\displaystyle {\begin{array}{lll}\mathbb{E}(X)&=&\sum _{i=1}^{n}\sum _{i=1}^{m_{i}}(j^{k}-(j-1)^{k})\\&=&{\frac {m}{m}}[(1^{k}+(2^{k}-1^{k})+\ldots +(m_{1}^{k}-(m_{1}-1)^{k}))\\&&\;+\;(1^{k}+(2^{k}-1^{k})+\ldots +(m_{2}^{k}-(m_{2}-1)^{k}))+\ldots \\&&\;+\;(1^{k}+(2^{k}-1^{k})+\ldots +(m_{n}^{k}-(m_{n}-1)^{k}))]\\&=&\sum _{i=1}^{n}m_{i}^{k}=F_{k}\end{array}}}
      \end{equation}

      \paragraph{}
      En cuanto al coste espacial del algoritmo, se puede demostrar tal y como indican \emph{Alon}, \emph{Matias} y \emph{Szedgedy} en su artículo original \cite{alon1996space} que este sigue el orden descrito en la ecuación \eqref{eq:complexity_frecuency_moments} puesto que es necesario almacenar $a_p$ y $r$ lo cual requiere de $log(n) + log(m)$ bits de memoria, además de $S_1 x S_2$ variables aleatorias para mantener $X$.

      \begin{equation}
      \label{eq:complexity_frecuency_moments}
        {\displaystyle O\left({\dfrac {k\log {1 \over \varepsilon }}{\lambda ^{2}}}n^{1-{1 \over k}}\left(\log n+\log m\right)\right)}
      \end{equation}

  \section{Conclusiones}
  \label{sec:streaming_conclusions}

    \paragraph{}
    Tal y como se ha ilustrado a lo largo del capítulo, los \emph{algoritmos para streaming} son una solución adecuada tanto a nivel conceptual como práctica para problemas en los cuales el tamaño del conjunto de datos de entrada es tan elevado que no se puede hacer frente mediante estrategias clásicas. Por contra, estas soluciones presentan dificultades debido al elevado peso de la componente matemática y estadística que presentan. Además, la imprecisión en sus resultados restringe su uso en casos en los cuales la precisión es un requisito imprescindible por lo que tan solo deben ser utilizados en casos en los cuales no existan otras soluciones que calculen una solución con las restricciones de tiempo y espacio impuestas.

    \paragraph{}
    En este capítulo se ha realizado una introducción superficial acerca de este modelo, sin embargo las implementaciones y descripciones que se muestran tan solo gozan de importancia a nivel teórico y conceptual. Por lo tanto, en el capítulo \ref{chap:summaries} se continua la exposición de técnicas de tratamiento de grandes cantidades de datos desde una perspectiva más práctica hablando de las \emph{estructuras de datos de resumen}. Se describen en especial detalle las estructuras basadas en \emph{sketches}, que internamente utilizan \emph{algoritmos para streaming}.

  \chapter{Estrategias de Sumarización}
  \label{chap:summaries}

    \section{Introducción}
    \label{sec:summaries_intro}

      \paragraph{}
      El gran crecimiento tecnológico que se está llevando a cabo en la actualidad a todos los niveles está propiciando además un aumento exponencial en cuanto a la cantidad de información que se genera. La reducción de costes en cuanto a la instalación de sensores que permiten recoger información de muchos procesos productivos, así como la obtención de meta-datos a partir del uso de internet y las redes sociales por parte de los usuarios hace que el ritmo de crecimiento en cuanto a cantidad información generada por unidad de tiempo haya crecido a un ritmo frenético.

      \paragraph{}
      Una de las razones que han facilitado dicha tendencia es la disminución de costes de almacenamiento de información, a la vez que han aumentado las capacidades de cómputo necesarias para procesarla. Sin embargo, debido al crecimiento exponencial de los conjuntos de datos, es necesario investigar nuevas técnicas y estrategias que permitan obtener respuestas satisfactorias basadas en la gran cantidad de información de la que se dispone en un tiempo razonable.

      \paragraph{}
      Tradicionalmente, la investigación en el campo de las \emph{bases de datos} se ha centrado en obtener respuestas exactas a distintas consultas, tratando de hacerlo de la manera más eficiente posible, así como de tratar de reducir el espacio necesario para almacenar la información. \emph{Acharya y otros} proponen en el artículo \emph{Join synopses for approximate query answering} \cite{acharya1999join} el concepto de \emph{Approximate Query Processing}. Dicha idea se expone en la sub-sección \ref{sec:aproximate_query_processing}.

      \subsection{Approximate Query Processing}
      \label{sec:aproximate_query_processing}

        \paragraph{}
        El \emph{procesamiento de consultas aproximado}, (\emph{Approximate Query Processing} o \textbf{AQP}) se presenta como una estrategia de resolución de consultas basada en conceptos y propiedades estadísticas que permiten una gran reducción en la complejidad computacional y espacial necesaria para la resolución de las mismas por una base de datos. Por contra, dicha reducción a tiene como consecuencia la adicción de un determinado nivel de imprecisión en los resultados a la cual se denomina error. Se pretende que dicho error pueda ser acotada en un intervalo centrado en el valor verdadero con una desviación máxima determinada por $\epsilon$, y que la pertenencia de la solución a este intervalo se cumpla con una probabilidad $\delta$. Al igual que en el anterior capítulo, en este caso también se presta especial importancia a la minimización del error relativo, lo cual consigue que las soluciones mediante el \emph{procesamiento de consultas aproximado} sean válidas tanto para consultas de tamaño reducido como de gran tamaño.

      \paragraph{}
      Durante el resto del capítulo se describen y analizan distintas estrategias que permiten llevar a cabo implementaciones basadas en \emph{procesamiento de consultas aproximado} centrando especial atención en los \emph{Sketches} por su cercanía respecto del \emph{Modelo en Streaming} descrito en el capítulo \ref{chap:streaming}. En la sección \ref{sec:summaries_types} se realiza una descripción a partir de la cual se pretende aclarar las diferencias entre las distintas estrategias de sumarización de grandes conjuntos de datos. Las estrategias que se describen son \emph{muestreo probabilístico}, mantenimiento de un \emph{histograma}, utilización de \emph{wavelets} y por último se describen en profundidad conceptos referidos a \emph{Sketches}. En las secciones \ref{sec:bloom_filter}, \ref{sec:count_min_sketch}, \ref{sec:count_sketch} y \ref{sec:hyper_log_log} se habla del \emph{Bloom-Filter} \emph{Count-Min Sketch}, \emph{Count Sketch} y \emph{HyperLogLog} respectivamente.

    \section{Tipos de Estrategias de Sumarización}
    \label{sec:summaries_types}

      \paragraph{}
      Para el diseño de soluciones basadas en \emph{procesamiento de consultas aproximado} existen distintas estrategias, cada una de las cuales presentan ventajas e inconvenientes por lo que cada una de ellas es más conveniente para una determinada tarea, sin embargo en ocasiones surgen solapamientos entre ellas tal y como se pretende mostrar en esta sección. Dichas descripciones han sido extraídas del libro \emph{Synopses for massive data} \cite{cormode2012synopses} redactado por \emph{Cormode y otros}. En las secciones \ref{sec:sampling}, \ref{sec:histogram}, \ref{sec:wavelet} y \ref{sec:sketch} se habla de \emph{Sampling}, \emph{Histogram}, \emph{Wavelet} y \emph{Sketches} respectivamente.

      \subsection{Sampling}
      \label{sec:sampling}

        \paragraph{}
        El \emph{Sampling} o \emph{muestreo probabilístico} es la estrategia más consolidada de entre las que se presentan. Las razones se deben a su simplicidad conceptual así como su extendido uso en el mundo de la estadística. Uno de los primeros artículos en que se trata el muestreo aplicado a bases de datos es \emph{Accurate estimation of the number of tuples satisfying a condition} \cite{piatetsky1984accurate} redactado por \emph{Piatetsky-Shapiro} y \emph{Connell}. La intuición en que se basa esta solución es la selección de un sub-conjunto de elementos denominado \emph{muestra} extraída del conjunto global al cual se denomina \emph{población}. Una vez obtenida la \emph{muestra} del conjunto de datos global, cuyo tamaño es significativamente menor respecto del global (lo cual reduce drásticamente el coste computacional), se realizan los cálculos que se pretendía realizar sobre toda la \emph{población} para después obtener un estimador del valor real que habría sido calculado al realizar los cálculos sobre el conjunto de datos global.

        \paragraph{}
        Para que las estrategias de sumarización de información obtengan resultados válidos o significativos respecto del conjunto de datos, es necesario que se escojan adecuadamente las instancias de la \emph{muestra}, de manera que se maximice la similitud del resultado respecto del que se habría obtenido sobre toda la población. Para llevar a cabo dicha labor existen distintas estrategias, desde las más simples basadas en la selección aleatoria sin reemplazamiento como otras mucho más sofisticadas basadas en el mantenimiento de \emph{muestras} estratificadas. Sea $R$ la población y $|R|$ el tamaño de la misma. Denominaremos $t_j$ al valor $j$-ésimo de la población y $X_j$ al número de ocurrencias del mismo en la \emph{muestra}. A continuación se describen distintas técnicas de muestreo:

        \begin{itemize}

          \item \textbf{Selección Aleatoria Sin Reemplazamiento}: Consiste en la estrategia más simple de generación de \emph{muestras}. Se basa en la selección aleatoria de un valor entero $r$ en el rango $[1, |R|]$ para después añadir el elemento localizado en la posición $r$ de la \emph{población} al sub-conjunto de la \emph{muestra}. Este proceso se repite durante $n$ veces para generar una \emph{muestra} de tamaño $n$. A modo de ejemplo se muestra el estimador para la operación \emph{SUMA} en la ecuación \eqref{eq:sum_with_replacement}, además se muestra la fórmula de la desviación para dicho estimador en la ecuación \eqref{eq:sum_with_replacement_deviation}.
            \begin{align}
            \label{eq:sum_with_replacement}
              Y &= \frac{|R|}{n}\sum_jX_jt_j \\
            \label{eq:sum_with_replacement_deviation}
              \sigma^2(Y) &= \frac{|R|^2\sigma^2(R)}{n}
            \end{align}

          \item \textbf{Selección Aleatoria Con Reemplazamiento}: En este caso se supone que la selección de una instancia de la población tan solo se puede llevar a cabo una única vez, por lo tanto se cumple que $\forall X_j \in {0,1}$. La selección se lleva a cabo de la siguiente manera: se genera de manera aleatoria un valor entero $r$ en el rango $[1, |R|]$ para después añadir el elemento localizado en la posición $r$ de la \emph{población} al sub-conjunto de \emph{muestra} si este no ha sido añadido ya, sino volver a generar otro valor $r$. Después repetir dicha secuencia durante $n$ veces para generar una \emph{muestra} de tamaño $n$. Al igual que en la estrategia anterior, en este caso también se muestra el estimador para la operación \emph{SUMA} en la ecuación \eqref{eq:sum_without_replacement}. Nótese que el cálculo es el mismo que en el caso de la estrategia sin reemplazamiento. Sin embargo, la varianza obtenida a partir de dicha estrategia es menor tal y como se muestra en la ecuación \eqref{eq:sum_without_replacement_deviation}.
            \begin{align}
            \label{eq:sum_without_replacement}
              Y &= \frac{|R|}{n}\sum_jX_jt_j \\
            \label{eq:sum_without_replacement_deviation}
              \sigma^2(Y) &= \frac{|R|(|R| - n)\sigma^2(R)}{n}
            \end{align}

          \item \textbf{Bernoulli y Poisson}: Mediante esta alternativa de muestreo se sigue una estrategia completamente distinta a las anteriores. En lugar de seleccionar la siguiente instancia aleatoriamente de entre todas las posibles, se decide generar $|R|$ valores aleatorios $r_j$ independientes en el intervalo $[0,1]$ de tal manera que si $r_j$ es menor que un valor $p_j$ fijado a priori, la instancia se añade al conjunto de \emph{muestra}. Cuando se cumple que $\forall i, j \ p_i = p_j$ se dice que es un muestreo de \emph{Bernoulli}, mientras que cuando no se cumple dicha condición se habla de muestreo de \emph{Poisson}. El cálculo del estimador para la \emph{SUMA} en este caso es muy diferente de los ilustrados anteriormente tal y como se muestra en la ecuación \eqref{eq:sum_bernoulli_poisson}. La desviación de este estimador se muestra en la ecuación \eqref{eq:sum_bernoulli_poisson_deviation}, que en general presenta peores resultados (mayor desviación) que mediante otras alternativas, sin embargo, esta posee la cualidad de poder aplicar distintos pesos a cada instancia de la población, lo que puede traducirse en que una selección adecuada de los valores $p_j$, lo cual puede mejorar significativamente la precisión de los resultados si estos se escogen de manera adecuada.
            \begin{align}
            \label{eq:sum_bernoulli_poisson}
              Y &= \sum_{i \in muestra }\frac{t_i}{p_i} \\
            \label{eq:sum_bernoulli_poisson_deviation}
              \sigma^2(Y) &= \sum_i(\frac{1}{p_i}-1)t_i^2
            \end{align}

          \item \textbf{Muestreo Estratificado}: El muestreo estratificado trata de minimizar al máximo las diferencias entre la distribución del conjunto de datos de la \emph{población} respecto de la \emph{muestra} que se pretende generar. Para ello existen distintas alternativas entre las que se encuentra una selección que  actualiza los pesos $p_j$ tras cada iteración, lo que reduce drásticamente la desviación de la \emph{muestra}, sin embargo produce un elevado coste computacional para su generación. Por tanto, existen otras estrategia más intuitivas basada en la partición del conjunto de datos de la \emph{población} en sub-conjuntos disjuntos que poseen la cualidad de tener varianza mínima a los cuales se denomina \emph{estratos}. Posteriormente, se selecciona mediante cualquiera de los métodos descritos anteriormente una \emph{muestra} para cada \emph{estrato}, lo cual reduce en gran medida la desviación típica global del estimador.

        \end{itemize}

        \paragraph{}
        La estrategia de sumarización de información mediante \emph{muestreo} tiene como ventajas la independencia de la complejidad con respecto a la dimensionalidad de los datos (algo que como se ilustrará con en posteriores secciones no sucede con el resto de alternativas) además de su simplicidad conceptual. También existen cotas de error para las consultas, para las cuales no ofrece restricciones en cuanto al tipo de consulta (debido a que se realizan sobre un sub-conjunto con la misma estructura que el global). El muestre es apropiado para conocer información general acerca del conjunto de datos. Además, presenta la cualidad de permitir su modificación y adaptación en tiempo real, es decir, se pueden añadir o eliminar nuevas instancias de la muestra conforme se añaden o eliminan del conjunto de datos global.

        \paragraph{}
        Sin embargo, en entornos donde el ratio de adicciones/eliminaciones es muy elevado el coste computacional derivado del mantenimiento de la muestra puede hacerse poco escalable. El \emph{muestreo} es una buena alternativa para conjuntos de datos homogéneos, en los cuales la presencia de valores atípicos es irrelevante. Tampoco obtiene buenos resultados en consultas relacionadas con el conteo de elementos distintos. En las siguientes secciones se describen alternativas que resuelven de manera más satisfactoria estas dificultades y limitaciones.

      \subsection{Histogram}
      \label{sec:histogram}

        \paragraph{}
        Los \emph{histogramas} son estructuras de datos utilizadas para sumarizar grandes conjuntos de datos mediante el mantenimiento de tablas de frecuencias, por lo que tienen un enfoque completamente diferente al que siguen las estrategias de \emph{muestreo} de la sección anterior. En este caso, el concepto es similar a la visión estadística de los histogramas. Consiste en dividir el dominio de valores que pueden tomar las instancias del conjunto de datos en intervalos o contenedores disjuntos entre si de tal manera que se mantiene un conteo del número de instancias pertenecientes a cada partición.

        \paragraph{}
        Durante el resto de la sección se describen de manera resumida distintas estrategias de estimación del valor de las particiones, así como las distintas estrategias de particionamiento del conjunto de datos. Para llevar a cabo dicha tarea, a continuación se describe la notación que se ha seguido en esta sección: Sea $D$ el conjunto de datos e $i \in [1,M]$ cada una de las categorías que se presentan en el mismo. Denotaremos por $g(i)$ al número de ocurrencias de la categoría $i$. Para referirnos a cada uno de las particiones utilizaremos la notación $S_j$ para $j \in [1, B]$. Nótese por tanto que $M$ representa el cardinal de categorías distintas mientras que $B$ representa el cardinal de particiones utilizadas para \say{comprimir} los datos. La mejora de eficiencia en cuanto a espacio se consigue debido a la elección de $B \ll M$

        \paragraph{}
        Cuando se hablamos de \emph{esquemas de estimación} nos estamos refiriendo a la manera en que se almacena o trata el contenido de cada una de las particiones $S_j$ del histograma. La razón por la cual este es un factor importante a la hora de caracterizar un histograma es debida a que está altamente ligada a la precisión del mismo.

        \begin{itemize}

          \item \textbf{Esquema Uniforme}: Los esquemas que presuponen una distribución uniforme de las instancias dentro del contenedor se subdividen en dos categorías: \begin{enumerate*} [label=\itshape\alph*\upshape)]
        			\item \emph{continous-value asumption} que presupone que todas las categorías $i$ contenidas en la partición $S_j$ presentan el mismo valor para la función $g(i)$ y
        			\item \emph{uniform-spread asumption} que presupone que el número de ocurrencias de la partición $S_j$ se localiza distribuido uniformemente al igual que en el caso anterior, pero en este caso entre los elementos de un sub-conjunto $P_j$ generado iterando con un determinado desplazamiento $k$ sobre las categorías $i$ contenidas en $S_j$.
        		\end{enumerate*} El segundo enfoque presenta mejores resultados en el caso de consultas de cuantiles que se distribuyen sobre más de una partición $S_j$

          \item \textbf{Esquema Basado en Splines}: En la estrategia basada en splines se presupone que los valores se distribuyen conforme una determinada función lineal de la forma $y_j = a_jx_j + b_j$ en cada partición $S_j$ de tal manera que el conjunto total de datos $D$ puede verse como una función lineal a trozos y continua, es decir, los extremos de la función en cada partición coinciden con el anterior y el siguiente. Nótese que en este caso se ha descrito la estrategia suponiendo el uso de una función lineal, sin embargo esta puede extenderse a funciones no lineales.

          \item \textbf{Esquema Basado en Árboles}: Consiste en el almacenamiento de las frecuencias de cada partición $S_j$ en forma de árbol binario, lo cual permite seleccionar de manera apropiada el nivel del árbol que reduzca el número de operaciones necesarias para obtener la estimación del conteo de ocurrencias según el la amplitud del rango de valores de la consulta. La razón por la cual se escoje un árbol binario es debida a que se puede reducir en un orden de $2$ el espacio necesario para almacenar dichos valores manteniendo únicamente los de una de las ramas de cada sub-árbol. La razón de ello es debida a que se puede recalcular el valor de la otra mediante una resta sobre el valor almacenado en el nodo padre y la rama que si contiene el valor.

          \item \textbf{Esquema Heterogéneo}: El esquema heterogéneo se basa la intuición de que la distribución de frecuencias de cada una de las particiones $S_j$ no es uniforme y tiene peculiaridades propias, por lo tanto sigue un enfoque diferente en cada una de ellas tratanto de minimizar al máximo la tasa de error producida. Para ello existen distintas heurísticas basadas en distancias o teoría de la información entre otros.

        \end{itemize}

        \paragraph{}
        Una vez descritas distintas estrategias de estimación del valor de frecuencias de una determinada partición $S_j$, el siguiente paso para describir un \emph{histograma} es realizar una descripción acerca de las distintas formas de generación de las particiones o contenedores. Para tratar de ajustarse de manera más adecuada a la distribución de los datos se puede realizar un \emph{muestreo} a partir del cual se generan las particiones. A continuación se describen las técnicas más comunes para la elaboración de dicha tarea:

        \begin{itemize}

          \item \textbf{Particionamiento Heurístico}: Las estrategias de particionamiento heurístico se basan en el apoyo sobre distintas presuposiciones que en la práctica han demostrado comportamientos aceptables en cuanto al nivel de precisión que se obtiene en los resultados, sin embargo, no proporcionan ninguna garantía desde el punto de vista de la optimalidad. Su uso está ampliamente extendido debido al reducido coste computacional. Dentro de esta categoría las heurísticas más populares son las siguientes:
            \begin{itemize}

              \item \textbf{Equi-Width}: Consiste en la división del dominio de categorías $[1,M]$ en particiones equi-espaciadas unas de otras. Para dicha estrategia tan solo es necesario conocer \emph{a-priori} el rango del posible conjunto de valores. Es la solución con menor coste computacional, a pesar de ello sus resultados desde el punto de vista práctico son similares a otras estrategias más sofisticadas cuando la distribución de frecuencias es uniforme.

              \item \textbf{Equi-Depth}: Esta estrategia de particionamiento requiere conocer la distribución de frecuencias \emph{a-priori} (o una aproximación que puede ser obtenida mediante alguna estrategia como el muestreo). Se basa en la división del dominio de valores de tal manera que las particiones tengan la misma frecuencia. Para ello se crean particiones de tamaños diferentes.

              \item \textbf{Singleton-Bucket}: Para tratar de mejorar la precisión esta estrategia de particionamiento se basa en la utilización de dos particiones especiales, las cuales contienen las categorías de mayor y menor frecuencia respectivamente para después cubrir el resto de categorías restante mediante otra estrategia (generalmente \emph{equi-depth}) lo cual se basa en la idea de que estas particiones especiales contendrán los valores atípicos y el resto de la muestra será más uniforme.

              \item \textbf{Maxdiff}: En este caso, el método de particionamiento se basa en la idea de utilizar los puntos de mayor variación de frecuencias mediante la medida $|g(i+1) - g(i)|$, para dividir el conjunto de categorías en sus respectivas particiones, de tal manera que las frecuencias contenidas en cada partición sean lo más homogéneas posibles entre sí.

            \end{itemize}
          \item \textbf{Particionamiento con Garantías de Optimalidad}: En esta categoría se enmarcan las estrategias de generación de particiones que ofrecen garantías de optimalidad a nivel de la precisión de resultados en las consultas. Para ello se utilizan técnicas de \emph{Programación Dinámica} (DP), de tal manera que la selección de las particiones se presenta como un problema de \emph{Optimización}. Sin embargo, dichas estrategias conllevan un elevado coste computacional que muchas veces no es admisible debido al gran tamaño del conjunto de datos que se pretende sumarizar. Como solución ante dicha problemática se han propuesto distintas estrategias que se basan en la resolución del problema de optimización, pero sobre una \emph{muestra} del conjunto de datos global, lo cual anula las garantías de optimalidad pero puede ser una buena aproximación si la muestra seleccionada es altamente representativa respecto de la población.

          \item \textbf{Particionamiento Jerárquico}: Las estrategias de particionamiento jerárquico se basan en la utilización de particiones siguiendo la idea de utilizar un árbol binario. Por lo tanto, las particiones no son disjuntas entre ellas, sino que se contienen unas a otras. Esto sigue la misma idea que se describió en el apartado de \emph{Esquemas de estimación Basados en Árboles}. Apoyándose en esta estrategia de particionamiento se consigue que las consultas de rangos de frecuencias tengan un coste computacional menor en promedio (aún en el casos en que el rango sea muy amplio). En esta categoría destacan los histogramas \emph{nLT} (n-level Tree) y \emph{Lattice Histograms}. Estos últimos tratan de aprovechar las ventajas a nivel de flexibilidad y precisión que presentan los histogramas, además de las estrategias jerárquicas de sumarización en que se apoyan las \emph{Wavelets} tal y como se describe en la siguiente sección.

        \end{itemize}

        \paragraph{}
        Las ideas descritas en esta sección sobre los \emph{histogramas} son extrapolables conforme se incrementa la dimensionalidad de los datos. En el caso de los esquemas de estimación, esto sucede de manera directa. Sin embargo, en el caso de los esquemas de particionamiento surgen distintos problemas debido al crecimiento exponencial tanto a nivel de espacio como de tiempo conforme aumenta el número de dimensiones de los datos por lo cual se convierten en soluciones impracticables para conjuntos de datos con elevada dimensionalidad

        \paragraph{}
        Los \emph{Histogramas} representan una estrategia sencilla, tanto a nivel de construcción como de consulta, la cual ofrece buenos resultados en un gran número de casos. Dichas estructuras han sido ampliamente probadas para aproximación de consultas relacionadas con suma de rangos o frecuencias puntuales. Tal y como se ha dicho previamente, su comportamiento en el caso unidimensional ha sido ampliamente estudiado, sin embargo, debido al crecimiento exponencial a nivel de complejidad conforme las dimensiones del conjunto de datos aumentan estas estrategias son descartadas en muchas ocasiones. Los \emph{Histogramas} requieren de un conjunto de parámetros fijados \emph{a-priori}, los cuales afectan en gran medida al grado de precisión de los resultados (pero cuando se seleccionan de manera adecuada esta solución goza de una gran cercanía al punto de optimalidad), por tanto, en casos en que la estimación de dichos valores necesarios \emph{a-priori} se convierte en una labor complicada, existen otras técnicas que ofrecen mejores resultados.

      \subsection{Wavelet}
      \label{sec:wavelet}

        \paragraph{}
        Las estructuras de sumarización denominadas \emph{Wavelets}, a diferencia de las anteriores, han comenzado a utilizarse en el campo del \emph{procesamiento de consultas aproximado} desde hace relativamente poco tiempo, por lo que su uso no está completamente asentado en soluciones comerciales sino que todavía están en fase de descubrimiento e investigación. Las \emph{Wavelets} (u \emph{ondículas}) se apoyan en la idea de representar la tabla de frecuencias del conjunto de datos como una función de ondas discreta. Para ello, se almacenan distintos valores (dependiendo del tipo de \emph{Wavelet}) que permiten reconstruir la tabla de frecuencias tal y como se describirá a continuación cuando se hable de la \emph{transformada de Haar}, la mejora de eficiencia en cuanto a espacio a partir de esta estructura de sumarización se basa en el mantenimiento aproximado de los valores que permiten reconstruir el conjunto de datos.

        \paragraph{}
        A continuación se describe la \emph{transformada de Haar}, a partir de la cual se presentan las distintas ideas en que se apoyan este tipo de estructuras de sumarización. En los últimos años se ha trabajado en estrategias más complejas como la \emph{Daubechies Wavelet} \cite{akansu2001multiresolution} de \emph{Akansu y otros} o la \emph{transformada de Wavelet basada en árboles duales completos} \cite{selesnick2005dual} de \emph{Selesnick y otros}.

        \subsubsection{Haar Wavelet Transform}

          \paragraph{}
          La \emph{Haar Wavelet Transform} (\textbf{HWT}) consiste en una construcción de carácter jerárquico que colapsa las frecuencias de las distintas categorías de manera pareada recursivamente hasta llegar a un único elemento. Por tanto, la idea es la similar a la creación de un árbol binario desde las hojas hasta la raiz. Esta estrategia es similar a la que siguen los \emph{Histogramas jerárquicos} de la sección anterior. Además, se aprovecha de manera equivalente al caso de los histogramas jerárquicos para optimizar el espacio, consistente en almacenar únicamente la variación de uno de los nodos hoja con respecto del padre, lo cual permite reconstruir el árbol completo mediante una simple operación.

          \paragraph{}
          Para simplificar el entendimiento de la construcción de la \emph{transformada de Haar} se describe un ejemplo Extraído del libro \emph{Synopses for massive data} \cite{cormode2012synopses} de \emph{Cormode y otros}. Supongamos los valores de frecuencias recogidos en $A = [2,2,0,2,3,5,4,4]$. Para construir la transformada realizaremos la media de los elementos contiguos dos a dos recursivamente, de tal manera que para los distintos niveles obtenemos los resultados de la tabla \ref{table:wavelet_example}. Además, se muestran los coeficientes de detalle, los cuales se obtienen tras calcular la diferencia entre el primer y segundo elemento contiguos del nivel anterior.

          \begin{table}[H]
            \centering
            \begin{tabular}{| c | c | c |}
              \hline
              Nivel & Medias & Coeficientes de Detalle  \\ \hline \hline
              $3$ & $[2,2,0,2,3,5,4,4]$ & $-$         \\ \hline
              $2$ & $[2,1,4,4]$         & $[0,-1,-1,0]$ \\ \hline
              $1$ & $[3/2,4]$           & $[1/2, 0]$    \\ \hline
              $0$ & $[11/4]$            & $[-5/4]$      \\
              \hline
            \end{tabular}
            \caption{Ejemplo de construcción de \emph{Haar Wavelet Transform}}
            \label{table:wavelet_example}
          \end{table}

          \paragraph{}
          Nótese que a partir de la media de nivel 0 que denominaremos $c_0 = 11/4$ así como el conjunto de coeficientes de detalle, que denotaremos por $c_1 = -5/4, c_2 = 1/2, ..., c_7 = 0$ y los coeficientes de detalle es posible reconstruir la tabla de frecuencias $A$.

        \paragraph{}
        Una vez entendida la estrategia de construcción en que se apoya la \emph{transformada de Haar}, se puede apreciar que esta no ofrece ventajas a nivel de coste de almacenamiento respecto del conjunto de frecuencias respecto del cual ha sido construida. Sin embargo, posee la siguiente cualidad, sobre la cual se apoya esta estrategia de sumarización: \emph{Para las categorías contiguas en que la variación de frecuencias es muy reducida, los coeficientes de detalle tienden a aproximarse a $0$}.

        \paragraph{}
        Por la razón descrita en el parrafo anterior, se intuye que dichos coeficientes de detalle pueden ser obviados, de tal manera que el espacio utilizado para el almacenamiento de la \emph{Wavelet} se convierte en sub-lineal ($o(N)$), en lugar de lineal ($O(N))$ respecto del espacio del conjunto de datos. Para elegir qué coeficientes de detalle se pueden utiliza estrategias que tratan de minimizar el error. Comúnmente, las \emph{Wavelets} han sido construidas a partir del \emph{error quadrático medio} o \emph{norma-$L_2$}, la cual se describe en la ecuación \eqref{eq:l2_error}. Sin embargo, distintos estudios como el realizado en el artículo \emph{Probabilistic wavelet synopses} \cite{garofalakis2004probabilistic} de \emph{Garofalakis y otros} muestran como esta medida del error obtiene malos resultados cuando se aplica a la sumarización de datos mediante \emph{Wavelets}.

        \paragraph{}
        Por tanto, se proponen otras medidas de error como la minimización del máximo error absoluto o relativo, que se ilustran en las ecuaciones \eqref{eq:abs_error} y \eqref{eq:rel_error}. También se propone como alternativa la minimización de la \emph{norma-$L_p$} que se muestra en la ecuación \eqref{eq:lp_error}. Dicha medida de error es una generalización del \emph{error cuadrático medio} (caso $p=2$) a cualquier valor de $p \geq 0$. Por último se muestra en la ecuación \eqref{eq:lp_w_error} el caso del cálculo del error mediante la \emph{norma-$L_p$} con pesos o ponderada, lo cual permite fijar el grado de importancia para cada categoría en la representación mediante \emph{Wavelets} permitiendo aumentar la precisión de las categorías más significativas.

        \begin{align}
        \label{eq:l2_error}
          ||A - \widetilde{A} ||_2  &= \sqrt{\sum_{i}(A[i]-\widetilde{A}[i])^2} \\
        \label{eq:abs_error}
          max_i\{absErr_i\} &= max_i\{|A[i]-\widetilde{A}[i]|\} \\
        \label{eq:rel_error}
          max_i\{relErr_i\} &= max_i\bigg\{\frac{|A[i]-\widetilde{A}[i]|}{|A[i]|} \bigg\} \\
        \label{eq:lp_error}
          ||A - \widetilde{A} ||_{p}  &= (\sum_{i}(A[i]-\widetilde{A}[i])^p)^{\frac{1}{p}} \\
        \label{eq:lp_w_error}
          ||A - \widetilde{A} ||_{p,w}  &= (\sum_{i}w_i \cdot(A[i]-\widetilde{A}[i])^p)^{\frac{1}{p}}
        \end{align}

        \paragraph{}
        Al igual que en el caso de los \emph{Histogramas}, las \emph{Wavelets} presentan problemas de eficiencia cuando se usan en entornos en los cuales el conjunto de datos está compuesto por una gran número de atributos. Por tanto, se dice que sufren la \emph{Maldición de la dimensionalidad} (\emph{Curse of Dimensionality}), que provoca un crecimiento de orden exponencial en el coste tanto en espacio como en tiempo.

        \paragraph{}
        Tal y como se puede apreciar, esta estrategia es muy similar a la basada en \emph{Histogramas}, dado que ambas se apoyan en el almacenamiento de valores que tratan de describir o resumir la tabla de frecuencias de los datos de manera similar. Sin embargo, mientras que en el caso de los \emph{Histogramas} estos destacan cuando se pretende conocer la estructura general de los datos, las \emph{Wavelets} ofrecen muy buenos resultados cuando se pretenden conocer valores atípico o extremos (a los cuales se denomina \emph{Heavy Hitters}).

        \paragraph{}
        Por su estrategia de construcción, las \emph{Wavelets} permiten sumarizar una mayor cantidad de información utilizando un espacio reducido. Además, en el caso de la \emph{transformada de Haar}, que posee la característica de linealidad, se puede adaptar de manera sencilla al \emph{modelo en Streaming}. Tal y como se ha dicho en el párrafo anterior, las desventajas de esta alternativa derivan en gran medida de los problemas relacionados con el incremento de la dimensionalidad de los datos.

      \subsection{Sketch}
      \label{sec:sketch}

        \paragraph{}
        Las estructuras de sumarización conocidas como \emph{Sketches} son las que menos tiempo de vida tienen de entre las descritas, por lo tanto, aún están en una fase muy temprana por lo que su uso en sistemas reales todavía es anecdótico. Sin embargo poseen distintas características por las que se piensa que en el futuro las convertirán en estrategias muy importantes el el ámbito del \emph{procesamiento aproximado de consultas}. Los \emph{Sketches} se amoldan perfectamente al modelo en streaming del cual se habla en el capítulo anterior en la sección \ref{sec:streaming_model}. Este modelo se amolde perfectamente a muchos sucesos cambiantes que se dan en la actualidad y que requieren de la obtención de analíticas. Un ejemplo de ello es un sistema de transacciones financieras, que suceden con una frecuencia elevada y para las cuales sería apropiado obtener métricas en tiempo real para la mejora de la toma de decisiones. También encajan de manera apropiada en el modelo de transacciones de una base de datos, la cual es modificada constantemente mediante inserciones, modificaciones y eliminaciones.

        \paragraph{}
        Los \emph{Sketches} se corresponden con estructuras de datos que funcionan manteniendo estimadores sobre cada una de las instancias que han sido procesadas hasta el momento, es decir, realizan una modificación interna por cada entrada. Esto se opone a las estrategias descritas en anteriores secciones, que procesan una parte del conjunto de datos o requieren que el mismo tenga un carácter estático. Estas modificaciones pueden enmarcarse bajo distintas especificaciones del modelo en streaming tal (serie temporal, modelo de caja registradora o modelo de molinete) teniendo que realizar pequeñas adaptaciones en algunos algunos \emph{Sketches}, pero siendo muy poco escalable la implementación del modelo de molinete (que permite eliminaciones) en otros.

\begin{figure}
          \centering
          \includegraphics[width=0.5\textwidth]{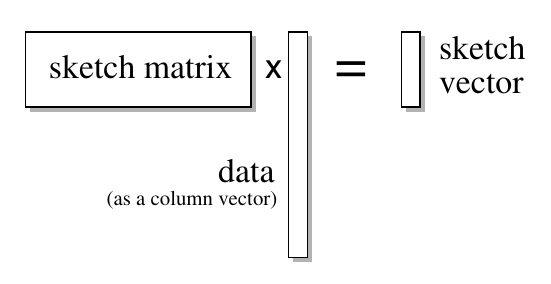}
          \caption{Abstracción del \emph{Sketch Lineal}. \emph{(extraída de \cite{cormode2012synopses})}}
          \label{fig:linear_sketch}
        \end{figure}

        \paragraph{}
        Los \emph{Sketches Lineales} son un sub-conjunto de \emph{Sketches} que pueden ser vistos como una transformación lineal de la estructura de sumarización, la cual puede ser interpretada como un vector de longitud $1*n$ al cual denominaremos $S$. Para generar dicho vector es nos apoyamos en la matriz de tamaño $n*m$ a la cual denominaremos $A$ y que representa la transformación lineal del conjunto de datos, el cual puede ser codificado como un vector al cual denominaremos $D$ de tamaño $1xm$. Dicha transformación lineal se representa en la ecuación \eqref{eq:linear_sketch} y de manera gráfica en la figura \ref{fig:linear_sketch}.

        \begin{equation}
        \label{eq:linear_sketch}
          A  * D = S
        \end{equation}

        \paragraph{}
        Puesto que la transformación que se muestra es de carácter lineal, entonces posee la propiedad de que siendo $S_1$ y $S_2$ dos \emph{Sketches} lineales, entonces se cumple la propiedad de que $S_1 + S_2 = S_{sum}$, lo que puede entenderse como la combinación de los estimadores de los dos \emph{Sketches}. Por tanto, se puede decir que el cada actualización se procesa como una transformación de la nueva instancia codificada en el vector $D$ para después combinar el \emph{Sketch} resultante con el generado anteriormente.

        \paragraph{}
        Por motivos de eficiencia a nivel de espacio, la matriz de transformación $A$ no se utiliza de manera explicita, sino que por contra, en muchos casos se utilizan distintas \emph{funciones hash} (las cuales se describen en la sección \ref{sec:hash_functions}), que realizan la transformación de la misma forma que la matriz $A$, solo que el coste computacional para utilizarlas es significativamente menor que el coste necesario para el almacenamiento y multiplicación de una matriz de tal tamaño.

        \paragraph{}
        Tal y como se puede apreciar en la figura \ref{fig:linear_sketch} la ventaja de sumarización de esta estrategia se obtiene en la reducción de tamaño obtenida en la transformación lineal. Posteriormente, cada estructura de \emph{Sketch} tiene asociadas un conjunto de \emph{consultas o queries} para las cuales extraer la información valiosa contenida en ellas. Puesto que la forma en que se llevan a cabo dichas consultas varía según la alternativa a la que nos estamos refiriendo, estas técnicas se expondrán en profundidad en sus correspondientes secciones.

        \paragraph{}
        Siguiendo con las restricciones descritas anteriormente, existe una implementación trivial de las estructuras de \emph{Sketch} consistente en mantener una tabla de frecuencias sobre cada posible instancia de entrada de tal manera que a partir de esta pueden resolver consultas como sumas de frecuencias, conteo de elementos distintos, búsqueda del máximo o mínimo, media o varianza del conjunto de datos. Sin embargo, esta solución no proporciona ninguna ventaja a nivel de espacio, por tanto, la tarea es encontrar técnicas que permitan \say{colapsar} dicha tabla de frecuencias para tratar de minimizar dicho coste asumiendo un cierto grado de imprecisión.

        \paragraph{}
        Existen dos categorías principales según la naturaleza de las consultas que el \emph{Sketch} puede responder. Dichas categorías se describen a continuación:

        \begin{itemize}

          \item \textbf{Estimación de Frecuencias}: Los \emph{Sketches} basados en estimación de frecuencias se corresponden con aquellos que tratan de recoger estimadores que se aproximen a la frecuencia de cada elementos $f(i)$. En esta categoría se enmarcan el \emph{Count-Min Sketch} y \emph{Count Sketch}.

          \item \textbf{Elementos Distintos}: Los \emph{Sketches} basados en el conteo de elementos distintos (y que también pueden responder consultas referidas a la presencia de un determinado elemento) se basan en el mantenimiento de distintos indicadores que permiten aproximar dichos valores. Entre los más destacados en esta categoría se encuentran el \emph{Bloom Filter} y el \emph{HyperLogLog}.

        \end{itemize}

        \paragraph{}
        A pesar de que todos los \emph{Sketches} siguen una estructura básica similar, existen distintos parámetros diferenciadores que caracterizan unas alternativas frente a otras entre las que destacan las consultas soportadas, el tamaño de los mismos (algunos necesitan menor espacio para obtener un grado de precisión similar a otros), el tiempo de procesamiento de cada nueva actualización (el cual se pretende que sea muy reducido debido al modelo en streaming), el tiempo de resolución de consultas o la estrategia de inicialización del mismo.

        \paragraph{}
        La extensión de los \emph{Sketches} sobre conjuntos de datos de carácter multidimensional puede resolverse utilizando funciones hash que mapeen una entrada multidimensional sobre un espacio unidimensional para después utilizarlos de la manera que para los casos de una dimensión. Esta solución no presenta problemas en la estimación de frecuencias puntuales o búsqueda de máximos o mínimos, sin embargo, no es factible para consultas de sumas de rangos utilizando \emph{sketches} multidimensionales o asumiendo la independencia entre dimensiones y manteniendo una estructura de \emph{sketch} por cada dimensión.

        \paragraph{}
        Los \emph{Sketches} presentan un gran nivel de simplicidad en sus implementaciones, lo que les hace muy eficientes y válidos para modelos que requieren de actualizaciones constantes como el descrito en el modelo en streaming. La dificultad de los mismos se basa en los conceptos matemáticos subyacentes, que complica en gran medida la demostración de sus características de convergencia hacia el valor real que estiman. Generalmente para reducir su espacio se apoyan en la utilización de funciones hash con distintas propiedades, pero que también se pueden calcular eficientemente. La mayor limitación se refiere al rango de consultas que cada tipo de \emph{Sketch} puede resolver de manera eficiente, estando especialmente limitados en el ámbito de múltiples consultas anidadas.

        \paragraph{}
        Los \emph{Sketches} son un ámbito interesante dentro del mundo de la investigación, con una visión de futuro muy prometedora que les coloca como la mejor solución a medio-largo plazo para la tarea del \emph{procesamiento de consultas aproximadas} por su reducido coste computacional, tanto a nivel de espacio como de tiempo de procesamiento. Actualmente universidades prestigiosas como la \emph{Universidad de California, Berkeley} o el \emph{MIT} están trabajando en el diseño de bases de datos que se apoyan fuertemente en el uso de estas estructuras. Dicha universidad está trabajando en una base de datos a la cual denominan \emph{BlinkDB} y la cual describen en el artículo \emph{BlinkDB: queries with bounded errors and bounded response times on very large data} \cite{agarwal2013blinkdb}

    \paragraph{}
    En las siguientes secciones se describen en profundidad distintas estructuras de datos basadas en \emph{Sketches} tratando de prestar especial importancia en una visión conceptual de la misma pero sin dejar de lado la parte práctica de las mismas. Además, se trata de incluir demostraciones acerca de la precisión de las mismas respecto del espacio utilizado para su almacenamiento respecto del tamaño del conjunto de datos global.

    \section{Bloom Filter}
    \label{sec:bloom_filter}

      \paragraph{}
      El \emph{Bloom-Filter} se corresponde con la primera estructura de datos basada en la idea de \emph{Sketch} por procesar cada nueva entrada de la misma forma y tener un coste a nivel de espacio sub-lineal ($o(N)$) respecto del cardinal de posibles valores que pueden ser procesados. Esta estructura de datos fue diseñada inicialmente por \emph{Burton Howard Bloom} la cual describe en el artículo \emph{Space/time trade-offs in hash coding with allowable errors} \cite{bloom1970space} publicado en 1970.

      \paragraph{}
      La funcionalidad que suministra dicha estructura de datos es la consulta acerca de presencia de un determinado elemento. Para la implementación descrita en el artículo inicial no se permiten incrementos ni eliminaciones, tan solo la existencia o inexistencia del elemento, por tanto esta estructura de datos se enmar en el marco de \emph{serie temporal} del \emph{modelo en streaming}. El \emph{Bloom-Filter} asegura la inexistencia de falsos negativos (elementos que se presupone que no han sido introducidos cuando en realidad si que lo fueron) pero por contra, se admite una determinada tasa de falsos positivos (elementos que se presupone que han sido introducidos cuando en realidad no lo fueron).

      \paragraph{}
      Debido al tiempo de vida de dicha estrategia, su uso está muy asentado y se utiliza en distintos sistemas para tareas en las cuales se pretende optimizar el tiempo, pero sin embargo son admisibles fallos. Es comúnmente utilizado en bases de datos comerciales para limitar el número de accesos al disco durante búsquedas de elementos. Esto se lleva a cabo manteniendo un \emph{Bloom-Filter} al cual se consulta sobre la presencia del elemento a buscar y en el caso de que la respuesta sea negativa, entonces se prescinde de dicho acceso a la unidad de almacenamiento. Mientras que si es positiva se realiza el acceso a disco. Nótese por tanto que en algunos ocasiones dicho acceso será innecesario, sin embargo las consecuencias de dicha tarea tan solo tienen consecuencias negativas a nivel de eficiencia y no de resultados.

      \paragraph{}
      Una vez descrita la funcionalidad que proporciona el \emph{Bloom-Filter} así como un caso práctico de uso del mismo ya se tiene una idea a grandes rasgo acerca del mismo. Por tanto, lo siguiente es describir su composición. La estructura de datos se basa en el mantenimiento de un \emph{mapa de bits} $S$ de longitud $n$ ($|S| = n$) de tal manera que se cumpla que $n \ll m$ siendo $m$ el cardinal de elementos distintos que se pueden presentar en la entrada. Además, se utilizan $k$ funciones hash denominadas $h_1, h_2,..., h_j,..., h_k$ que distribuyen de manera independiente el elemento $i$-ésimo en $S$. El mapa de bits $S$ se inicializa con todas sus posiciones tomando el valor $0$.

      \paragraph{}
      El modo de funcionamiento del \emph{Bloom-Filter} se lleva a cabo de la siguiente manera: Cada vez que se presenta un nuevo elemento en la entrada (al cual denominaremos $i \in [1, m]$) se tomaran los $k$ valores de las funciones funciones hash indicadas anteriormente y se asignará el valor $1$ a dichas posiciones del mapa de bits $S$. Es decir, se realiza la operación descrita en la ecuación \eqref{eq:bloom_filter_update}. La consulta acerca de la presencia del elemento $i$-ésimo se realiza por tanto, consultando dichas posiciones, de tal manera que si $\forall j \in [1,k], \ S[h_j(i)]$ toma el valor valor $1$ se dice que el elemento ha aparecido en la entrada mientras que si alguna posición es distina de $1$ (toma el valor $0$) se dice que el elemento no ha sido introducido.

      \begin{align}
      \label{eq:bloom_filter_update}
        S[h_j(i)] = 1 && \forall j \in [1,k]
      \end{align}

      \paragraph{}
      Nótese que la restricción acerca de la presencia de un determinado elemento es mucho más débil que la de no existencia. La razón de ello se debe a que pueden existir colisiones entre las distintas funciones hash que conviertan dicho resultado en érroneo. Sin embargo, tal y como se ha indicado anteriormente, en caso de que el \emph{Bloom-Filter} indique la no presencia del elemento dicha respuesta será válida. A continuación se realiza un análisis acerca del índice de error para el caso de los falsos positivos la cual se basa en el artículo \emph{Network applications of bloom filters: A survey} \cite{broder2004network} de \emph{Broder} y \emph{Mitzenmacher}. Nótese que dicho análisis depende de 3 parámtros: el tamaño $n$ del mapa de bits, el cardinal $m$ del conjunto de posibles elementos en la entrada y el número $k$ de funciones hash utilizadas.

      \begin{figure}
        \centering
        \includegraphics[width=0.5\textwidth]{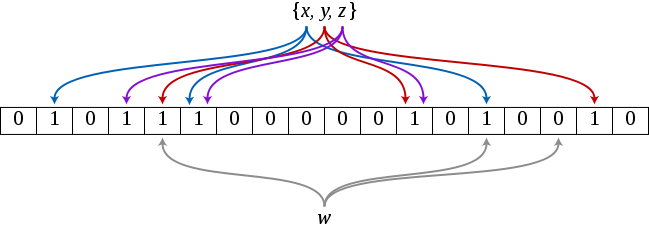}
        \caption{Modo de funcionamiento del \emph{Bloom Filter}, que inserta los valores $\{ x, y, z\}$ y comprueba la existencia de $w$. La imagen ha sido extraída de \cite{wiki:Bloom_filter}.}
        \label{fig:bloom_filter}
      \end{figure}

      \paragraph{}
      Para el análisis se presupone que las funciones hash $h_j$ son totalmente aleatorias, es decir, distribuyen los valores de manera totalmente uniforme en el intervalo $[1,n]$ y son totalmente independientes entre sí. La probabilidad $p'$ de que cualquier posición de $S$ siga siendo igual a cero después de la aparición de $l$ elementos distintos se muestra en la ecuación \eqref{eq:bloom_filter_p_estimator} siendo la base del lado derecho de la igualdad la probabilidad de que cada función hash mantenga con el valor $0$ una posición.

      \begin{equation}
      \label{eq:bloom_filter_p_estimator}
       p' = \bigg(1-\frac{1}{m}\bigg)^{kl}
      \end{equation}

      \paragraph{}
      La probabilidad de que un elemento no introducido en la entrada tome todas sus correspondientes posiciones de $S$ con el valor $1$ se da con probabilidad $(1-p)^k$ y dado que $p'$ es un buen estimador para $p$, entonces podemos aproximar la tasa de falsos positivos. Tras desarrollar dicha expresión se puede llegar a la ecuación \eqref{eq:bloom_filters_false_positives}, que aproxima de manera apropiada la tasa de falsos positivos.

      \begin{equation}
      \label{eq:bloom_filters_false_positives}
        f = (1-e^{-kn/m})^k
      \end{equation}

      \paragraph{}
      El \emph{Bloom-Filter} es una buena aproximación para los casos en que es necesario reducir el sobre coste de comprobación de accesos a medios costosos, sin embargo, su utilización puede utilizarse en filtros anti-spam u otros entornos en que una determinada tasa de error sea admisible. Nótese que este tipo de \emph{Sketches} no puede agruparse en el sub-conjunto de \emph{Sketches Lineales} puesto que la operación de asignar el valor $1$ no es lineal.

    \section{Count-Min Sketch}
    \label{sec:count_min_sketch}

      \paragraph{}
      El \emph{Count-Min Sketch} es otra estructura de datos con la característica de presentar un coste espacial de carácter sub-lineal ($o(N)$) respecto del cardinal de posibles elementos de entrada. El \emph{Count-Min Sketch} (en adelante \emph{CM Sketch} por abreviación) fue descrito por primera vez por \emph{Cormode} y \emph{Muthukrishnan} en el artículo \emph{An improved data stream summary: the count-min sketch and its applications} \cite{cormode2005improved} publicado en 2005. Nótese por tanto que el tiempo de vida de dicha estructura de datos es muy corto.

      \paragraph{}
      El propósito del \emph{CM Sketch} es la estimación de frecuencias para el rango de posibles valores que se presentan en la entrada. En la formulación inicial se enmarcaba modelo de caja registradora, que tan solo permite adicciones, sin embargo posteriormente se han propuesto mejoras para ampliar su uso en entornos en que también se permitan eliminaciones (modelo en molinete). Tal y como se verá a continuación, el nombre de este \emph{Sketch} se refiere a las operaciones de operaciones que utiliza durante su funcionamiento, es decir, el conteo o suma y la búsqueda del mínimo.

      \paragraph{}
      La estimación de frecuencias del \emph{CM Sketch}, tal y como se puede apreciar debido al carácter sub-lineal del mismo, no garantiza la exactitud en de los mismos, sino que al igual que en el caso del \emph{Bloom-Filter} devuelve como resultado una aproximación. En este caso dicha aproximación garantiza que el resultado estimado siempre será mayor o igual que el verdadero, es decir, realiza una sobre-estimación del valor real. La razón por la cual el \emph{CM Sketch} utiliza la operación de búsqueda del mínimo es para tratar de reducir dicho efecto en la medida de lo posible.

      \paragraph{}
      Debido al corto periodo de vida su uso todavía no está asentado en sistemas reales, sin embargo existen una gran variedad de situaciones que se enmarcan sobre el modelo en streaming en las cuales la precisión de frecuencias no tienen grandes efectos perjudiciales. Un ejemplo de ello podría ser el conteo de visitas a un determinado sitio web formado por distintas páginas. La estimación sobre el número de visitas para cada una de estas páginas podría llevarse a cabo utilizando esta estructura de datos, ya que es una tarea que puede admitir una determinada tasa de error y requiere de actualizaciones constantes.

      \paragraph{}
      Una vez descrita la funcionalidad que suministra el \emph{CM Sketch} se describirá la estructura interna del mismo: Esta estructura de datos está formada por una matriz $S$ de estructura bidimensional con $w$ columnas y $d$ filas, de tal manera que se mantienen $w * d$ contadores. Cada una de las $d$ filas tiene asociada una función hash $h_j$ que distribuye elementos pertenecientes al dominio $[1, m]$ (siendo $m$ el cardinal de elementos distintos) sobre la fila de dicha matriz, es decir, sobre $[1,w]$. Para que las estimaciones acerca de la precisión del \emph{CM Sketch} sean correctas cada función hash $h_j$ debe cumplir la propiedad de ser independiente del resto además de seguir una distribución uniforme. En cuanto a la inicialización de la estructura de datos, todas las posiciones toman el valor cero, esto se muestra en la ecuación \eqref{eq:count_min_sketch_init}.

      \begin{align}
      \label{eq:count_min_sketch_init}
        S[k,j] = 0 && \forall k \in [1,w], \ \forall j \in [1,d]
      \end{align}

\begin{figure}
        \centering
        \includegraphics[width=0.5\textwidth]{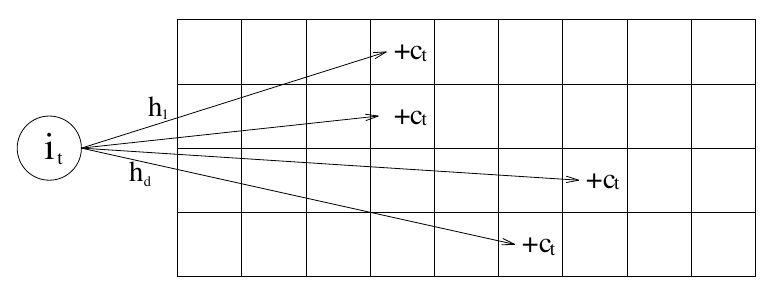}
        \caption{Modo de funcionamiento del \emph{Count-Min Sketch} durante el proceso de inserción de un nuevo elemento. La imagen ha sido extraída de \cite{cormode2005improved}.}
        \label{fig:count_min_sketch}
      \end{figure}

      \paragraph{}
      En cuanto al modo de funcionamiento del \emph{CM Sketch}, tal y como se ha indicado anteriormente, se enmarca en el modelo de caja registradora, es decir, cada actualización se corresponde con una tupla formada por el identificador del elemento al que se refiere y un determinado valor positivo que indica el incremento $c$ al cual se refiere la actualización. El funcionamiento es similar al del \emph{Bloom-Filter} en el sentido de que por cada nueva entrada se realizan $d$ actualizaciones (una para cada fila). La operación de conteo se refiere a dicha actualización, que se ilustra de manera matemática en la ecuación \eqref{eq:count_min_sketch_update}. Una representación gráfica de dicha operación se muestra en la figura \ref{fig:count_min_sketch}. Nótese por tanto que el coste de actualización es de $O(d)$. Nótese que esta operación es de carácter lineal por lo que el \emph{CM Sketch} se agrupa en el sub-conjunto de \emph{Sketches Lineales}.

      \begin{align}
      \label{eq:count_min_sketch_update}
        S[j, h_j(i)] = S[j, h_j(i)] + c && \forall j \in [1,d]
      \end{align}

      \paragraph{}
      En cuanto al proceso de consulta sobre la frecuencia del elemento $i$-ésimo, esto se lleva a cabo recogiendo el valor almacenado en cada fila y devolviendo el menor, de tal manera que se pretende reducir el efecto de las colisiones generados sobre las funciones hash al distribuir elementos sobre un espacio de menor tamaño, ya que se presupone que $w * d \ll m$ para que la ventaja a nivel de espacio se produzca. Dicha operación se muestra en la ecuación \eqref{eq:count_min_sketch_estimate}

      \begin{equation}
      \label{eq:count_min_sketch_estimate}
        \widetilde{f}(i) = min_{j \in [1,d]}\{S[j, h_j(i)]\}
      \end{equation}

      \paragraph{}
      A nivel de análisis sobre el coste espacial necesario para almacenar en memoria el \emph{CM Sketch}, este es dependiente del grado de precisión que se pretenda asegurar mediante el uso del mismo. Sin embargo, el análisis no se realiza sobre el cardinal de posibles elementos que pueden aparecer en la entrada, sino que depende del sumatorio del conteo de estos. Este valor se denomina $F_1$ (o primer momento de frecuencia), que se define tal y como se indica en la ecuación \eqref{eq:count_min_sketch_sum}.

      \begin{equation}
      \label{eq:count_min_sketch_sum}
        F_1 = \sum_{i=1}^m f(i)
      \end{equation}

      \paragraph{}
      Por tanto, la precisión de esta estructura de datos probabilística se indica diciendo que $\widetilde{f}(i)$ tendrá un error máximo de $\epsilon N$ que el cual se cumplirá con probabilidad $1-\delta$. Estos parámetros se fijan en el momento de la inicialización del \emph{CM Sketch} fijando el número de filas y columnas de tal manera que $d = log(1/\delta)$ y $w=2/\epsilon$. La demostración acerca de la elección de estos valores puede encontrarse en el artículo original \cite{cormode2005improved}.

      \paragraph{}
      En cuanto a la estimación de frecuencias obtenida por el \emph{CM Sketch}, en sentido estricto se dice que esta es sesgada respecto de su valor real. La razón se debe a que siempre se produce una sobre-estimación, a pesar de tratar de reducir los efectos negativos de la misma tratando de seleccionar el mínimo de las estimaciones siempre será mayor o igual (nunca menor). Para tratar de solventar esta problemática se han propuesto distintas heurísticas que tratan de contrarrestar esta problemática en el modelo de caja de registradora (tan solo se permiten inserciones).

      \paragraph{}
      Sin embargo, dicho problema no sucede en el caso del modelo general o de molinete, sobre el cual si que están permitidas las eliminaciones. En este caso es más apropiado utilizar como estimación la mediana de los valores almacenados en cada columna, puesto que se escoge el mínimo para un elemento que tan solo ha recibido actualizaciones negativas probablemente este será muy diferente del valor real. La razón por la cual se escoge la mediana y no la media es por sus propiedades de resistencia ante valor atípicos u \emph{outliers}. En este caso la demostración se puede llevar a cabo apoyándose en la desigualdad de Chernoff descrita en la sección \ref{sec:chernoff_inequality}. Esta demostración se puede encontrar en el artículo \emph{Selection and sorting with limited storage }\cite{munro1980selection} de \emph{Munro} y \emph{Paterson}.

      \paragraph{}
      El \emph{Count-Min Sketch} consiste en una estrategia apropiada para el conteo de ocurrencias en un espacio de orden sub-lineal. Además, la implementación del mismo es simple en contraposición con otras estrategias más sofisticadas. En posteriores secciones se hablará del \emph{Count Sketch}, que proporcionan valores de precisión equivalentes en espacios de menor tamaño añadiendo una mayor complejidad conceptual.

    \section{Count Sketch}
    \label{sec:count_sketch}

      \paragraph{}
      El \emph{Count Sketch} se refiere a una estructura muy similar al \emph{Count-Min Sketch} ya que sigue la misma estructura e ideas muy similares en para la actualización y consulta de valores, sin embargo proporciona ventajas a nivel de coste espacial respecto del anterior reduciendo en un orden de dos el espacio necesario para mantener la estructura de datos. La definición acerca del \emph{Count Sketch} se indicó por primera vez en el trabajo \emph{Finding frequent items in data streams} \cite{charikar2002finding} desarrollado por \emph{Charikar y otros}. Nótese que este trabajo fue llevado a cabo en el año 2002, es decir, es anterior a la publicación del referido al \emph{Count-Min Sketch} (2005). La razón por la cual se ha seguido este orden y no el temporal se debe a que el \emph{Count Sketch} puede ser visto como una mejora sobre el descrito en la sección anterior, lo que simplifica el entendimiento de los mismos.

      \begin{figure}
        \centering
        \includegraphics[width=0.5\textwidth]{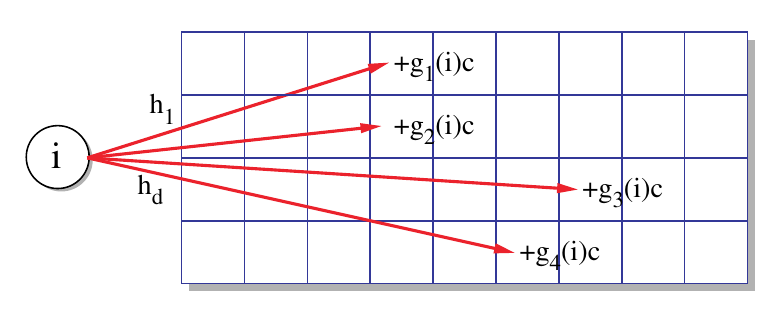}
        \caption{Modo de funcionamiento del \emph{Count Sketch} durante el proceso de inserción de un nuevo elemento. La imagen ha sido extraída de \cite{cormode2012synopses}.}
        \label{fig:count_sketch}
      \end{figure}

      \paragraph{}
      El \emph{Count Sketch} se basa en los mismos elementos para su funcionamiento que su homónimo más simple, por lo tanto se utilizará la definición utilizada previamente de la matriz $S$ compuesta por $w$ columnas y $d$ filas además de las funciones hash $h_1, h_2, ..., h_j,..., h_d$ independientes entre si, y cuya distribución es uniforme. La variación surge a nivel de las operaciones que se realizan en el proceso de actualización y consulta sobre la estructura de datos. El funcionamiento del \emph{Count Sketch} puede ser visto de varias formas, primero describiremos la versión extendida del mismo para después describir una versión que reduce a la mitad el espacio necesario para su mantenimiento en memoria.

      \begin{equation}
      \label{eq:count_sketch_hash_2}
        h'_j(i) =
          \begin{cases}
            h_j(i) - 1, & h_j(i) \ mod \ 2 = 0\\
            h_j(i) + 1, & h_j(i) \ mod \ 2 = 1
          \end{cases}
      \end{equation}

      \paragraph{}
      A continuación se describe la versión extendida, que se apoya en la utilización de la función hash $h'_j(i)$, la cual se define en la ecuación \eqref{eq:count_sketch_hash_2}, que consiste en una variación sobre la función hash base $h_j(i)$. Esta función se utiliza únicamente en el momento de la estimación, es decir, no se utiliza para la actualización tras la llegada de elementos. La intuición en la que se basa su utilización es la reducción del sesgo producido por las colisiones de otros elementos en cada fila, que siempre sobre-estima los resultados (en promedio la sobre-estimación de cada fila es de $\frac{\sum_{k=1}^wS[j,k]}{w}$). De esta manera se pretende contrarrestar dicho efecto por lo que ya no es apropiada la utilización de la búsqueda del mínimo, sino que se puede llevar a cabo la estimación mediante el uso de la mediana, tal y como se indica en la ecuación \eqref{eq:count_sketch_estimation_1}.

      \begin{equation}
      \label{eq:count_sketch_estimation_1}
        \widetilde{f}(i) = median_{j \in [1,d]}\{S[j, h_j(i)] - S[j, h'_j(i)]\}
      \end{equation}

      \paragraph{}
      La versión reducida de esta alternativa se apoya en la utilización de $d$ funciones hash uniformes e independientes entre sí con dominio de entrada $[1,m]$ e imagen en el sub-conjunto $\{-1,1\}$. Si se sustituye el método de actualización de elementos por el descrito en la ecuación \eqref{eq:count_sketch_update} donde $i$ identifica el $i$-ésimo elemento y $c$ la variación del mismo correspondiente a dicha actualización, entonces para conseguir la misma precisión que en el caso anterior se requiere de la mitad de espacio. Puesto que esto es equivalente a mantener $S[j, h_j(i)] - S[j, h'_j(i)]$ en una única posición de $S$, entonces la estimación de la frecuencia se sustituye por la ecuación \eqref{eq:count_sketch_estimation_2}.

      \begin{equation}
      \label{eq:count_sketch_update}
        S[j,h_j(i)] = S[j,h_j(i)] +  g_j(i)c
      \end{equation}

      \begin{equation}
      \label{eq:count_sketch_estimation_2}
      \widetilde{f}(i) = median_{j \in [1,d]}\{S[j, h_j(i)]\}
      \end{equation}

      \paragraph{}
      Para el análisis sobre la precisión del \emph{Count Sketch}, se utiliza un enfoque similar al del \emph{CM Sketch}. Sin embargo, en este caso el valor $\epsilon$ se escoge teniendo en cuanto el segundo momento de frecuencias, es decir, $F_2 = \sum_{i=1}^mf(i)^2$, de tal manera que para estimación obtenida sobre cada fila, esta se encuentra en el intervalo $[f(i) - \sqrt{F_2/w}, f(i) + \sqrt{F_2/w}]$. La estimación global por tanto tiene un error máximo de $\epsilon\sqrt{F_2}$ con probabilidad $1- \delta$.

      \paragraph{}
      Para que se cumpla dicha estimación del error el tamaño de la matriz $S$ debe escogerse de tal forma que el número de columnas sea $w = O(1/\epsilon^2)$ y el número de filas sea $d = O(log(1/\delta))$. La demostración completa acerca de la elección puede encontrarse en el artículo original \cite{charikar2002finding}. Nótese la diferencia en cuanto a la estimación del error del \emph{Count Sketch} con respecto del \emph{CM Sketch}, se debe a que en el primer caso se utiliza el segundo momento de frecuencias ($F_2$) mientras que en el \emph{CM Sketch} la estimación de la tasa de error se apoya en el primer momento de frecuencias ($F_1$).

      \paragraph{}
      En esta implementación del \emph{Count Sketch} tan solo se ha impuesto la restricción a nivel de funciones hash de que estas sean de carácter \emph{2-universales}, lo cual es una condición suficiente para estimar frecuencias puntuales. Sin embargo, esta estructura de datos se puede extender para la estimación del segundo momento de frecuencias ($F_2$) mediante la técnica descrita en la sección \ref{sec:streaming_frecuency_moment_aproximation} extraída del artículo de \emph{Alon y otros} \cite{alon1996space}. Para la estimación de $F_2$ basta con calcular la mediana de las sumas de los elementos de cada fila al cuadrado. Esto se indica en la ecuación \eqref{eq:ams_sketch_f2}. Sin embargo, para que esta solución sea correcta desde el punto de vista de la varianza del estimador es necesario que las funciones hash utilizadas sean \emph{4-universales} (la razón se debe al cálculo de las potencias de grado 2). Cuando se cumple esta propiedad entonces la técnica se denomina \textbf{AMS Sketch}.

      \begin{equation}
      \label{eq:ams_sketch_f2}
        \widetilde{F}_2 = median_{j \in [1,d]}\{\sum_{k=1}^w S[j, k]^2\}
      \end{equation}

      \paragraph{}
      A nivel práctico estas dos alternativas ofrecen resultados muy similares cuando se imponen las mismas restricciones a nivel de espacio. Sin embargo, ambos poseen una característica destacada: En muchos casos, la estimación obtenida es mucho más precisa puesto que muchas distribuciones de probabilidad tienen formas asimétricas, es decir, algunos elementos presentan una frecuencia de aparición elevada mientras que otros aparecen en contadas ocasiones, lo cual mejora la precisión de los estimadores obtenidos por estos \emph{sketches}.

    \section{HyperLogLog}
    \label{sec:hyper_log_log}

      \paragraph{}
      En esta sección se habla del \emph{Sketch} \emph{HyperLogLog}, que en este caso tiene tanto un enfoque como una utilidad distintas a las alternativas descritas en secciones anteriores. El \emph{HyperLogLog} se utiliza para tratar de estimar de manera precisa el número de elementos distintos que han aparecido sobre un stream de elementos. Esta estructura de datos es una evolución de la versión básica de la cual se cual se habló en el capítulo de \emph{Algoritmos para Streaming} en la sección \ref{sec:streaming_flajolet_martin_algorithm}.

      \paragraph{}
      Esta versión fue ampliada posteriormente en \emph{Loglog counting of large cardinalities} \cite{durand2003loglog} y para finalmente en 2007 se presentar el \emph{HyperLogLog}. La descripción del \emph{HyperLogLog} se indica en el trabajo \emph{Hyperloglog: the analysis of a near-optimal cardinality estimation algorithm} \cite{flajolet2007hyperloglog} llevado a cabo por \emph{Flajolet}. Para las explicaciones indicadas en esta secciones se ha seguido el artículo \emph{HyperLogLog in practice: algorithmic engineering of a state of the art cardinality estimation algorithm} \cite{heule2013hyperloglog} en el cual se exponen distintas mejoras sobre el \emph{HyperLogLog} que se describirán posteriormente.

      \paragraph{}
      Al igual que se indicó en la descripción del \emph{Algorimo de Flajolet-Martin} (Sección \ref{sec:streaming_flajolet_martin_algorithm}), la tarea que el \emph{HyperLogLog} pretende resolver es el conteo de elementos distintos o cardinalidad de un conjunto de elementos. Para ello, esta estrategia se apoya en la idea de mapear cada elemento que se presente en la entrada sobre una función hash binaria que cumpla una distribución uniforme de probabilidad. Bajo estas condiciones, la intuición en que se basa el \emph{HyperLogLog} es que en el sub-espacio de destino de la función hash, el $50\%$ de los elementos se codificarán de tal manera que el bit menos significativo sea un $0$, el $25\%$ se codificarán de tal manera que los dos bits más significativos sean $00$ y así sucesivamente, de tal manera que $1/2^k$ tendrán $k$ ceros como bits más significativos.

      \paragraph{}
      Por tanto, el \emph{HyperLogLog} se basa en la utilización de la función $\rho(x)$, que indica el número de ceros contiguos más significativos mas uno de en la representación binaria generada por la función hash. Hasta este punto, la estrategia es equivalente al \emph{Algorimo de Flajolet-Martin}, sin embargo, a continuación se indica una alternativa a esta a partir de la cual se obtiene una estimación con varianza mucho más reducida. Para ello en lugar de tratar el stream de elementos $S$ de manera conjunta, se divide dicho stream en $m$ sub-streamns de tal manera que todos tengan longitudes similares a los cuales denotaremos por $S_i$ con $i \in [1,m]$.

      \begin{equation}
      \label{eq:hyper_log_log_counter}
        M[i] = max_{x in S_i}\rho(x)
      \end{equation}

      \paragraph{}
      Para mantener la cuenta de cada estimador se utiliza un vector $M$ que almacena dicho valor. La estrategia de construcción del mismo se muestra en la ecuación \eqref{eq:hyper_log_log_counter}. En cuanto a la estimación de resultados, se utiliza la estrategia que se describe en la ecuación \eqref{eq:hyper_log_log_estimation}, que se apoya en el término $\alpha_m$ descrito en la ecuación \eqref{eq:hyper_log_log_alpha}. Nótese que el último término se corresponde con una media armónica de la cada contador. La elección de dichos operadores así como el término $\alpha_m$ se describe en el artículo original \cite{flajolet2007hyperloglog} y la razón por la cual se escogen de dicha manera es la de tratar la varianza de la estimación y, por tanto, mejorar su precisión.

      \begin{equation}
      \label{eq:hyper_log_log_estimation}
        card(S) = \alpha_m \cdot m^2 \cdot \bigg( \sum_{j=1}^{m}2^{-M[j]}\bigg)
      \end{equation}

      \begin{equation}
      \label{eq:hyper_log_log_alpha}
        \alpha_m = \bigg(m \cdot \int_0^\infty \bigg(log_2\bigg(\frac{2+u}{1+u}\bigg)\bigg)^m du\bigg)^{-1}
      \end{equation}

      \paragraph{}
      En cuanto a las mejoras propuestas en \cite{heule2013hyperloglog} sobre el \emph{HyperLogLog} original, estas se basan en la utilización de una función hash de \emph{64 bits} (en lugar de 32 como la propuesta original), la utilización de un estimador diferente cuando se cumple que $card(S) < \frac{5}{2}m$ y una estrategia de representación dispersa que trata de aprovechar en mayor medida el espacio en memoria.

    \section{$L_p$-Samplers}
    \label{sec:lp_samplers}

      \paragraph{}
      En esta sección se describe el concepto de \emph{$L_p$-Sampler}, para lo que se han seguido las ideas expuestas en \emph{Tight bounds for Lp samplers, finding duplicates in streams, and related problems} \cite{jowhari2011tight} de \emph{Jowhari y otros}. \emph{$L_p$-Samplers} se basan en estructuras de datos que procesan un stream de elementos definidos sobre el conjunto $N= \{1, 2, ...,i, ...n\}$ sobre el cual están permitidas tanto adicciones como eliminaciones, es decir, se enmarcan dentro del modelo en molinete (\emph{turnstile model}). Denotaremos por $x_i \in x$ al número de ocurrencias del elemento $i$-ésimo sobre el stream.

      \paragraph{}
      Los \emph{$L_p$-Samplers} devuelven un sub-conjunto de elementos $N'$ Extraído del conjunto global $N$ de tal manera que cada elemento es seleccionado con probabilidad $\frac{|x_i|^p}{||x||_p^p}$ siendo $p\geq0$. De esta manera, se obtienen una muestra del conjunto global que mantiene la misma proporción de elementos con respecto del global pero en un espacio menor utilizando como medida de estimación para el muestreo la norma $p$.

      \paragraph{}
      La dificultad del problema se basa en el mantenimiento de la muestra teniendo en cuenta que los elementos pueden ser tanto añadidos como eliminados de la misma según varía su cuenta $x_i$ conforme se procesa el stream.

      \paragraph{}
      De esta manera, cuando escogemos $p =1$ se seleccionarán los elementos pertenecientes a la muestra con una probabilidad proporcional al número de ocurrencias de los mismos en el stream. En \cite{jowhari2011tight} se describe un algoritmo para \emph{$L_p$-Sampler} con $p \in [0,2]$, ajustándose a las restricciones de espacio del modelo en streaming. Para ello se apoyan en la utilización del \emph{Count-Sketch} (sección \ref{sec:count_sketch}). A continuación se describe en más detalle el caso de los \emph{$L_0$-Samplers}.

      \subsection{$L_0$-Samplers}
      \label{sec:l0_samplers}

      \paragraph{}
      Los \emph{$L_0$-Samplers} siguen la misma idea de muestreo descrita en esta sección. En este caso utilizan la norma $0$, por lo tanto, seleccionan los elementos pertenecientes a la muestra con probabilidad uniforme de entre los que han aparecido en el stream. Este es por tanto el caso más básico del problema debido a que tan solo es necesario mantener un contador que indique si el elemento está presente. Los \emph{$L_0$-Samplers} son de gran utilidad en el contexto de grafos, ya que encajan de manera adecuada en dicho modelo, permitiendo mantener un sub-conjunto de aristas del grafo basándose únicamente en si estas han aparecido o no en el stream.

      \paragraph{}
      Esto es sencillo cuando solo están permitidas adicciones, ya que una vez aparecido el elemento este no desaparecerá. Sin embargo, en el caso de las eliminaciones la tarea se vuelve más complicada ya que tal y como se ha indicado en el párrafo anterior, es necesario mantener la cuenta de ocurrencias para saber si el elemento a desaparecido o únicamente ha decrementado su cuenta.

      \paragraph{}
      A continuación se describe la estructura básica de estos algoritmos siguiendo el trabajo desarrollado por \emph{Cormode y Fermain} en \emph{On unifying the space of L0-sampling algorithms} \cite{cormode2013unifying}. Estos indican que todas las propuestas de \emph{$L_0$-Samplers} se basan en tres fases:

      \begin{itemize}
        \item \emph{Sampling}: Se mantiene el conteo de ocurrencias en una estructura de datos auxiliar con ideas similares a las del \emph{Count-Sketch}.
        \item \emph{Recovery}: Se realizan peticiones a la estructura de datos auxiliar para recuperar el conteo de ocurrencias de cada elemento en el stream (saber si ha apararecido en el mismo).
        \item \emph{Selection}: Se realiza una selección de entre los elementos aparecidos en el stream siguiendo una distribución uniforme de probabilidad.
      \end{itemize}

    \section{Conclusiones}
    \label{sec:summaries_conclusions}

      \paragraph{}
      En este capítulo se han descrito distintas estrategias y técnicas para tratar de estimar indicadores estadísticos que ayuden a entender y comprender mejor conjuntos de datos de tamaño masivo. Estos estimadores tratan de agilizar la obtención de métricas sobre el conjunto de datos que debido a gran tamaño, muchas veces hace impracticable su obtención de manera determinística por el coste en complejidad tanto espacial como temporal. Para ello, se han descrito distintas soluciones, desde la extracción de muestras, así como histogramas o wavelets, hasta técnicas más novedosas y que encajan de manera más apropiada en el entorno cambiante sobre el que en muchas ocasiones se trabaja.

      \paragraph{}
      Por contra, las soluciones basadas en \emph{Sketches} tienen un escaso tiempo de vida, por lo que todavía se encuentran en fase de investigación. A pesar de ello existen soluciones para la comprobación de la aparición de elementos mediante el \emph{Bloom Filter}, la estimación de frecuencias mediante el \emph{Count-Min Sketch} o el \emph{Count Sketch}, y el conteo de elementos distintos a partir del \emph{HyperLogLog}.

      \paragraph{}
      Se cree que el conjunto de estructuras de datos probabilísticas basadas en \emph{Sketches} aumentará conforme el paso del tiempo y en el futuro se diseñarán alternativas más sofisticadas que permitan estimar un gran número de parámetros.

  \chapter{Algoritmos aplicados a Grafos}
  \label{chap:graphs}

    \section{Introducción}
    \label{sec:graphs_intro}

      \paragraph{}
      En las últimas décadas se han dedicado grandes esfuerzos en el estudio de sucesos basados en interelaciones entre elementos. Esto puede entenderse como el análisis de una red de interconexiones modelada como flechas que relacionan un conjunto de puntos. Una gran cantidad de situaciones de la vida cotidiana puede ser representada de esta manera, mediante la cual, se consigue un formalismo matemático enriquecedor sobre el cual resolver distintos problemas que surgen sobre dichas redes.

      \paragraph{}
      Estas redes de interconexiones se pueden apreciar en ámbitos muy dispares, como las ciudades y las carreteras que conectan unas con otras, las personas y las amistades entre si, los teléfonos y las llamadas de unos a otros o las páginas web de internet y los enlaces para navegar de unas a otras. El estudio de estas situaciones es de gran interés en las sociedades modernas, permitiendo una mejora del rendimiento a partir de la extracción de información poco obvia mediante distintas técnicas, lo cual conlleva reducción de costes y un aumento del grado de satisfacción para los usuarios de dichos servicios.

      \paragraph{}
      Sin embargo, utilizar un lenguaje de representación más enriquecedor conlleva sobrecostes asociados que en otros modelos más simples no se dan, lo cual requiere de técnicas sofisticadas para tratar de hacer frente a la resolución de los problemas que se pretende resolver. Además, el crecimiento exponencial en la infraestructura tecnológica ha traido como consecuencia un gran aumento a nivel de tamaño en estos. Algunos ejemplos destacados en el ámbito de internet se dan el protocolo IPv6, que cuenta con $2^128$ posibles direcciones hacia las que poder comunicarse, o el caso de la red social \emph{Facebook}, que cuenta con 1 trillón de relaciones de amistad, tal y como se indica en \emph{One trillion edges: Graph processing at facebook-scale} \cite{ching2015one}.

      \paragraph{}
      Conforme el tamaño de las redes aumenta, una estrategia razonable es la de utilizar soluciones aproximadas para la resolución de los problemas que se plantean sobre ellas, a través de los cuales se pretende encontrar una solución cercana a la exacta (admitiendo una desviación máxima de $\epsilon$, que se cumple con probabilidad $\delta$), lo que otorga como ventaja una reducción significativa tanto desde la perspectiva del coste temporal como del espacial en la resolución del problema.

      \paragraph{}
      En este capítulo se pretende realizar una descripción acerca de las distintas alternativas posibles para tratar de agilizar la resolución de problemas sobre redes de tamaño masivo utilizando técnicas aproximadas. Por tanto, primero se ha realizado una descripción formal sobre el modelo de representación de grafos en la sección \ref{sec:graph_formalism}. A continuación, se ha descrito un modelo sobre el cual diseñar algoritmos que resuelvan problemas aplicados a grafos tratando de aprovechar al máximo el espacio en memoria (que se considera limitada) y reduciendo el número de accesos sobre el espacio de almacenamiento mediante el \emph{modelo en semi-streaming}, del cual se habla en la sección \ref{sec:semi_streaming_model}. El siguiente tema que se trata en este capítulo se refiere a técnicas que tratan de reducir la complejidad de una red de tamaño masivo mediante la eliminación de relaciones entre sus punto mediante la utilización de \emph{Spanners} y \emph{Sparsifiers} en la sección \ref{sec:spanners_sparsifiers}. Después, se ha realizado una breve descripción acerca de distintos problemas bien conocidos para los cuales se han encontrados soluciones sobre el \emph{modelo en semi-streaming} en la sección \ref{sec:graph_problems}. Finalmente se ha realiza una breve conclusión en la sección \ref{sec:graph_conclusions}.

    \section{Definición Formal}
    \label{sec:graph_formalism}

      \paragraph{}
      En esta sección se describen los conceptos básicos necesarios para entender el estudio de problemas modelados como \emph{Grafos}. Para la descripción formal sobre dichas estructuras se han utilizado las notas de clase de la asignatura de \emph{Matemática Discreta} \cite{matematicaDiscreta2016notes} impartida en la \emph{Universidad de Valladolid} así como las de la asignatura equivalente (\emph{Discrete Mathematics CS 202} \cite{aspnes2013notes}) impartida por \emph{Aspnes} en la \emph{Universidad de Yale}.

\begin{figure}
        \centering
        \includegraphics[width=0.4\textwidth]{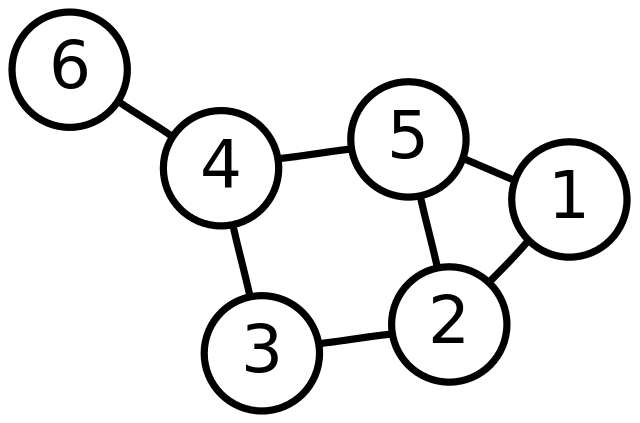}
        \caption{Ejemplo de \emph{Grafo No Dirigido}. (Extraído de \cite{wiki:Graph_(discrete_mathematics)})}
        \label{img:graph_example}
      \end{figure}

      \paragraph{}
      La \textbf{Teoría de Grafos} (\emph{Graph Theory}) es la disciplina encargada del estudio de estructuras compuestas por vértices y aristas desde una persepectiva matemática. Los vértices representan objetos o elementos, mientras que las aristas se corresponden con las relaciones que se dan entre vértices. Un grafo $G$ se define por tanto como la tupla del conjunto de vértices $V = \{ v_1, _2, ..., v_n \}$ y el conjunto de aristas $E = \{ e_1, e_2, ..., e_m \}$, de tal manera que $e_j = (v_{i_1}, v_{i_2})$ representa el arista que conecta el vértice $v_{i_1}$ con el vértice $v_{i_2}$. Nótese por tanto, que el grafo está compuesto por $n$ vértices y $m$ aristas. El grafo $G$ se puede modelizar por tanto como $G = (V, E)$. En la figura \ref{img:graph_example} se muestra una representación gráfica de un determinado grafo no dirigido compuesto por $6$ vértices y $7$ aristas.

      \paragraph{}
      Aquellos grafos para los cuales la arista $e_j$ representa los dos sentidos de la relación, es decir, $v_{i_1}$ está relacionado con $v_{i_2}$ y $v_{i_2}$ está relacionado con $v_{i_1}$ se denominan \emph{Grafos no nirigidos}, mientras que en los casos en que esta relación no es recíproca se habla de \emph{Grafos dirigidos}. Cuando cada arista $e_j$ tiene asociado un determinado peso $w_j \in W =  \{ w_1, w_2, ..., w_m\}$ se dice entonces que $G$ es un \emph{grafo ponderado} y se denota como $G=(V, E, W)$, mientras que cuando se presupone que todas las aristas tienen el mismo peso $W$ se omite de la notación.

      \paragraph{}
      Cuando un vértice denominado $v_{i_1} \in V$ está directamente relacionado con otro $v_{i_2} \in V$, es decir, existe una arista $e_j \in E$ que los conecta ($e_j = (v_{i_1}, v_{i_2})$) se dice que son $e_j$ es \emph{incidente} sobre dichos vértices. De la misma manera se dice que $v_{i_1}$ y $v_{i_2}$ son \emph{adjacentes} entre sí.

      \paragraph{}
      Respecto del conjunto de aristas incidentes sobre cada vértice, se denomina \emph{grado} al cardinal dicho conjunto, diferenciando en los grafos no dirigidos entre \emph{in-grado} a las de entrada y \emph{out-grado} a las de salida. Se utiliza la notación $d(v_i)$ para referirse al grado del vértice $i$-ésimo, $d^+(v_i)$ al \emph{in-grado} y $d^-(v_i)$ al \emph{out-grado} de dicho vértice. Nótese por tanto, que se cumple la siguiente propiedad: $d(v_i) = d^+(v_i) + d^-(v_i)$.

      \paragraph{}
      Un \emph{camino} es un conjunto de aristas $P_p = \{ e_{k_1}, e_{k_2}, ..., e_{k_p}\}$, tales que el arista k-ésimo tiene como vértice de destino el mismo que utiliza el arista $k+1$ como vértice origen. Nótese que el valor $p$ indica la \emph{longitud} del camino. Cuando el vértice de destino de la arista $e_{k_p}$ es el mismo que el de origen de $e_{k_1}$ se denomina \emph{ciclo} y se denomina $C_p$.

      \paragraph{}
      Se denota como $K_n$ al grafo compuesto por $n$ vértices y $n*(n-1)$ aristas, de tal manera que estas conectan cada vértice con todos los demás. Los grafos que cumplen esta propiedad se denominan grafos completos de grado $n$. Nótese que el cardinal de aristas se reduce a la mitad en el caso de los grafos no dirigidos.

\begin{figure}
        \centering
        \includegraphics[width=0.4\textwidth]{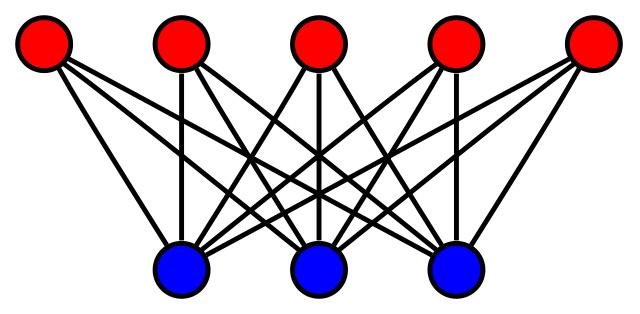}
        \caption{Ejemplo de \emph{Grafo Bipartito}. (Extraído de \cite{wiki:Graph_(discrete_mathematics)})}
        \label{img:bipartite_graph_example}
      \end{figure}

      \paragraph{}
      Cuando al estudiar la estructura de un grafo, se comprueba que el conjunto de vértices puede dividirse en dos sub-conjuntos disjuntos $V_1$ y $V_2$, de tal manera que para todas las aristas $e_j = (v_{i_1}, v_{i_2})$ el vértice $v_{i_1}$ se encuentra en el sub-conjunto $V_1$ y el vértice $v_{i_2}$ se encuentra en el sub-conjunto $V_2$, entonces se habla de un \emph{Grafo Bipartito}. Un ejemplo de esta situación se muestra en la figura \ref{img:bipartite_graph_example}. Nótese que el concepto de grafo bipartito puede extenderse fácilmente a más de dos sub-conjuntos, denominandose entonces \emph{Grafo k-partito}. Estos grafos son de gran utilidad para modelizar algunos problemas tales como los que se dan en empresas como \emph{Netflix}, para los cuales $V_1$ puede estar formado por el conjunto de usuarios mientras que $V_2$ representa el contenido multimedia. Por tanto, cada arista puede entenderse como una visualización del usuario $v_{i_1} \in V_1$ sobre el contenido $v_{i_2} \in V_2$ .

      \paragraph{}
      Un \emph{sub-grafo} $H$ es el grafo compuesto por un sub-conjunto de vectores y aristas del grafo $G$. Nótese que en el caso de que se eliminen vértices, es necesario eliminar también todas sus aristas incidentes. Esto se puede indicar de manera matemática de la siguiente forma: $H \subseteq G$ por lo que $H_V \subseteq G_V$ y $H_E \subseteq G_E$.

      \paragraph{}
      Desde el punto de vista de las transformaciones que se pueden realizar sobre grafos, se denomina \emph{isomorfismo} a una transformación biyectiva sobre el grafo $G =(V_G, E_G)$ al grafo $H = (V_H, E_H)$ que se realiza sobre los vértices de manera que $f: V_G \rightarrow V_H$ y por cada arista $(e_1, e_2) \in E_G$, entonces $(f(e_1), f(e_2)) \in E_H$ de tal manera que la estructura de $G$ se conserva en $H$. Entonces se dice que $G$ y $H$ son isomorfos entre si. Si se modifica la condición de biyectividad del \emph{isomorfismo} y tan solo se requiere la propiedad de inyectividad entonces se habla de \emph{homomorfismo}. Por tanto, esto puede ser visto como una transformación sobre la cual no se mantiene la estructura de $G$, entonces $H$ ya no es equivalente a $G$, sino que es un \emph{sub-grafo} de este.

      \paragraph{}
      Las transformaciones son interesantes desde el punto de vista del tratamiento de grafos de tamaño masivo, dado que a partir de estas se trata de reducir la complejidad cuando el tamaño de estos los hace intratables. Por lo tanto, en este caso interesa conseguir transformaciones que reduzcan el tamaño del grafo, pero que tratan de mantener su estructura lo más semejante posible. Destacan transformaciones conocidas como \emph{Spanners} (de los que se hablará en la sección \ref{sec:spanners}) y \emph{Sparsifiers} (en la sección \ref{sec:sparsifiers}).

      \subsection{Métodos de Representación}
      \label{sec:representation_methods}

        \paragraph{}
        Existen diversas estrategias para representar un grafo sobre estructuras de datos, las cuales presentan distintas ventajas e inconvenientes, tanto a nivel de espacio como de tiempo de acceso. Dichas estrategias se escogen teniendo en cuenta la estructura del grafo. En esta sección se habla de matrices de adyacencia, de listas de adyacencia y de la matriz laplaciana, la cual contiene un conjunto de propiedades interesantes que pueden ser muy útiles en la resolución de algunos problemas.

        \subsubsection{Matriz de Adyacencia}
        \label{sec:adjacency_matrix}

          \paragraph{}
          Se denomina matriz de adyacencia $A$ del grafo $G = (V,E)$ compuesto por $n=|V|$ vértices, $m=|E|$ aristas y $W$ el conjunto de pesos de los aristas a una matriz de tamaño $n*n$ construida tal y como se indica en la ecuación \eqref{eq:adjacency_matrix}. Nótese que esta definición es válida tanto para grafos ponderados como para no ponderados suponiendo que $w_k = 1, k \in [1, m]$.

          \begin{equation}
          \label{eq:adjacency_matrix}
            A_{i,j} =
              \begin{cases}
                w_k,  & (v_i, v_j) = e_k \in E\\
                0,    &\text{en otro caso}
              \end{cases}
          \end{equation}

          \paragraph{}
          Esta estrategia de representación es apropiada cuando se trabaja sobre grafos altamente conexos (con un gran número de aristas), ya que el número de posiciones nulas en la matriz es reducido y la estructura matricial indexada proporciona un tiempo de acceso de $O(1)$ a cada posición. Sin embargo, se requiere de $O(n^2)$ para almacenar dicha matriz, algo admisible cuando $n^2 \approx m$ para lo que esta representación se acerca a su óptimo espacial.

        \subsubsection{Lista de Adyacencia}
        \label{sec:adjacency_list}

          \paragraph{}
          La alternativa a la codificación del grafo como una matriz de adyacencia se conoce como \emph{lista de adyacencia}. En este caso, se mantiene una lista (posiblemente ordenada u otras estrategias estructuradas como listas enlazadas o árboles, para tratar de reducir los tiempos de acceso) para almacenar el conjunto de aristas $E$. Nótese por tanto, que en este caso la codificación es óptima a nivel de espacio $O(m)$, no existiendo una estrategia de representación que pueda almacenar la estructura del mismo de manera exacta utilizando un tamaño menor. Sin embargo, tal y como se puede intuir, el tiempo de acceso es de $O(m)$ en el peor caso. Esta solución es apropiada para grafos muy dispersos (aquellos en que $n \ll m$).

        \subsubsection{Matriz Laplaciana}
        \label{sec:laplacian_matrix}

          \paragraph{}
          La \emph{matriz laplaciana} consiste en una estrategia de representación de grafos que cumple importantes propiedades, a partir de las cuales se facilita en gran medida la resolución de distintos problemas, entre los que se encuentran \emph{árboles de recubrimiento} (sección \ref{sec:minimum_spanning_tree}), aunque también se utiliza en la estimación de la distribución de probabilidad de \emph{paseos aleatorios} (sección \ref{sec:random_walks_overview}).

          \paragraph{}
          El método de construcción de la \emph{matriz laplaciana} se indica en la ecuación \eqref{eq:laplacian_matrix}, el cual genera una matriz $L$ de tamaño $n * n$ (al igual que en el caso de la matriz de adyacencia) que aporta información sobre el grafo subyacente. La diagonal de la matriz laplaciana contiene el grado del vértice referido a dicha posición, mientras que el resto de las celdas se construyen colocando el valor $-w_k$ cuando existe un arista entre el vértice $v_i$ y el vértice $v_j$ ( $e_k = (v_i,v_j) \in E$ ) y $0$ cuando no existe. La matriz laplaciana también puede entenderse como $L = D - A$ donde $D$ representa una matriz diagonal de tamaño $n*n$ que almacena el grado del vértice $v_i$ en la posición $D_{i,i}$ y $0$ en otro caso, mientras que $A$ se corresponde con la matriz de adyacencia descrita anteriormente.

          \begin{align}
          \label{eq:laplacian_matrix}
            L_{{i,j}} = {\begin{cases}
              d(v_{i})&{\mbox{if}}\ i=j\\
              -1&{\mbox{if}}\ i\neq j\ {\mbox{and}}\ v_{i}{\mbox{ is adjacent to }}v_{j}\\
              0&{\mbox{otherwise}}\end{cases}}
          \end{align}

          \paragraph{}
          A partir de esta representación se facilita la resolución de distintos problemas tal y como se ha indicado anteriormente. También existen distintas variaciones respecto de la descrita en esta sección, entre las que se encuentran la \emph{matriz laplaciana normalizada} o la \emph{matriz laplaciana de caminos aleatorios}, de la cual se hablará en el capítulo \ref{chap:pagerank} para el cálculo del ranking de importancia \emph{PageRank}.

      \paragraph{}
      En la práctica, cuando se trabaja con grafos de tamaño masivo, es muy común que estos sean muy dispersos. Algunos ejemplos de ello son el \emph{grafo de la web} (grafo dirigido), en el cual cada vértice representa sitio web y cada arista un enlace hacia otro sitio web. Es fácil comprobar que este grafo es altamente disperso ya que es muy poco probable que un sitio web contenga enlaces al resto de sitios de la red cuando se habla de millones de vértices. Algo similar ocurre en el caso de redes sociales como \emph{Facebook} (grafo no dirigido debido a relaciones de amistad) o \emph{Twitter} (grafo dirigido debido a relaciones seguimiento). Por tanto, la representación mediante listas de adyacencia es una herramienta útil a nivel conceptual pero que en la práctica no se utiliza por la inviabilidad derivada del gran coste espacial para su almacenamiento.

    \section{Modelo en Semi-Streaming}
    \label{sec:semi_streaming_model}

      \paragraph{}
      Al igual que sucede en el caso de conjuntos de datos de carácter numérico y categoríco, en el modelo de grafos, también es necesario hacer frente al elevado tamaño del problema mediante nuevas estrategias de diseño de algoritmos. El \emph{modelo en streaming}, del cual se habló en la sección \ref{sec:streaming_model} es una buena alternativa para tratar de agilizar la búsqueda de soluciones que varían con respecto del tiempo, además de trabajar sobre un espacio reducido lo cual aprovecha en mayor medida las capacidades del hardware subyacente. Tal y como se indicó anteriormente, esta estrategia permite reducir el número de acceso sobre el dispositivo de almacenamiento tratando de trabajar únicamente con los estimadores que se mantienen en la memoria del sistema, cuyo tiempo de acceso es mucho más reducido.

      \paragraph{}
      Sin embargo, en el caso de los problemas referidos a grafos, este modelo presenta mayores dificultades, tanto a nivel de espacio como del número de accesos sobre el stream de datos por la estructura enlazada de la representación de grafos. Por lo tanto, se han definido variaciones sobre el \emph{modelo en streaming} original para tratar de hacer frente a los requisitos característicos de este tipo de problemas. En los artículos \emph{On graph problems in a semi-streaming model} \cite{feigenbaum2005graph} y \emph{Analyzing graph structure via linear measurements} \cite{ahn2012analyzing} los autores describen dicho modelo, al cual denominan \textbf{Modelo en Semi-Streaming}.

      \paragraph{}
      Este se corresponde con una relajación del \emph{modelo en streaming} estándar, el cual permite mayor libertad tanto a nivel de espacio como de pasadas permitidas sobre el stream de datos. Por esta razón, cuando el número de pasadas $p$ es superior a 1 ($p > 1$), entonces ya no es posible su uso en entornos en que se requiere que el procesamiento de la información y las consultas sean en tiempo real, algo que, por contra, si sucedía en el caso del \emph{modelo en streaming} definido anteriormente. Por lo tanto, el \emph{modelo en semi-streaming} se presenta como una estrategia de diseño de algoritmos que trata de obtener estimaciones sobre grafos en un espacio reducido y con un buen planteamiento a nivel de accesos a disco cuando el grafo completo no puede mantenerse en la memoria del sistema de manera completa.

      \paragraph{}
      El \emph{modelo en semi-streaming} impone la forma en que el grafo es recibido bajo la idea de stream de datos, lo cual se describe a continuación: Sea $G = (V, E)$ un grafo dirigido (su adaptación al caso de grafos no dirigidos es trivial) compuesto por $n = |V|$ vértices y un número indeterminado de arístas desde el punto de vista del algoritmo que procesará el stream. Se presupone que se conoce \emph{a-priori} el número de vértices que forman el grafo, mientras que el stream consiste en el conjunto de tuplas que representan las aristas. El conjunto de aristas $E$ se define como $E = \{ e_{i_1}, e_{i_2}, ..., e_{i_j}, ..., e_{i_m} \}$, tal y como se ha hecho en secciones anteriores. Por tanto el grafo $G$ está formado por $m = |E|$ aristas (es necesario remarcar que el algoritmo que procesa el stream no conoce dicho valor). Dichas aristas son procesadas en un orden desconocido de manera secuencial marcado por el conjunto de permutaciones arbitrarias $\{ i_1, i_2, i_j, i_m \}$ sobre $[1, m]$.

      \paragraph{}
      Una vez descrita la estrategia de procesamiento sobre el \emph{modelo en semi-streaming}, lo siguiente es indicar las unidades de medida sobre las cuales realizar el análisis sobre la complejidad de los algoritmos que se desarrollan sobre este modelo. En \cite{feigenbaum2005graph}, los autores definen $S(n,m)$ como el espacio utilizado para procesar el stream de aristas, $P(n,m)$ el número de pasadas sobre dicho stream y $T(n,m)$ el tiempo necesario para procesar cada arista. Sobre esta contextualización se requiere que para que un algoritmo sea válido en el \emph{modelo en semi-streaming} $S(n,m)$ esté contenido en el orden de complejidad $O(n \cdot polylog(n))$. Tal y como se puede apreciar, esta restrición a nivel de espacio es mucho más relajada que la impuesta sobre el \emph{modelo en streaming} estándar, que requiere de $o(N)$. Sin embargo, es necesario tener en cuenta que gran cantidad de problemas sobre grafos requieren un coste espacial de $O(n^2)$, por lo que tratar de encontrar una solución en $O(n \cdot polylog(n))$ representa una tarea compleja, pero conlleva una mejora significativa.

      \paragraph{}
      Al igual que sucede con en el caso del \emph{modelo en streaming} estándar, al utilizar un orden espacial menor del necesario para encontrar soluciones exactas, las soluciones encontradas admiten la existencia de una determinada tasa de error máxima delimitada por $\epsilon$, la cual se debe cumplir con una probabilidad $\delta$. Para conjuntos de datos sobre los cuales no es admisible la búsqueda de una solución exacta o para los cuales sea admisible una reducida tasa de error, esta estrategia de diseño de algoritmos se presenta por tanto como una alternativa acertada.

    \section{Spanners y Sparsifiers}
    \label{sec:spanners_sparsifiers}

      \paragraph{}
      Para tratar de agilizar los cálculos necesarios para la resolución de problemas sobre grafos, se han propuesto distintas alternativas, muchas de las cuales son aplicables únicamente a problemas concretos. Sin embargo, estas ideas se discutirán en la sección \ref{sec:graph_problems}. En esta sección se describen distintas técnicas utilizadas para tratar de \say{sumarizar} o disminuir el espacio necesario para almacenar un determinado grafo $G$ (al igual que ocurría con las técnicas descritas en el capítulo anterior para valores numéricos) transformándolo en un nuevo grafo $H$, de tal manera que la estructura del mismo siga siendo lo más similar a la del original respecto de distintas medidas.

      \paragraph{}
      La descripción de estas técnicas se ha basado en las ideas recogidas en los artículos \emph{Graph stream algorithms: a survey} \cite{mcgregor2014graph} de \emph{McGregor} y \emph{Graph sketches: sparsification, spanners, and subgraphs} \cite{ahn2012graph} de \emph{Ahn y otros}, así como las notas de la \emph{Clase 11} de la asignatura \emph{Randomized Algorithms} \cite{harvey2011randomized} impartida en la \emph{Universidad de Columbia Británica}. Tal y como se ha indicado en el párrafo anterior, estas técnicas consisten en la eliminación de vértices y/o aristas, de tal manera que la distorsión producida tras dichas modificaciones sea mínima.

      \paragraph{}
      Existe un enfoque trivial para la reducción del tamaño en grafos, sin embargo, este tan solo ofrece resultados aceptables sobre grafos densos (aquellos similares al grafo completo $K_n$). La técnica consiste en la eliminación de aristas a partir de una determinada tasa de probabilidad $\alpha$. Mediante esta estrategia se reduce el espacio necesario para almacenar las aristas de $(1-\alpha)$. Tal y como se ha dicho, esta solución tan solo es válida en aquellos casos en que el grafo sea denso. Cuando el grafo es disperso, por contra, la eliminación de aristas mediante la selección uniforme puede producir una gran distorsión respecto del grafo original.

      \begin{figure}
        \centering
        \includegraphics[width=0.3\textwidth]{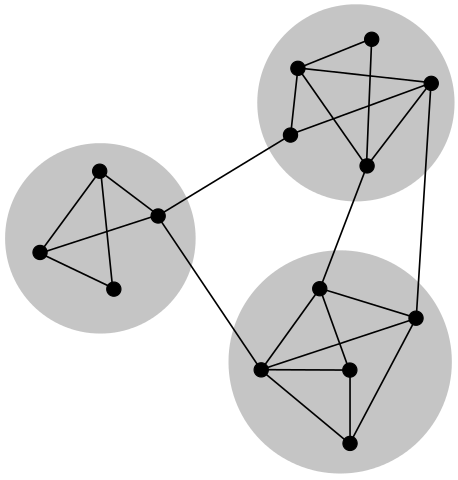}
        \caption{Ejemplo de \emph{Grafo Disperso}. (Extraído de \cite{wiki:Community_structure})}
        \label{img:graph_community_structure}
      \end{figure}

      \paragraph{}
      En la figura \ref{img:graph_community_structure} se muestra un grafo disperso, que además forma 3 comunidades (conjuntos de vértices altamente conexos entre sí). Nótese que en aquellos casos en que el grafo posea una estructura similar a la del de la figura, ya no es conveniente utilizar la estrategia descrita en el párrafo anterior, puesto que para tratar de preservar la estructura del mismo, no todas las aristas tienen la misma relevancia. Esto puede entenderse de manera sencilla si al comprobar la variación del grafo al eliminar una de las aristas que conectan dos comunidades. Dicha variación estructural es significativamente mayor que la ocurrida tras eliminar una arista que conecta dos vértices pertenecientes a una misma comunidad.

      \paragraph{}
      Por esta razón, distintos autores han trabajado en técnicas para tratar de mantener la estructura del sub-grafo generado lo más semejante posible a la del grafo original. Para ello, las estrategias más populares se basan en la utilización de \emph{Spanners} y \emph{Sparsifiers}, los cuales se describirán a continuación en las secciones \ref{sec:spanners} y \ref{sec:sparsifiers} respectivamente. Estas técnicas han sido ampliamente estudiadas, sin embargo, en este caso se pretende orientar la descripción de las mismas sobre el \emph{modelo en semi-streaming} del cual se habló anteriormente. En este modelo se han encontrado soluciones eficientes para \emph{grafos no dirigidos}. Por contra, para el caso de los \emph{grafos dirigidos}, la búsqueda de soluciones eficientes para este problema continua siendo un problema abierto.

      \subsection{Spanners}
      \label{sec:spanners}

        \paragraph{}
        Se denomina \emph{Spanner} a un determinado sub-grafo que mantiene las propiedades de distancia entre todos sus vértices respecto del original con una tasa máxima de variación acotada por $\alpha$. Por tanto, se denomina \emph{$\alpha$-spanner} del grafo $G = (V, E)$ a un sub-grafo $H = (V, E')$ donde $E' \subset E$, construido de tal manera que se cumpla la ecuación \eqref{eq:alpha_spanner} representando $d_G(v_{i_1},v_{i_2})$ la distancia del camino más corto entre los vértices $v_{i_1}$ y $v_{i_2}$ sobre el grafo $G$.

        \begin{equation}
        \label{eq:alpha_spanner}
          \forall v_{i_1}, v_{i_2} \in V, \ d_G(v_{i_1},v_{i_2}) \leq d_H(v_{i_1},v_{i_2}) \leq \alpha \cdot d_G(v_{i_1},v_{i_2})
        \end{equation}

        \paragraph{}
        Tal y como se indica en la ecuación \eqref{eq:alpha_spanner}, lo que se pretende acotar mediante esta estrategia es, por tanto, el error desde el punto de vista de la distancia entre vértices. Tal y como se puede intuir, mediante el cumplimiento de esta propiedad se soluciona el problema descrito anteriormente que surge sobre grafos dispersos, dado que si se elimina el único arista que conecta dos comunidades distintas, entonces la distancia del camino más corto entre los vértices de comunidades distintas variará en gran medida, posiblemente superando la acotada del valor $\alpha$.

        \paragraph{}
        Para la construcción de un \emph{$\alpha$-spanner} sobre el modelo en streaming existe un algoritmo sencillo que resuelve este problema para el caso del modelo de caja registradora (sección \ref{sec:streaming_cash_register}), es decir, en el cual tan solo estén permitidas adicciones. Esta estrategia se ilustra en el algoritmo \ref{code:basic_spanner}. En este caso, la solución es trivial y consiste en añadir únicamente al conjunto de aristas $E'$ del grafo $H$, aquellas cuya distancia del camino más corto entre sus vértices en $H$ sea mayor que $\alpha$, con lo cual se garantiza la propiedad del \emph{$\alpha$-spanner}.

        \paragraph{}
        \begin{algorithm}
          \SetAlgoLined
          \KwResult{$E'$ }
          $E' \gets \emptyset$\;
          \For{cada $(u, v) \in E$}{
            \If{$d_H(u,v) > \alpha$}{
              $E' \gets E' \cup \{(u,v)\}$\;
            }
          }
          \caption{Basic Spanner}
          \label{code:basic_spanner}
        \end{algorithm}

        \paragraph{}
        Sin embargo, esta técnica requiere del cálculo del camino más corto entre los vértices $u$ y $v$ en cada actualización, lo cual genera un coste de $O(card(E'))$ en tiempo, o el mantenimiento de una estructura de datos auxiliar que almacene dichas distancias, lo que requiere de un $O(n^2)$ en espacio. El mejor resultado encontrado para este problema es la construcción de un \emph{$(2k-1)$-spanner} utilizando $O(n^(1+1/k)$ de espacio. Esta solución se ha demostrado que es óptima respecto de la precisión utilizando tan solo una pasada sobre el stream de aristas. La descripción completa de la misma se encuentra en el trabajo \emph{Streaming and Fully Dynamic Centralized Algorithms for Constructing and Maintaining Sparse Spanners} \cite{elkin2007streaming} de \emph{Elkin}.

        \paragraph{}
        Para el caso general, en el cual están permitidas tanto adicciones como eliminaciones (modelo en molinete descrito en la sección \ref{sec:streaming_turnstile}) la solución básica no es trivial. El algoritmo que se describe en \cite{elkin2007streaming} se basa en la generación de árboles incrementales que se construyen a partir de la selección aleatoria de vértices. Sin embargo, esto no es sencillo cuando se permiten las eliminaciones. Por lo tanto, para la adaptación de dichas técnicas al modelo en molinete una solución es la utilización de \emph{$L_0$-Samplers}, que se describieron en la sección \ref{sec:lp_samplers}, lo que requiere de múltiples pasadas sobre el stream de aristas.

        \paragraph{}
        Tal y como se ha visto en esta sección, la construcción de \emph{Spanners} añade un sobrecoste a la resolución de problemas sobre grafos, junto con una determinada tasa de error desde el punto de vista de la distancia entre vértices. Sin embargo, dichos inconvenientes se ven recompensados en muchos casos por la reducción en tiempo y espacio de la resolución del problema sobre el sub-grafo resultante.

      \subsection{Sparsifiers}
      \label{sec:sparsifiers}

        \paragraph{}
        Otra alternativa para la generación del sub-grafo $H$ construido sobre el mismo conjunto de vértices y un sub-conjunto de aristas respecto del grafo $G$ son los \emph{Sparsifiers}. En este caso, en lugar de tratar de mantener la propiedad de la distancia del camino más corto entre pares de vértices, se basa en la minimización del número de cortes mínimos (eliminación de aristas) para separar el conjunto de vértices del grafo en dos sub-conjuntos disjuntos para cada par de vértices. A este concepto se lo denomina \emph{corte mínimo} (o \emph{Minimun Cut}) y se denota por $\lambda_{u,v}(G)$ para indicar el número de aristas a eliminar para formar dos sub-conjuntos disjuntos $V_{1}, V_{2}$ de tal manera que $u\in V_{1}$ y $v \in V_{2}$. Nótese por tanto, que en un grafo dirigido el \emph{corte mínimo} puede tomar valores en el intervalo $[1, n\cdot(n-1)]$.  En la ecuación \eqref{eq:sparsifier_cut} se muestra la definición formal de \emph{Sparsifier} donde $A$ representa todas las combinaciones de pares de vértices y $\epsilon$ la desviación máxima permitida por el \emph{Sparsifier}, de tal manera que el resultado se corresponde con un \emph{$(1 +\epsilon)$-Sparsifier}.

        \begin{figure}
          \centering
          \includegraphics[width=0.5\textwidth]{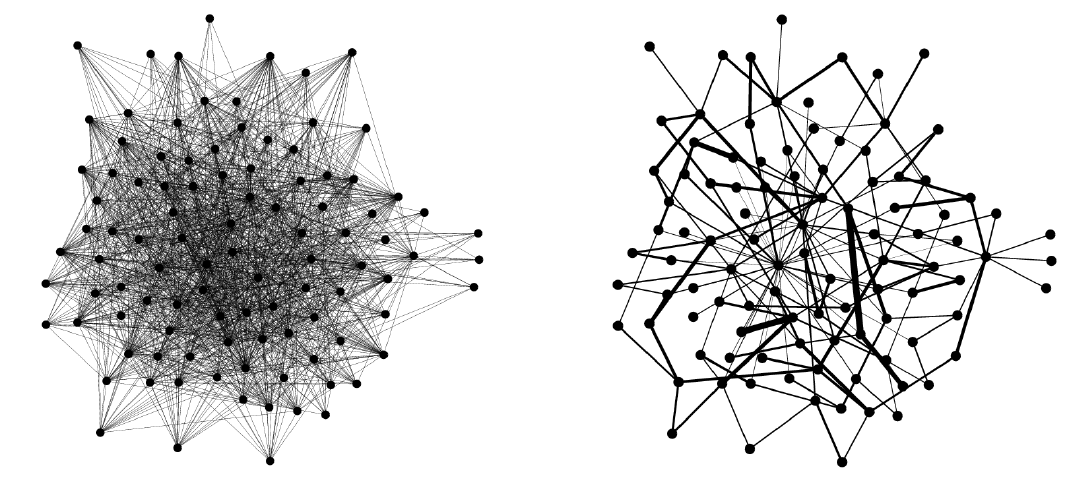}
          \caption{Ejemplo de \emph{Sparsifier}. (Extraído de \cite{harvey2011randomized})}
          \label{img:graph_community_structure}
        \end{figure}

        \begin{equation}
        \label{eq:sparsifier_cut}
          \forall A \in V \  (1-\epsilon)\lambda_A(G)\leq\lambda_A(H)\leq(1+\epsilon)\lambda_A(G)
        \end{equation}

        \paragraph{}
        A continuación se describe una definición para entender las ideas subyacentes en que se basan los \emph{Sparsifiers}. Para construir un \emph{$(1 +\epsilon)$-Sparsifier} de $G$, el valor $p$ debe ser escogido tal y como se indica en la ecuación \eqref{eq:sparsifier_p} donde $\lambda(G)$ representa el corte mínimo de $G$. La demostración se encuentra en el artículo \emph{On graph problems in a semi-streaming model} \cite{feigenbaum2005graph} desarrollado por \emph{Feigenbaum y otros}. Por tanto, el problema se reduce al mantenimiento de un \emph{$L_{0}$-Sampler} que devuelva un sub-conjunto de aristas seleccionados con probabilidad $p$ del stream de aristas.

        \begin{equation}
        \label{eq:sparsifier_p}
          p \geq min\{6\lambda(G)^{-1}\epsilon^{-2}log(n),1\}
        \end{equation}

        \paragraph{}
        A continuación se describe una estrategia sencilla de construcción de \emph{$(1 +\epsilon)$-Sparsifiers} para grafos no dirigidos. Esta se basa en la selección de aristas con probabilidad $p$ definido tal y como se indicó anteriormente en la ecuación \eqref{eq:sparsifier_p}.

        \paragraph{}
        \begin{algorithm}
          \SetAlgoLined
          \KwResult{$E'$ }
          $E' \gets \emptyset$\;
          \For{cada $(u, v) \in E$}{
            $r \gets Uniform(0,1)\;$ \\
            \If{$r < p$}{
              $E' \gets E' \cup \{(u,v)\}$\;
            }
          }
          \caption{Basic Sparsifier}
          \label{code:basic_sparsifier}
        \end{algorithm}

        \paragraph{}
        Otro enfoque más general para la construcción de \emph{Sparsifiers} se basa en la utilización de la \emph{Matriz Laplaciana} (definida en la sección \ref{sec:laplacian_matrix}) del grafo $H$, que denotaremos como $L_{H}$. A los \emph{Sparsifiers} construidos manteniendo la propiedad definida en la ecuación \eqref{eq:sparsifier_spectral} se los denomina \emph{Spectral Sparsifiers}. Desarrollando la ecuación obtenida si restringimos el rango de valores de $x$ al sub-conjunto $\{ 0, 1\}^n$ entonces podemos obtener la ecuación \ref{eq:sparsifier_cut}. Dado que el vector $x$ modeliza el conjunto $A$ en dicha ecuación. Por tanto, los \emph{Spectral Sparsifiers} pueden ser vistos como una generalización de los anteriores.

        \begin{equation}
        \label{eq:sparsifier_spectral}
          \forall x \in \mathbb{R} \leq (1 - \epsilon) x^TL_{G}x \leq x^TL_{H}x \leq (1 + \epsilon) x^TL_{G}x
        \end{equation}

        \paragraph{}
        Mediante la estrategia de los \emph{Spectral Sparsifiers}, además de aproximar el corte mínimo $\lambda(G)$ se pueden aproximar otras propiedades como propiedades de los paseos aleatorios (que se describen en la sección \ref{sec:random_walks_overview}). En el trabajo \emph{Twice-Ramanujan Sparsifiers} \cite{batson2012twice} redactado por \emph{Batson y otros} se describe una estrategia de construcción de \emph{$(1 +\epsilon)$-Spectral Sparsifiers} en un espacio de $O(\epsilon^{-2}n)$.

      \paragraph{}
      Las estrategias descritas en esta sección para la reducción de la complejidad de grafos ajustándose al \emph{modelo en semi-streaming} proporcionan una ventaja significativa a nivel de espacio mediante la redución del conjunto de aristas para la resolución de otros problemas a partir de ellas. Por contra, estas generan una determinada tasa de error.

      \paragraph{}
      Cuando se ha hablado de optimalidad en esta sección, ha sido desde el punto de vista de los \emph{grafos no dirigidos}, dado que para el caso de los \emph{grafos dirigidos}, aún no se han encontrado métodos de generación de \emph{Spanners} o \emph{Sparsifiers} óptimos debido a la mayor complejidad que estos conllevan. En la siguiente sección se describen distintos problemas para los cuales se ha encontrado una solución sobre el \emph{modelo en semi-streaming} y sus correspondientes restricciones.

    \section{Problemas sobre Grafos}
    \label{sec:graph_problems}

      \paragraph{}
      El modelo de representación de grafos proporciona un marco de trabajo flexible sobre el cual se pueden modelizar un gran número de situaciones. Un ejemplo característico de esta situación son las redes sociales, en las cuales cada persona representa un vértice en el grafo mientras que las aristas representan las relaciones entre estas. El conjunto de relaciones generadas a partir de los enlaces entre páginas web (\emph{Web Graph}) también son un claro ejemplo de problemas que se pueden representar mediante un grafo. Sin embargo, este modelo permite representar otras muchas situaciones como planificación de tareas donde cada vértice representa una tarea y las aristas marcan el orden de precedencia de las mismas. Los grafos también se pueden utilizar para modelizar el problema de encontrar determinados objetos o estructuras en imágenes.

      \paragraph{}
      Muchos de estos problemas pueden extenderse sobre un entorno dinámico, en el cual la estructura del grafo varía con respecto del tiempo. Algunos ejemplos son la conexión con nuevos amigos que se conectan entre sí en un grafo que modele redes sociales, la eliminación de enlaces entre webs en el caso del grafo de la red (\emph{Web Graph}), cambios de última hora debido a factores externos en cuanto a planificación de tareas, o la extensión del reconocimiento de estructuras en vídeos, que pueden ser vistos como la variación del estado de la imagen con respecto del tiempo.

      \paragraph{}
      Por estas razones, en esta sección se pretende realizar una descripción acerca de distintos problemas aplicados a grafos sobre el \emph{modelo en semi-streaming}. Los problemas que se describen, generalmente son de carácter básico. Sin embargo, son de vital importancia puesto que a partir de ellos pueden plantearse soluciones a otros de mayor complejidad.

      \paragraph{}
      El resto de la sección se organiza de la siguiente manera: En el apartado \ref{sec:bipartite_matchings} se describirá el problema de \emph{Verificación de Grafos Bipartitos}, a continuación se habla del \emph{problema de Conteo de Triángulos} en el apartado \ref{sec:counting_triangles}, posteriormente se expone el problema de encontrar el \emph{Árbol Recubridor Mínimo} en el apartado \ref{sec:minimum_spanning_tree}, en el apartado \ref{sec:graph_connected_components} se discute sobre el problema de \emph{Componentes Conectados} y finalmente, en el apartado \ref{sec:random_walks_overview} se realiza una breve introducción acerca de los \emph{Paseos Aleatorios}, que será extendida en profundidad en el capítulo siguiente, destinado exclusivamente al \emph{Algoritmo PageRank}. Se vuelve a remarcar que estos problemas se describen desde la persepectiva del \emph{modelo en semi-streaming}.

      \subsection{Verificación de Grafos Bipartitos}
      \label{sec:bipartite_matchings}

        \paragraph{}
        En la sección \ref{sec:graph_formalism} en la cual se realizó una descripción formal acerca de las definiciones sobre grafos que se utilizarían durante el resto del capítulo se hablo de \emph{Grafos Bipartitos}, que tal y como se indicó, estos se refieren a aquellos grafos que cumplen la propiedad de formar dos sub-conjuntos disjuntos de vértices que se relacionan entre si mediante aristas incidentes sobre 2 vértices cada uno perteneciente a un sub-conjunto distinto. Tal y como se indicó anteriormente, en la figura \ref{img:bipartite_graph_example} se muestra un ejemplo de grafo bipartito.

        \paragraph{}
        En el trabajo \emph{On graph problems in a semi-streaming model}\cite{feigenbaum2005graph} (en el cual se expone por primera vez el modelo en semi-streaming) los autores ilustran un algoritmo con un coste espacial de $O(n \cdot log(n))$ procesando cada arista en $O(1)$ y realizando $O(log(1 / \epsilon)) \epsilon$ pasadas sobre el stream de aristas que indica si se cumple la propiedad de grafo bipartito.

        \paragraph{}
        El algoritmo se basa en lo siguiente: Una fase inicial de construcción de un \emph{Matching Maximal} (selección de aristas de tal manera que ninguna sea indicente del mismo vértice que otra). Esta tarea se realiza sobre la primera pasada del stream de aristas. El resto de pasadas se basan en la adicción de aristas para hacer conexo el grafo de tal manera que se no se formen ciclos. Si esta estrategia es posible para todas las aristas del stream, entonces se cumple la propiedad de grafo bipartito.

        \paragraph{}
        Existen otras estrategias para probar que un grafo cumple la propiedad de grafo bipartito. Cuando se hable de problemas de conectividad en la sección \ref{sec:graph_connected_components} se expondrá otra estrategia para la validación de la propiedad de grafos bipartitos.

      \subsection{Conteo de Triángulos}
      \label{sec:counting_triangles}

        \paragraph{}
        Uno de los problemas que se ha tratado en profundidad sobre el \emph{modelo en semi-streaming} aplicado a grafos es el \emph{conteo de triángulos}. Este problema se define como el número de tripletas de vértices conectadas entre sí mediante tres aristas de tal manera que estas sean incidentes a dichos vértices. Una modelización más precisa se define a continuación sobre el grafo $G=(V,E)$.

        \paragraph{}
        Sean $v_{i_1},v_{i_2},v_{i_3} \in V$ tres dístintos vértices del grafo $G$ y $e_{j_1}, e_{j_2} e_{j_3} \in E$ tres aristas distintos de dicho grafo. Entonces, se denomina triángulo a la tripleta $\{e_{j_1}, e_{j_2} e_{j_3}\}$ si se cumple que $e_{j_1} = (v_{i_1},v_{i_2})$, $e_{j_2} = (v_{i_2},v_{i_3})$ y $e_{j_3} = (v_{i_3},v_{i_1})$. Al cardinal del conjunto de tripletas distintas sobre el grafo $G$ se lo denomina $T_3$. Esta definición puede extenderse a figuras de con mayor número de vértices modificando el valor $3$ por otro mayor. Sin embargo, estos casos son de menor interés puesto que aportan información similar sobre la estructura del grafo pero requieren un mayor coste computacional para su cálculo.

        \paragraph{}
        Se han encontrado estrategias para soluciones aproximadas al problema del \emph{conteo de triángulos} sobre las restricciones del \emph{modelo en semi-streaming}. La primera propuesta expuesta en \emph{Reductions in streaming algorithms, with an application to counting triangles in graphs} \cite{bar2002reductions} por \emph{Bar-Yossef y otros} fue la modelización del problema como el conteo de aristas que unen todas aquellas tripletas de nodos, denotadas como $x_{\{v_{i_1},v_{i_2},v_{i_3}\}}$ cuyo valor es $3$.

        \paragraph{}
        Mediante la estimación de momentos de frecuencia (de los cuales se habló en la sección \ref{sec:streaming_frecuency_moment_aproximation}) se puede calcular el número de triángulos distintos tal y como se indica en la ecuación \ref{eq:graph_triangles_frecuency}. En el trabajo \cite{bar2002reductions} se muestra una descripción acerca del algoritmo para encontrar una \emph{$(1 + \epsilon)$-estimación} basada en esta técnica con un coste computacional de $O(\alpha^{-2})$ en espacio donde $\alpha = \epsilon / (8 \cdot m \cdot n)$.

        \begin{equation}
        \label{eq:graph_triangles_frecuency}
          T_3 = F_0 + -1.5F_1 + 0.5 F_2
        \end{equation}

        \paragraph{}
        Otra propuesta es la estimación de $T_3$ mediante el uso de un \emph{$L_0$-Sampler} (del cual se habló en la sección \ref{sec:lp_samplers}). De esta técnica se habla en \cite{ahn2012graph} y presupone que si el cumplimiento de la propiedad $x_{\{v_{i_1},v_{i_2},v_{i_3}\}} = 3$ se denota por $Y$ y representa un \emph{proceso de Bernoulli}, entonces esta se puede extender a una distribución binomial al probar todas las combinaciones posibles. Los autores exponen que $\mathbb{E}[Y] = T_3/F_0$ y dado que existen técnicas para la estimación de $F_0$ (como el algoritmo de \emph{Flajolet-Martin}), entonces es posible obtener $T_3$.

        \paragraph{}
        La razón por la cual resulta interesante el conteo de triángulos es que se utiliza para el cálculo del \emph{coeficiente de agrupamiento} que denotaremos como $C$. La forma de calcular dicha propiedad se muestra en la ecuación \eqref{eq:clustering_coefficient_graph}. A partir de dicho indicador se puede obtener una estimación sobre la estructura del grafo, es decir, si es altamente conexo o por contra es disperso.

        \begin{equation}
        \label{eq:clustering_coefficient_graph}
          C = \frac{1}{n} \sum_{v\in V} \frac{T_3(v)}{\binom{deg(v)}{2}}
        \end{equation}

      \subsection{Árbol Recubridor Mínimo}
      \label{sec:minimum_spanning_tree}

        \paragraph{}
        El problema de la búsqueda del \emph{árbol recubridor mínimo} (\emph{Minimum Spanning Tree}) es uno de los casos más estudiados en problemas de grafos. Se refiere a la búqueda del sub-grafo $H = (V, E')$ respecto del grafo $G=(V, E)$ formado por todos los vértices del grafo $G$ y un sub-conjunto de aristas del mismo.

\begin{figure}
          \centering
          \includegraphics[width=0.3\textwidth]{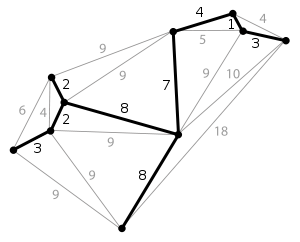}
          \caption{Ejemplo de \emph{Árbol Recubridor Mínimo}. (Extraído de \cite{wiki:Minimum_spanning_tree})}
          \label{img:minimum_spanning_tree}
        \end{figure}

        \paragraph{}
        La propiedad que debe cumplir el sub-conjunto de aristas $E'$ es que tan solo debe existir un único camino para llegar desde un vértice cualquiera al resto de vértices del grafo y, además, el sumatorio de los pesos de las aristas contenidas en dicho camino debe ser el mínimo respecto de todos los posible en $G$. En la figura \ref{img:minimum_spanning_tree} se muestra un ejemplo del  \emph{árbol recubridor mínimo} sobre un grafo ponderado

        \paragraph{}
        Nótese que para que exista un \emph{árbol recubridor mínimo} el grafo $G$ debe ser conexo. En el caso de que $G$ no sea conexo entonces se habla de \emph{bosque recubridor mínimo} (\emph{Minimum Spanning Forest}). El problema del \emph{árbol recubridor mínimo} no siempre tiene una solución única, sino que pueden encontrarse distintos sub-grafos $H_i$ que cumplan la propiedad utilizando sub-conjuntos de aristas diferentes. El caso más característico de esta situación es cuando se busca el \emph{árbol recubridor mínimo} sobre un grafo no ponderado, es decir, aquel en el cual todas las aristas tienen el mismo peso.

        \paragraph{}
        Este problema se ha definido comúnmente sobre grafos no dirigidos, por su mayor simplicidad y aplicación práctica. Sin embargo, la modelización es extensible a grafos dirigidos, para los cuales el problema es mucho más complejo dado que por cada vértice es necesario mantener un arista de entrada y otro de salida.

        \paragraph{}
        La versión estática del problema del \emph{árbol recubridor mínimo} ha sido ampliamente estudiada en la literatura, existiendo un gran número de alternativas, entre las que destaca históricamente el \emph{Algoritmo de Kruskal} descrito en el artículo \emph{On the shortest spanning subtree of a graph and the traveling salesman problem} \cite{kruskal1956shortest} cuya complejidad temporal es de $O(n + log(m))$ sobre un espacio de $O(n+m)$.

        \paragraph{}
        También se han propuesto soluciones basadas en algoritmos probabilistas, que ofrecen garantías de optimalidad en su solución. La versión más eficiente encontrada hasta la fecha se describe en el trabajo \emph{An optimal minimum spanning tree algorithm} \cite{pettie2002optimal} de \emph{Pettie y otros}. Sin embargo, debido a las técnicas que utiliza (mediante reducción a \emph{Árboles de Decisión}) no se puede representar mediante una función, pero sí que se ha demostrado su optimalidad. El problema también se ha estudiado desde la perspectiva de la parelelización. En el trabajo \emph{Fast shared-memory algorithms for computing the minimum spanning forest of sparse graphs} \cite{bader2006fast} \emph{Bader y otros} muestran un algoritmo que mediante el multiprocesamiento consiguen soluciones $5$ veces más eficientes que la versión optimizada secuencial.

        \paragraph{}
        Una vez descrito el problema y su estado actual sobre el modelo estático, el resto del apartado se basará en la descripción del mismo sobre las restricciones del \emph{modelo en semi-streaming}. Para ello, se realiza una descripción del algoritmo exacto propuesto por \emph{Ahn y otros} en \emph{Analyzing graph structure via linear measurements} \cite{ahn2012analyzing}. Dicha estrategia es capaz de encontrar el \emph{árbol recubridor mínimo} en $O(log(n)/log(log(n)))$ pasadas sobre el stream de aristas ajustándose a las restricciones espaciales de $O(polylog(n))$.

        \paragraph{}
        El planteamiento del algoritmo en semi-streaming se basa inicialmente en el \emph{algoritmo de Boruvka’s} (en \cite{wiki:Boruvkas_algorithm} se encuentra una breve descripción), cuyo planteamiento es el siguiente: se realizan $O(log(n))$ de tal manera que en cada una de ellas se seleccionan las aristas de menor peso que conecten vértices aún no marcados como conectados. De manera incremental se construye el \emph{árbol recubridor mínimo} hasta que todos los vértices del grafo están conectados.

        \paragraph{}
        Tal y como se indica en el documento, el \emph{algoritmo de Boruvka’s} puede emularse fácilmente sobre el modelo en semi-streaming emulando cada fase en $O(log(n))$ pasadas por el stream (para encontrar el arista de menor peso), lo cual conlleva $O(log(log(n)))$ pasadas en total. La idea en que se basa el algoritmo para reducir el número de pasadas sobre el streaming es realizar la búsqueda del arista de menor peso de cada nodo en \say{paralelo}, teniendo en cuenta las limitaciones espaciales ($O(polylog(n))$) lo cual es posible en $O(log(n)/log(log(n)))$.

        \paragraph{}
        El problema del \emph{árbol recubridor mínimo} tiene aplicaciones prácticas en muchos entornos reales, como el diseño de redes de telecomunicaciones o de transporte. También se han encontrado reducciones sobre el \emph{Problema del Viajante} y el \emph{Problema del corte mínimo}. Otros usos son el análisis de estructuras de grafos, problemas de clustering o reconocimiento de estructuras sobre imágenes.

      \subsection{Componentes Conectados}
      \label{sec:graph_connected_components}

        \paragraph{}
        El problema de \emph{Componentes Conectados} se refiere a la búsqueda del cardinal de sub-conjuntos de vértices para los cuales existe un camino entre todos los vértices del sub-conjunto. Mediante este resultado se puede determinar por tanto si existe un camino que entre dos vértices simplemente comprobando si pertenecen al mismo sub-conjunto de componentes conectados. Nótese por tanto que un grafo conexo tan solo posee un único componente conectado que contiene todos los vértices del grafo.

        \paragraph{}
        Se denota como $cc(G)$ al cardinal de componentes conectados del grafo $G$. El algoritmo clásico para resolver este problema se describe a continuación. Este requiere de $O(log(n))$ fases para su finalización y se basa en la fusión de vértices conectados. En cada fase se fusiona cada vértice con otro que posea una arista incidente a el para después crear un \emph{súper-vértice} con el cual se relacionen las aristas de los dos vértices fusionados. Tras realizar esta operación repetidas veces, el algoritmo termina cuando ya no es posible fusionar mas vértices. El número de \emph{súper-vértices} resultantes, por tanto es equivalente a $cc(G)$.

        \paragraph{}
        En el trabajo \emph{Analyzing graph structure via linear measurements} \cite{ahn2012analyzing} se describe un algoritmo para realizar dicha tarea sobre el \emph{modelo en semi-streaming} en una única pasada y con una complejidad espacial de $O(n \cdot log(n))$. Para ello, se basa en la construcción de sketches a partir del \emph{$L_0$-Samplers} denotados como $S_1, S_2, ..., S_t$ con $t = O(log(n))$. Esto conlleva un coste espacial de $O(n\cdot t \cdot log(log(n)))$, por lo que es válido sobre el \emph{modelo en semi-streaming}. La idea es la construcción jerárquica de estos sketches, de tal manera que $S_1$ represente el sketch de las aristas de los vértices del grafo $G$, $S_2$ las aristas de los \emph{súper-vértices} generados en la primera iteración, y así sucesivamente, de tal manera que a partir de $S_t$ se puede obtener $cc(G)$ al igual que en el algoritmo básico.

        \paragraph{}
        En \cite{ahn2012analyzing}, además, se extiende dicho algoritmo para el problema de \emph{$k$-aristas conectadas}, que indica el número mínimo de aristas incidentes sobre cada vértice. Las aplicaciones prácticas tanto de este problema como el de \emph{componentes conectados} son de utilizad para conocer la estructura del grafo. Un ejemplo práctico se da en grafos referidos a redes sociales, en las cuales se pretende encontrar el número de agrupaciones (o \emph{clusters}) de usuarios aislados del resto, así como el número mínimo de amistades que presenta cada usuario.

      \subsection{Paseos Aleatorios}
      \label{sec:random_walks_overview}

        \paragraph{}
        Un paseo aleatorio se define como la distribución de probabilidad referida a la realización de un camino de longitud $l$ desde un determinado vértice de inicio fijado \emph{a-priori}. Suponiendo que en cada vértice se tiene una distribución de probabilidad sobre sus aristas incidentes, entonces este problema está íntimamente relacionado con las \emph{cadenas de Markov}.

        \paragraph{}
        El algoritmo PageRank se refiere a la obtención del estado estacionario de la distribución de probabilidad de los paseos aleatorios con algunas modificaciones. Tanto los \emph{paseos aleatorios} como las cadenas de Markov se describen en detalle en el capítulo \ref{chap:pagerank} destinado al algoritmo PageRank.

    \section{Conclusiones}
    \label{sec:graph_conclusions}

      \paragraph{}
      A lo largo del capítulo se han citado distintas estrategias mediante las cuales se pretende reducir el grado de complejidad para la resolución de problemas sobre grafos, lo cual implica una reducción del tiempo de computo y del espacio necesario. Estas técnicas están ganando una gran relevancia en la actualidad debido a la necesidad de obtener respuestas en un periodo corto de tiempo, lo cual permite mejorar la toma de decisiones.

      \paragraph{}
      Debido al dinamismo del entorno en que vivimos, en un gran número de ocasiones es más beneficioso encontrar soluciones aproximadas que a pesar de no otorgar el resultado óptimo, son obtenidas en un tiempo mucho menor, lo cual permite una rápida adaptación a los cambios. Por lo cual, a través de este punto de vista, se pueden obtener mayores beneficios en promedio, puesto que los sobrecostes temporales propiciados por el tiempo de respuesta en soluciones exactas muchas veces conllevan una rápida obsolescencia de los resultados.

      \paragraph{}
      Se cree que en los próximos años, el estudio e investigación en el ámbito de la resolución de problemas sobre grafos de tamaño masivo será creciente, al igual que sucedía en el caso numérico como se indicó en anteriores capítulos. A pesar de ello, debido a su mayor grado de dificultad, dichos avances son lentos y actualmente se encuentran en fases muy tempranas de desarrollo, pero tal y como se ha indicado, se espera que esta situación cambie en el futuro.

  \chapter{Algoritmo PageRank}
  \label{chap:pagerank}

    \section{Introducción}
    \label{sec:pagerank_intro}

      \paragraph{}
      El algoritmo \emph{PageRank} fue nombrado por primera vez en el trabajo \emph{The PageRank citation ranking: Bringing order to the web} \cite{page1999pagerank} publicado por \emph{Larry Page} y \emph{Sergey Brin}. La motivación del mismo fue tratar de realizar un \emph{ranking de importancia} (o relevancia) sobre los nodos de un \emph{grafo dirigido no ponderado} de tal manera que este esté basado únicamente en la estructura de dicho grafo.

      \paragraph{}
      La motivación de dicho trabajo fue la de tratar de mejorar el ranking de resultados del buscador de sitios web \emph{Google} en que estaban trabajando. Hasta la publicación de dicho trabajo, los sistemas de búsqueda se basaban en heurísticas como el número de ocurrencias de la palabra clave sobre la cual se basaba la búsqueda o el número de enlaces hacia dicha página.

      \paragraph{}
      Sin embargo, los rankings basados en este tipo de estrategias podían ser fácilmente manipulables con el fin de tratar de conseguir posicionarse en las primeras posiciones del sistema de búsqueda. Por ejemplo, una página web que se basara en la repetición de la misma palabra muchas veces, entonces aparecería en primer lugar en rankings basados en el número de ocurrencias para dicha palabra clave. En el caso de rankings basados en el número de enlaces hacia dicha página, tampoco sería complejo manipular el resultado creando un número elevado de páginas web que contuvieran links hacia la página para la cual se pretende mejorar su posicionamiento.

      \paragraph{}
      La solución propuesta por \emph{Page} y \emph{Brin} para tratar de solucionar dicha problemática se basa en la generación de un ranking sobre los sitios web basado en la estructura del grafo subyacente, de tal manera que los vértices (sitios web) sobre los cuales existan aristas (enlaces) que provengan de otros vértices relevantes, tendrán una puntuación mayor que la de vértices que cuyo sub-conjunto de aristas los relacione con otros vértices menos relevantes.

      \paragraph{}
      La idea en que se basa el ranking se refiere por tanto a que los sitios web a los cuales se puede acceder a partir de otros sitios web considerados como importantes, entonces deberán encontrarse en primeras posiciones. Esta idea se extiende de manera inductiva sobre todos los vértices del grafo, puesto que tal y como veremos en las siguientes secciones converge hacia un estado estable (o \emph{distribución estacionaria} desde el punto de vista estadístico)

      \paragraph{}
      Para tratar de facilitar la comprensión acerca de esta idea a continuación se expone un ejemplo: Supongamos que en una red social como \emph{Twitter} (la cual se puede entender como un conjunto de usuarios que se relacionan entre si mediante relaciones de seguimiento, por lo que se puede ver como un grafo dirigido no ponderado donde el conjunto de usuarios se refiere a los vértices y el conjunto de relaciones de seguimiento con las aristas) un usuario habitual (el cual no tiene un número elevado de seguidores) comienza a seguir a un antiguo amigo de la universidad, el cual tampoco tiene un gran número de seguidores.

      \paragraph{}
      La red social \emph{Twitter} envía una notificación a todos seguidores del usuario indicando que este ha empezado a seguir a su amigo de la universidad. Puesto que su número de seguidores es bajo dicha acción no tendrá una gran relevancia y muy probablemente nadie más comience a seguir al amigo de la universidad. Sin embargo, supongamos que nuestro usuario, en lugar de tener un conjunto reducido de seguidores, es una persona influyente en la red social, a la cual siguen millones de personas, entonces la notificación del nuevo seguimiento le llegará a muchos más usuarios y probablemente el amigo de la universidad verá como su número de seguidores aumenta notablemente.

      \paragraph{}
      A grandes rasgos, esta es la idea en que se basa el algoritmo \emph{PageRank}, es decir, la puntuación de un vértice del grafo se basa en la relevancia de los vértices que contienen aristas que apuntan hacia el propio vértice.

      \paragraph{}
      La idea inicial del algoritmo \emph{PageRank} era la de realizar un ranking basado en la estructura del grafo de la web (\emph{Web Graph}), sin embargo, tal y como se verá a lo largo del capítulo, los conceptos matemáticos en que se basa dicho ranking son extrapolables a cualquier entorno que se pueda representar a partir de una red o grafo. En el trabajo \emph{PageRank beyond the Web}\cite{gleich2015pagerank} \emph{Gleich} realiza un estudio acerca de los distintos entornos sobre los cuales se han aplicado estas ideas.

      \paragraph{}
      Entre otros, se han realizado trabajos sobre los cuales se ha aplicado el algoritmo \emph{PageRank} en áreas tan dispares como la Biología, para analizar las células más importantes a partir de las interrelaciones entre ellas. También se ha aplicado en el sector de la neurociencia por razones similares. En el caso de la literatura, se ha aplicado sobre el grafo generado a partir del sistema de citaciones de artículos de investigación. Otros ámbitos de aplicación han sido sistemas de planificación de tareas o incluso estudios acerca de resultados en deportes, para conocer los encuentros más relevantes.

      \paragraph{}
      El resto del capítulo se organiza como sigue: Lo primero será hablar de \emph{Paseos Aleatorios} en la sección \ref{sec:random_walks}, lo cual permitirá comprender en mayor medida las idea sobre las cuales se basa el ranking \emph{PageRank}. A continuación se realizará una definición formal acerca del problema y lo que se pretende resolver en la sección \ref{sec:pagerank_formal_definition}. Una vez entendido el problema sobre el que se está tratando, en la sección \ref{sec:pagerank_algorithm} se describen las formulaciones para resolver el problema desde un punto de vista \emph{Algebraico} (sección \ref{sec:pagerank_algorithm_algebraic}), \emph{Iterativo} (sección \ref{sec:pagerank_algorithm_iterative}) y \emph{basado en Paseos Aleatorios} (sección \ref{sec:pagerank_algorithm_random_walks}). El siguiente paso es discutir cómo se puede añadir personalización al ranking, lo cual se lleva a cabo en la sección \ref{sec:pagerank_algorithm_personalized}. Por último, se presentan distintas alternativas al algoritmo \emph{PageRank} en la sección \ref{sec:pagerank_alternativas}, en la cual se habla de \emph{HITS} (sección \ref{sec:hits}), \emph{SALSA} (sección \ref{sec:salsa}) y \emph{SimRank} (sección \ref{sec:simrank}). Finalmente se realiza un breve resumen del capítulo en la sección \ref{sec:pagerank_conclusions}.

    \section{Paseos Aleatorios}
    \label{sec:random_walks}

      \paragraph{}
      En esta sección se trata el concepto de \emph{Paseos Aleatorios}, el cual está íntimamente relacionado con el algoritmo \emph{PageRank}. Para el desarrollo de esta sección se ha utilizado como herramienta bibliográfica el libro \emph{Randomized algorithms} \cite{motwani2010randomized} de \emph{Motwani} y \emph{Raghavan}, concretamente se ha prestado especial atención al \emph{capítulo 6: Markov Chains and Random Walks}. El resto de la sección se basará en la descripción de propiedades relacionadas con \emph{Paseos Aleatorios} para finalizar ilustrando la relacion de estos con \emph{PageRank}

      \paragraph{}
      Lo primero que haremos será describir en qué consiste un \emph{Paseo Aleatorio}. Para ello nos referiremos al grafo $G=(V,E)$ dirigido y no ponderado, el cual esta compuesto por $n = card(V)$ vértices y $m=card(E)$ aristas. Un paseo aleatorio se refiere entonces a un camino de longitud $l$ con origen en el vértice $v_{i_1}$ y final en $v_{i_l}$. Para que dicho camino constituya un paseo aleatorio, cada paso debe haber sido generado seleccionando de manera uniforme el siguiente vértice a visitar de entre los adyacentes al vértice actual. Nótese que este comportamiento puede ser visto como el promedio del modo en que los usuarios navegan por internet, de tal manera que acceden a páginas web mediante los enlaces que encuentran en la página que están visualizando. A continuación se describen las \emph{Cadenas de Markov} por su relación como herramienta de estudio para los paseos aleatorios.

      \subsection{Cadenas de Markov}
      \label{sec:markov_chains}

        \paragraph{}
        Para el estudio de paseos aleatorios, es apropiado utilizar la abstracción conocida como \emph{Cadenas de Markov}, las cuales están íntimamente relacionadas con el concepto de grafo y máquina de estados. Una \emph{Cadena de Markov} $M$ se define como un proceso estocástico que consta de $n$ posibles estados, los cuales tienen asociadas un conjunto de probabilidades denotadas como $p_{ij}=\frac{A_{ij}}{d^-(i)}$ para indicar la probabilidad con la cual se pasará del estado $i$ al estado $j$.

        \paragraph{}
        Dichas probabilidades se pueden representar de manera matricial sobre una matriz de transiciones $P$ de tamaño $n*n$, de tal manera que la posición $(i,j)$ contenga el valor $p_{ij}$ construido tal y como se indica en el párrafo anterior. Notese por tanto, que $\sum_{j}p_{ij}=1$ para que la distribución de probabilidades sea válida, es decir, la suma de probabilidades para las transiciones sobre cada estado deben sumar $1$.

        \paragraph{}
        Supóngase que se realiza un paseo aleatorio sobre la \emph{Cadena de Markov} $M$ cuya longitud $l$ es muy elevada ($l \gg n^2$), entonces, es fácil intuir que se visitará más de una vez cada estado. Sin embargo, el ratio de veces que se visitará cada estado muy probablemente no se distribuirá de manera uniforme, sino que habrá estados que serán visitados muchas más veces que otros. Esto depende en gran medida de la matriz de transiciones $P$. A la distribución de probabilidad generada sobre el ratio de visitas sobre cada nodo tras aplicar un paseo aleatorio de longitud elevada se le conoce como \emph{distribución estacionaria} y se denota como $\pi$.

        \begin{figure}
          \centering
          \includegraphics[width=0.6\textwidth]{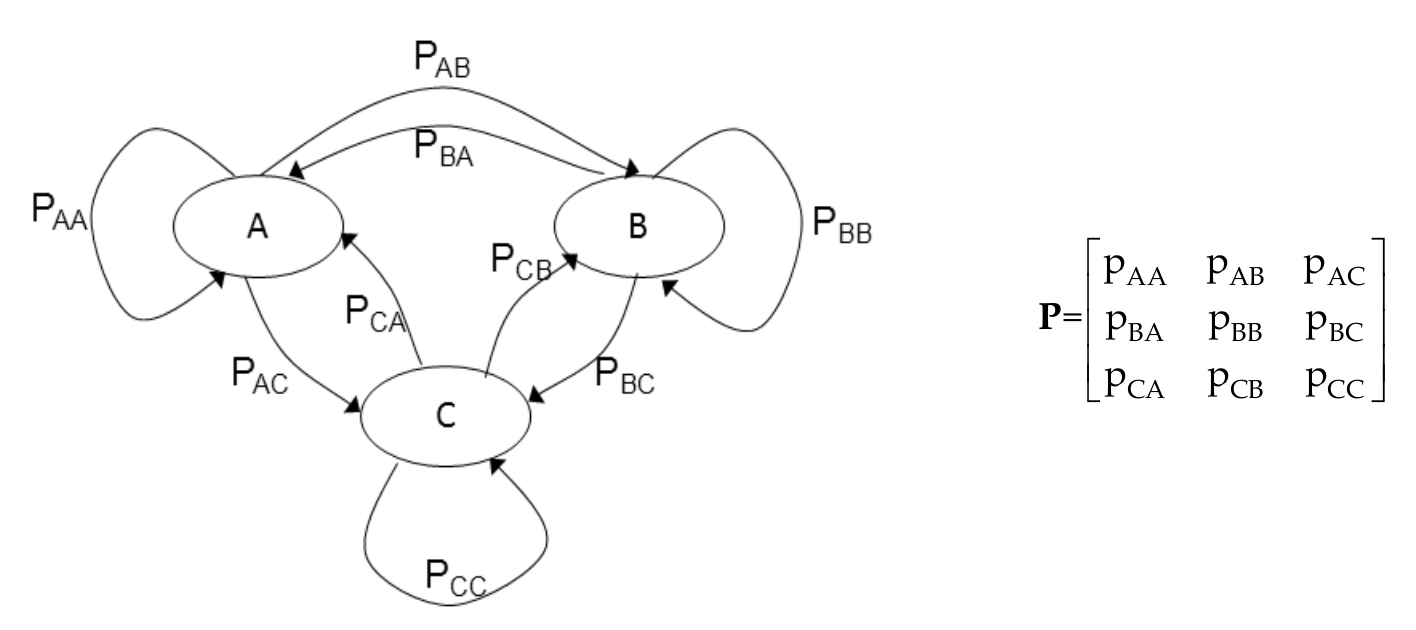}
          \caption{Ejemplo de \emph{Cadena de Markov}. (Extraído de \cite{sanchez2012wireless})}
          \label{img:markov_chain_example}
        \end{figure}

        \paragraph{}
        En la figura \ref{img:markov_chain_example} (Extraído de \cite{sanchez2012wireless}) se muestra una cadena de Markov constituida por 3 estados, tanto en su forma de grafo dirigido como de forma matricial.

        \paragraph{}
        La distribución estacionaria $\pi$ existe siempre que la \emph{Cadena de Markov} $M$ permita que a partir de un determinado estado $i$, se pueda llegar al menos a otro estado $j$. La razón se debe a que si se llega al estado $i$ y este no contiene más posibles estados de salida, entonces el resto de épocas se seguirá en el mismo estado. A los estados que poseen esta característica se los denomina sumideros. La segunda restricción para que se pueda calcular la distribución estacionaria $\pi$ es que la matriz de transiciones $P$ no debe ser periódica, es decir, no debe contener ciclos de probabilidad constante sobre los cuales el paseo aleatorio se quedaría iterando de manera indefinida.

        \paragraph{}
        Las definiciones descritas en este apartado se pueden extender de manera trivial al caso de grafos dirigidos sin más que utilizar cada vértice del grafo como un estado y construir la matriz $P$ de tal manera que $p_{ij}=\frac{A_{ij}}{d^-(i)}$ donde $d^-(i)$ representa el cardinal de aristas cuyo origen es el vértice $i$. Tal y como se verá posteriormente, el vector $\pi$ se corresponde con el resultado obtenido por el algoritmo \emph{PageRank} sobre una matriz $P$ de transiciones modificada.

      \subsection{Matriz Laplaciana de Paseos Aleatorios Normalizada}
      \label{sec:random_walk_normalized_laplacian_matrix}

        \paragraph{}
        En la sección \ref{sec:laplacian_matrix} se habló sobre la \emph{Matriz Laplacina}, la cual es una estrategia de representación, que ilustra distintas propiedades sobre el grafo subyacente. En este caso se describe una variación de la misma que es más apropiada para problemas relacionados con \emph{Paseos Aleatorios}. Esta se denota como $L^{{{\text{rw}}}}$ y se denomina \emph{Matriz Laplaciana de Paseos Aleatorios Normalizada}. La estrategia de construcción de la misma se indica en la ecuación \eqref{eq:random_walk_normalized_laplacian_matrix}. Esto consiste en asignar a la posición $(i,j)$ el opuesto de la probabilidad de transición del vértice $i$ al vértice $j$. Además, en la diagonal $(i,i)$ se asigna el valor $1$ cuando el grado del vértice es mayor que $0$.

        \begin{equation}
        \label{eq:random_walk_normalized_laplacian_matrix}
          L_{{i,j}}^{{{\text{rw}}}}:={
          \begin{cases}
            1&{\mbox{if}}\ i=j\ {\mbox{and}}\ d(v_{i})\neq 0\\
            -{P_{ij}}&{\mbox{if}}\ i\neq j\ {\mbox{and}}\ v_{i}{\mbox{ is adjacent to }}v_{j}\\
            0&{\mbox{otherwise}}.
          \end{cases}}
        \end{equation}

      \paragraph{}
      Una vez descritos los \emph{Paseos Aleatorios}, junto con las \emph{Cadenas de Markov} y la \emph{Matriz Laplaciana de Paseos Aleatorios Normalizada}, ya se está en condiciones suficientes como para describir de manera formal el \emph{PageRank} de un determinado grafo, que se realizará en la siguiente sección. Para ello, se indicarán las dificultades que surgen sobre este problema en grafos reales, así como las soluciones utilizadas para poder hacer frente a estas.

    \section{Definición Formal}
    \label{sec:pagerank_formal_definition}

      \paragraph{}
      Se define el \emph{PageRank} como la \emph{distribución estacionaria} $\pi$ de un determinado grafo dirigido no ponderado $G$ sobre el cual, la matriz de transiciones $P$, ha sido ligeramente modificada. Tal y como se ha visto en la sección anterior, la \emph{distribución estacionaria} consiste en la probabilidad de encontrarse en el estado $i$ durante un paseo aleatorio de longitud elevada sobre la \emph{Cadena de Markov} $M$. Tal y como se ha indicado, para que una cadena de Markov sea válida, entonces no deben existir estados \emph{sumideros} (no tienen posibles estados próximos).

      \paragraph{}
      Por estas razones, obtener la \emph{distribución estacionaria} de un grafo $G$, este no debe contener vértices \emph{sumideros}. La solución que proponen \emph{Page} y \emph{Brin} en \cite{page1999pagerank} para encontrar la \emph{distribución estacionaria} o \emph{PageRank} del grafo generado por los enlaces de la web (\emph{Web Graph}) es añadir un determinado índice de probabilidad sobre el cual los usuarios dejan de seguir los enlaces entre páginas web para acceder a otra distinta introduciendo la URL directamente. El apoyo en esta estrategia soluciona por tanto el problema de los vértices \emph{sumidero}, además de asemejarse en un mayor grado al comportamiento real de un usuario que navega por internet.

      \paragraph{}
      En \cite{page1999pagerank} \emph{Page} y \emph{Brin} proponen la modificación de la matriz de transiciones $P$ para la adaptación descrita en el párrafo superior, la cual se indica en la ecuación \eqref{eq:pagerank_transition_matrix}. De esta manera, se representa la situación en la cual un determinado usuario que llega a una página web sin enlaces hacia otras (\emph{sumidero}), accede a otra seleccionada de manera uniforme (esto se modeliza mediante el vector $p$ construido de tal manera que  $p_{i} = \frac{1}{n}, \ \forall i \in [1,n]$). Además, se añade un determinado índice $\beta$, que se corresponde con la probabilidad de que el usuario continúe seleccionando enlaces en la página actual o ,por contra, acceda a otra selecionandola de manera uniforme. Típicamente el valor $\beta$ se fija a $0.85$, sin embargo, admite cualquier valor contenido en el intervalo $[0,1]$.

      \begin{equation}
      \label{eq:pagerank_transition_matrix}
        p'_{ij} =
        \begin{cases}
          \beta * \frac{A_{ij}}{d^-(i)} + (1- \beta) * p_{i} & \mbox{if} \ d^-(i) \neq 0 \\
          p_{i}&\mbox{otherwise}
        \end{cases}
      \end{equation}

      \paragraph{}
      Tal y como se ha indicado anteriormente, el vector $p$ representa la distribución de probabilidad referida a los saltos que un usuario lleva a cabo entre sitios web sin tener en cuenta los enlaces del sitio web actual. En el párrafo anterior se ha indicado que este vector es construido siguiendo una distribución uniforme, por tanto, esto puede ser visto de tal manera que la probabilidad de saltar de un sitio web a otro es la misma. Sin embargo, dicha acción podría seguir una distribución de probabilidad distinta dependiendo de cada usuario de la red. Por tanto, en \cite{page1999pagerank} se habla de \emph{PageRank Personalizado} cuando el vector $p$ sigue una distribución de probabilidad distinta de la uniforme (desviada hacia los sitios web a los que más accede el usuario).

      \paragraph{}
      En el trabajo \emph{Topic-sensitive pagerank} \cite{haveliwala2002topic} \emph{Haveliwala} propone la generación de 16 \emph{distribuciones estacionarias} (\emph{PageRanks}) distintas mediante la personalización del vector $v$, para después realizar una combinación de estas y así conseguir que el ranking final sea personalizado.

      \paragraph{}
      Una vez descritas las transformaciones necesarias a realizar sobre la matriz de transiciones $P$ para que esta se adapte a la estructura de grafos con vértices \emph{sumidero}, y que además emule de manera más apropiada el comportamiento de un determinado usuario sobre el grafo de la web (\emph{Web Graph}), lo siguiente es explicar cómo se puede obtener la \emph{distribución estacionaria} o \emph{PageRank} del grafo. Para ello, a continuación se describe el \emph{Teorema de Perron–Frobenius}, que aporta una idea acerca de la manera en que se calcula, además de asegurar la convergencia de la matriz de transiciones hacia un estado estacionario del vector $\pi$.

      \subsection{Teorema de Perron–Frobenius}
      \label{sec:perron_frobenius_theorem}

        \paragraph{}
        El \emph{teorema de Perron–Frobenius} se refiere a la existencia de un \textbf{único} \emph{vector propio} (\emph{eigenvector}) para las matrices cuadradas reales positivas. Dicha descripción ha sido extraída del documento \emph{Notes on the perron-frobenius theory of nonnegative matrices} \cite{boyle2005notes} de \emph{Boyle} (profesor de matemáticas de la \emph{Universidad de Marylan}). En primer lugar es necesario describir los conceptos de \emph{vector propio} como de \emph{matriz cuadrada real positiva} para después ver que la \emph{distribución estacionaria} y la \emph{matriz de transiciones} referidos a una \emph{Cadena de Markov} $M$ pueden ser vistos de esta manera.

        \paragraph{}
        Una \emph{matriz cuadrada real positiva} $A$ es aquella formada por $n$ filas y $n$ columnas ($n*n$ celdas) para las cuales $\forall i,j \in [1,n]$ se cumple que $A_{ij} \in \mathbb{R} \geq 0$. Tal y como se puede apreciar, la \emph{matriz de transiciones modificada} $P'$ del grafo $G$ cumple esta propiedad ya que $\forall i,j \in [1,n] \ P'_{ij} \in [0,1]$.

        \paragraph{}
        En cuanto al concepto de \emph{vector propio} $\lambda$ de un matriz, se refiere a un vector de $n$ columnas ($1*n$) tal que cuando es multiplado por una determinada matriz $A$, el resultado sigue siendo el mismo. Es decir, se cumple que $\lambda = \lambda * A$. Notese por tanto, que esta idea es equivalente a la \emph{distribución estacionaria} desde el punto de vista de llegar a un estado estable.

        \paragraph{}
        El teorema de \emph{teorema de Perron–Frobenius} asegura por tanto, que para una \emph{matriz cuadrada real positiva} $A$ tan solo existe un único \emph{vector propio} $\lambda$ y el conjunto de valores de este se es estrictamente positivo, es decir, $\forall i \in [1,n] \ \lambda_i \geq 0$. La demostración de dicho teorema puede encontrarse en \cite{boyle2005notes}.

        \paragraph{}
        Además, en el caso de que la matriz $A$ haya sido normalizada por columna, es decir, se cumpla que $\forall i \in [1,n] \ \sum_j A_{ij} = 1$, entonces el autovector $\lambda$ también seguirá la propiedad de normalización ($\sum_i \lambda_i = 1$). Gracias a este resultado es posible calcular la \emph{distribución estacionaria} $\pi$ de una \emph{Cadena de Markov} como el \emph{vector propio} de su matriz de transiciones.

      \paragraph{}
      Una vez descrito el \emph{teorema de Perron–Frobenius} ya se está en condiciones suficientes para describir las distintas alternativas para calcular el \emph{PageRank} de un determinado grafo, el cual se calcula tal y como se ha indicado en esta sección, encontrando el \emph{vector propio} de la matriz de transición modificada de la cadena de Markov. Las distintas estrategias para obtener este resultado se describen en la siguiente sección.

    \section{Algoritmo Básico}
    \label{sec:pagerank_algorithm}

      \paragraph{}
      En esta sección se describe el método para obtener el vector \emph{PageRank} sobre un determinado grafo $G$. Para ello, es necesario fijar 3 parámetros los cuales se indican a continuación:
      \begin{itemize}
        \item Matriz de adyacencia $A$, que a partir de la cual se obtiene la estructura del grafo (Se habló de ella en la sección \ref{sec:adjacency_matrix}).
        \item El valor de probabilidad $\beta$ de seguir el paseo aleatorio a partir de la distribución del vértice actual (el cual se comentó en la sección anterior).
        \item El vector de personalización $p$ referido a la distribución de probabilidad de los saltos aleatorios entre vértices (también se habló en la sección anterior).
      \end{itemize}

      \paragraph{}
      Para calcular el \emph{vector propio} $\lambda$ existen distintas estrategias matemáticas. En esta sección se habla de dos estrategias, la primera de ellas basada en la resolución de un sistema de ecuaciones lineales mientras que la segunda se basa en acercamiento a la solución de manera iterativa. Tal y como se verá a continuación, la estrategia algebraica conlleva un coste computacional muy elevado, por lo que no es admisible sobre grafos de tamaño masivo. En estos casos se utiliza la estrategia iterativa u otras alternativas basadas en la generación de \emph{Paseos Aleatorios}.

      \subsection{Estrategia Algebraica}
      \label{sec:pagerank_algorithm_algebraic}

        \paragraph{}
        La idea de la estrategia algebraica se refiere a la búsqueda del vector $\lambda$ que resuelva la ecuación $\lambda = \lambda * P'$ como un sistema de ecuaciones lineales. Esto se puede llevar a cabo siguiendo el desarrollo de la ecuación \eqref{eq:pagerank_algorithm_algebraic_1}. Nótese que para ello no se utiliza la \emph{matriz de transiciones modificada} $P'$ explícitamente. En su lugar, esta es representada implícitamente a partir de las operaciones de \eqref{eq:pagerank_algorithm_algebraic_2} y \eqref{eq:pagerank_algorithm_algebraic_3}.

        \paragraph{}
        Para entender estas ecuaciones, lo primero es indicar la notación que se ha utilizado así como la interpretación de algunas operaciones: El símbolo $\boldsymbol{I}$ representa la matriz identidad ($1$'s en la diagonal y $0$'s en el resto) de tamaño $n$. El símbolo $d^-$ representa un vector columna de tamaño $n$ que representa en su posición $j$ el cardinal de aristas cuyo origen es el vértice $j$. Esto puede ser visto de la siguiente manera: $\forall j \in [1,n] \ \sum_i A_{ij} = d^{-}_{j}$. A nivel de operaciones es necesario resaltar el caso de la división $\frac{A}{d^-}$ por su carácter matricial. Esta se lleva a cabo realizando la división elemento a elemento por columnas.

        \begin{align}
          \label{eq:pagerank_algorithm_algebraic_1}
          \lambda =& \lambda * P' \\
          \label{eq:pagerank_algorithm_algebraic_2}
                  =& \lambda * \beta * \frac{A}{d^-} + (1-\beta)*p \\
          \label{eq:pagerank_algorithm_algebraic_3}
                  =& \bigg(\boldsymbol{I} - \beta * \frac{A}{d^-}\bigg)^{-1} * (1-\beta)*p
        \end{align}

        \paragraph{}
        A partir de las operaciones descritas en la ecuación \eqref{eq:pagerank_algorithm_algebraic_3}, se obtiene por tanto el \emph{vector propio} $\lambda$, que en este caso se refiere a la \emph{distribución estacionaria} de la \emph{cadena de Markov} descrita por la matriz de transiciones modificada $P'$, por lo que es equivalente al vector \emph{PageRank}. Sin embargo, el calculo del vector \emph{PageRank} siguiendo esta estrategia conlleva un elevado coste computacional derivado de la necesidad de invertir una matriz de tamaño $n*n$, algo inadmisible para grafos de tamaño masivo.

      \subsection{Estrategia Iterativa}
      \label{sec:pagerank_algorithm_iterative}

        \paragraph{}
        La \emph{estrategia iterativa} para el cálculo del vector \emph{PageRank} se basa en la aproximación a este mediante la multiplicación del mismo por la matriz de transiciones modificada $P'$ de manera repetida hasta que este llegue a un estado estable desde el punto de vista de una determinada norma vectorial.

        \paragraph{}
        El primer paso es calcular la \emph{matriz de transiciones modificada} $P'$ a partir de los tres parámetros de entrada $(A, \beta, v)$  siguiendo la definición de la ecuación \eqref{eq:pagerank_transition_matrix}. Una vez obtenida dicha matriz se está en condiciones de calcular el vector \emph{PageRank} siguiendo la idea del \emph{vector propio} único expuesta en la sección anterior.

        \paragraph{}
        El algoritmo para dicha tarea se muestra en la figura referida al \emph{Algorimo \ref{code:iterative_pagerank}}. Este toma como argumentos de entrada la matriz de transición aproximada $P'$ junto con un determinado valor de convergencia $\in (0,1)$, que condiciona la precisión del resultado así como el número de iteraciones necesarias para llegar a él.

        \paragraph{}
        \begin{algorithm}
          \SetAlgoLined
          \KwResult{$\pi(t)$ }
          $t \gets 0$\;
          $\pi(t) \gets \frac{1}{n}*\boldsymbol{1}$\;
          \Do{$||\pi(t) - \pi(t-1)|| > conv$}{
            $t \gets t + 1$\;
            $\pi(t) = \pi(t-1) * P'$\;
          }
          \caption{Iterative PageRank}
          \label{code:iterative_pagerank}
        \end{algorithm}

        \paragraph{}
        En cuanto al vector \emph{PageRank} $\pi$, este debe ser inicializado antes de comenzar el bucle de iteraciones. La única restriccione que se pide es que la suma de las posiciones del mismo sea $1$, es decir, que la norma 1 del sea igual a 1 ($||\pi||_1 = \sum_i \pi_i = 1$) para mantener la propiedad de distribución de probabilidad. Existen distintas heurísticas destinadas a reducir el número de iteraciones del algoritmo mediante la inicialización de este vector, sin embargo, en este caso se ha preferido inicializarlo siguiendo una distribución uniforme. Esta inicialización no determina el resultado, tan solo el tiempo de convergencia hacia este.

        \paragraph{}
        Tal y como se puede apreciar, el bucle de iteraciones consiste únicamente en la multiplicación del vector \emph{PageRank} por la matriz $P'$ junto con la actualización del índice referido a la iteración. El siguiente punto a discutir es el criterio de convergencia del resultado. Para ello existen distintas normas vectoriales, entre ellas la norma uno (descrita en el párrafo anterior) o la norma infinito ($||\pi||_{\infty}=max_i\{\pi_i\}$). En este caso se ha creído más conveniente la utilización de la \emph{norma 1} para el criterio de convergencia puesto que se pretenden reducir todos los valores del vector y no únicamente el máximo de todos ellos, por lo que de esta manera se consigue una aproximación más homogénea respecto  del \emph{PageRank Exacto}.

      \subsection{Estrategia Basada en Paseos Aleatorios}
      \label{sec:pagerank_algorithm_random_walks}

        \paragraph{}
        También existe una estrategia basada en paseos aleatorios para el cálculo del vector \emph{PageRank}, la cual sigue una estrategia muy diferente de las descritas en anteriores secciones. En este caso la idea principal se basa en realizar paseos aleatorios sobre los vértices del grafo siguiendo la distribución de probabilidades descrita a partir de la matriz de transiciones modificada $P'$ para después estimar el vector \emph{PageRank} a partir del número de veces que cada vértice ha aparecido en los paseos aleatorios.

        \paragraph{}
        La dificultad algorítmica en este caso se refiere a la generación de valores aleatorios a partir de una distribución ponderada (obtenida de la matriz de transiciones $P'$) ya que es necesario mantener $n$ generadores de valores aleatorios que indiquen el siguiente vértice a visitar en cada caso. Obviando esta dificultad, el resto del algoritmo es bastante simple y se basa en el mantenimiento del ratio de visitas sobre cada vértice conforme el paseo aleatorio aumenta su longitud. Como en el caso iterativo, también se basa en la repetición de la misma operación hasta llegar a un criterio de convergencia (idéntico a la solución iterativa).

        \paragraph{}
        La figura correspondiente al Algoritmo \ref{code:random_walks_pagerank}, que se basa en la actualización del vector \emph{PageRank} conforme se obtiene el resultado del próximo vértice a visitar. De esta manera en cada iteración se realizan $n$ paseos aleatorios de longitud $1$ que se combinan para llegar al estado estacionario. La dificultad conceptual en este caso se refiere tanto a la generación de números aleatorios ponderados y el mantenimiento de una media incremental normalizada sobre el vector \emph{PageRank} $\pi$. Tal y como se ha indicado en el párrafo anterior, el criterio de convergencia es equivalente al caso iterativo.

        \paragraph{}
        \begin{algorithm}
          \SetAlgoLined
          \KwResult{$\pi(t)$ }
          $t \gets 0$\;
          $\pi(t) \gets \frac{1}{n}*\boldsymbol{1}$\;
          \Do{$||\pi(t) - \pi(t-1)|| > conv$}{
            $t \gets t + 1$\;
            $\pi_u(t) \gets \frac{t}{t+n}*\pi_u(t) $\;
            \For{cada $v \in V$}{
              $u \gets randomWeighted(v)$\;
              $\pi_u(t) \gets \pi_u(t) + \frac{1}{t+n}$\;
            }
          }
          \caption{Random Walks PageRank}
          \label{code:random_walks_pagerank}
        \end{algorithm}

      \paragraph{}
      Las estrategias descritas en este capítulo para la obtención de la distribución estacionaria $\pi$ sobre la matriz de transiciones modificada $P'$ del grafo $G$ se han hecho desde una perspectiva básica. Existen gran cantidad de trabajos sobre distintas variaciones de estos métodos para tratar de reducir el tiempo de convergencia.

    \section{PageRank Personalizado}
    \label{sec:pagerank_algorithm_personalized}

      \paragraph{}
      Tal y como se ha indicado a lo largo del capítulo, el algoritmo \emph{PageRank} genera un ranking referido al ratio de importancia de cada vértice perteneciente a un determinado grafo dirigido no ponderado $G$. Dicha medida para el vértice $v$ se calcula como la probabilidad de que tras un paseo aleatorio por los vértices del grafo siguiendo las aristas del mismo, el vértice final del camino sea $v$.

      \paragraph{}
      Para hacer frente a aquellos casos en los cuales se llega a un vértice que no contiene aristas (\emph{vértice sumidero}) sobre las cuales acceder a otros vértices entonces es necesario realizar un salto hacia otro vértice del grafo sin tener en cuenta las aristas. Tal y como se ha indicado, este proceso se lleva a cabo tanto en los \emph{vértices sumidero}, para los cuales es necesario para poder calcular el ranking, como en el resto de vértices con probabilidad $1-\beta$.

      \paragraph{}
      En esta sección se habla de la distribución de probabilidad que sigue el proceso de \emph{saltar} de un vértice a otro. Dicha distribución se codifica mediante el vector $p = [ p_1, p_2,...,p_i,...,p_n ]$, cuya única restricción se refiere a la suma de valores, que debe ser 1, es decir, $p$ debe estar normalizado ($||p||_1 = 1$).

      \paragraph{}
      El caso básico se refiere a la distribución uniforme, para la cual todas las posiciones del vector toman el mismo valor, es decir, $\forall i \in [1,n], \ p_i = \frac{1}{n}$. De esta manera se modeliza la situación en la cual cuando se lleva a cabo un salto, todos los vértices tienen la misma probabilidad de ser el vértice de destino. Nótese por tanto, que en este caso no existe no existe ningún grado de personalización sobre el resultado, por lo que se conoce como \emph{PageRank General} ya que aporta una estimación acerca de la importancia de los vértices del grafo desde una perspectiva generalizada, lo cual es interesante desde el punto de vista de la obtención de analíticas del grafo.

      \paragraph{}
      La situación descrita en el párrafo anterior, sin embargo, no se asemeja al comportamiento real de un usuario que navega por la red. La razón se debe a que cuando alguien accede directamente a un sitio web escribiendo la url del mismo en su navegador favorito, este no selecciona una al azar de entre los millones de páginas posibles, sino que la escoge de entre un reducido sub-conjunto de ellas, que además no es uniforme, dado que los usuarios no visitan las páginas web el mismo número de veces.

      \paragraph{}
      A través de dicha idea, el resultado del algoritmo \emph{PageRank} cambia y en lugar de generar un ranking general del grafo $G$, ahora realiza un ranking sesgado otorgando una mayor puntuación a los vértices cercanos a los vértices de destino de los saltos, lo cual altera en gran medida el resultado anterior.

\begin{figure}
        \centering
        \includegraphics[width=0.4\textwidth]{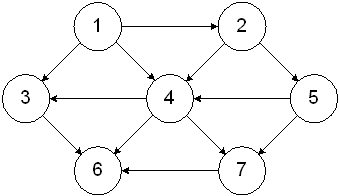}
        \caption{Ejemplo de \emph{Grafo Dirigido No Ponderado}. (Extraído de \cite{freedman2010graphs})}
        \label{img:directed_graph_example}
      \end{figure}

      \paragraph{}
      Para comprender esta idea se ha incluido en la figura \ref{img:directed_graph_example} un ejemplo de \emph{grafo dirigido no ponderado}. Supóngase que se calcula el \emph{PageRank} seleccionando el vector $p$ de tal manera que $p_3 =1$ y $\forall i \in [1,7] - [3], \ p_i = 0$. Mediante esta situación se consigue que todos los saltos (los referidos a vértices sumidero y los relizados con probabilidad $1-\beta$) tengan como vértice destino el $3$. Por esta razón con total seguridad el vértice $3$ tendrá la mejor puntuación \emph{PageRank} del grafo. Además, la puntuación del resto de vértices también se habrá modificado, con altas probabilidades de que el vértice $6$ se encuentre en segunda posición.

      \paragraph{}
      A través de la modificación de la distribución de probabilidad del vector $p$ se consigue por tanto la adaptación del ranking \emph{PageRank} desde una perspectiva de localidad, lo cual se convierte en una clasificación mucho más interesante desde el punto de vista del usuario, que en el caso del grafo de la web, conoce de manera mucho más fiel las páginas web que son importantes desde su perspectiva individual.

      \paragraph{}
      Sin embargo, la personalización del ranking \emph{PageRank} conlleva distintos problemas a nivel computacional. Esto se debe a que en este caso en lugar de tener que mantener un único ranking para todo el grafo es necesario mantener un ranking para cada usuario que pretenda consultar el pagerank del grafo. Este problema no es tan grave como parede puesto que el vector de la distribución estacionaria $\pi$ posee la propiedad de linealidad, lo cual simplifica dicha problemática, reduciendo el problema a la necesidad de mantener un vector $\pi$ para cada vértice del grafo, es decir, hay que generar $n$ vectores \emph{PageRank}.

      \paragraph{}
      Ahora supongamos que un determinado usuario desea calcular su propio ranking personalizado. La tarea se reduce a realizar una media ponderada sobre los vectores $\pi_i$ escogiendo los pesos según el vector $p$ de personalización para dicho usuario. A pesar de ello, esta tarea continua siendo compleja, tanto a nivel temporal como espacial, puesto que se incrementa el coste respecto del \emph{PageRank General} en un orden de $n$.

      \paragraph{}
      Debido al tamaño de grafos como el de la web (\emph{Web Graph}), cuyo número de vértices es tan elevado que no es posible mantener un vector \emph{PageRank} para cada uno de ellos en memoria. Esta razón dificulta la tarea de calcular el ranking personalizado para un determinado usuario en tiempo de consulta. Por tanto, se ha se han propuesto distintas soluciones para este problema.

      \paragraph{}
      Una de las primeras es la descrita en \cite{haveliwala2002topic}, que se basa en el mantenimiento de \emph{16} vectores \emph{PageRank} de referidos a temas distintos entre si, que después se combinan en tiempo de consulta para generar un ranking personalizado. En \cite{kamvar2003exploiting} los autores proponen el cálculo de vectores \emph{PageRank} teniendo en cuenta la estructura de bloques que se genera entre vértices cuyo ranking es similar. Esto se lleva a cabo a partir de un algoritmo de 3 fases (búsqueda de vértices similares, cálculo del vector \emph{PageRank} para los conjuntos resultantes y, en tiempo de consulta, combinación de los mismos para obtener el ranking personalizado).

      \paragraph{}
      En \cite{jeh2003scaling} se explica una técnica que permite calcular el vector \emph{PageRank} personalizado utilizando una técnica incrementa o jerárquica basada en \emph{Programación Dinámica}, lo cual permite hacer calcular el ranking basado en la combinación de cien mil vectores en tiempo de consulta. En \cite{sarlos2006randomize} se propone una solución semejante en un espacio menor mediante la utilización del \emph{Count-Min Sketch} (Sección \ref{sec:count_min_sketch}).En el trabajo \emph{Estimating pagerank on graph streams} \cite{sarma2011estimating} \emph{Sarma y otros} proponen un algoritmo basado en la generación de paseos aleatorios para estimar el ranking \emph{PageRank} sobre las restricciones que impone el \emph{modelo en semi-streaming}.

    \section{Alternativas a PageRank}
    \label{sec:pagerank_alternativas}

      \paragraph{}
      El algoritmo \emph{PageRank} se ha convertido en la alternativa más usada para el cálculo del ránking de vértices sobre un grafo basándose únicamente en la estructura del mismo (relaciones a partir de aristas). Su popularidad se debe en gran medida tanto a su sencillez desde el punto de vista algorítmico (a pesar de que su coste computacional es elevado), como a la gran fama del motor de búsquedas \emph{Google}, que desde sus comienzos otorgaba resultados muy buenos apoyándose en la utilización de dicho ranking.

      \paragraph{}
      Sin embargo, \emph{PageRank} no es la única alternativa sobre la que se ha trabajado en el ámbito del cálculo de importancia para los vértices de grafos. En esta sección se describen brevemente distintos algoritmos cuya finalidad es semejante. Además se hablará de una extensión del \emph{PageRank} conocida como \emph{SimRank} que obtiene el grado de similitud entre los vértices del grafo desde un punto de vista estructural. En la sección \ref{sec:hits} se describe \emph{HITS}, posteriormente en la sección \ref{sec:salsa} se describe \emph{SALSA} (una mejora respecto del anterior), y finalmente, en la sección \ref{sec:simrank} se hablará de \emph{SimRank}.

      \subsection{HITS}
      \label{sec:hits}

        \paragraph{}
        El algoritmo \emph{HITS} surge en paralelo junto con \emph{PageRank}, puesto que los artículos en los que se habla de dichos algoritmos fueron publicados en \emph{1999}. \emph{HITS} fue descrito por primera vez en el trabajo \emph{Authoritative sources in a hyperlinked environment} \cite{kleinberg1999authoritative} de \emph{Kleinberg}.

        \paragraph{}
        Este algoritmo, a diferencia del \emph{PageRank}, se basa en la generación del ranking en tiempo de consulta. Para ello utiliza un sub-conjunto de vértices inicial obtenido a partir del ranking del ranking de similitud basada en texto respecto de la consulta en cuestión. A partir del sub-grafo generado por este sub-conjunto. Las iteraciones del algoritmo se basan la generación de dos puntuaciones para cada vértice. Estas puntuaciones se conocen como \emph{Autority score} y \emph{Hub score}. Las cuales se construyen como la suma de \emph{Hub scores} de los vértices que apuntan hacia el vértice para el caso del \emph{Autority score} y la suma de \emph{Autority scores} de los vértices hacia los que apunta el vértice para el \emph{Hub score}. Dichos valores son normalizados en cada iteración. Este proceso se repite hasta llegar a un determinado grado de convergencia.

        \paragraph{}
        Las diferencias respecto del \emph{PageRank} por tanto se basan en la generación del ranking en tiempo de consulta en lugar de en tiempo de indexación o estático, por tanto este ranking varía respecto de cada búsqueda. En este caso, en lugar de obtener un único ranking se obtienen dos, indicando los vértices más importantes y indicando los vértices que más importancia generan. Además, en lugar de generar el ranking sobre el grafo completo, \emph{HITS} lo lleva a cabo sobre un sub-grafo del mismo (obtenido mediante búsqueda textual tal y como se ha indicado anteriormente).

      \subsection{SALSA}
      \label{sec:salsa}

        \paragraph{}
        El algoritmo \emph{SALSA} se refiere a una combinación de \emph{PageRank} y \emph{HITS} (descrito en la anterior sección), por tanto, tiene características semejantes de ambas alternativas. La descripción del mismo fue llevada a cabo en el documento \emph{SALSA: the stochastic approach for link-structure analysis}\cite{lempel2001salsa} de \emph{Lempel y Moran}.

        \paragraph{}
        Debido a que este algoritmo consiste en una combinación de los citados previamente, a continuación se indican las semejanzas que tiene respecto de estos. \emph{SALSA} se basa en la misma idea que \emph{HITS}, en el sentido de que genera dos ranking, referidos a autoridades y generadores de autoridad (\emph{Autority} y \emph{Hub}). Además, también realiza dicho ranking en tiempo de consulta utilizando un sub-grafo generado a partir del ranking de similitud textual. La diferencia respecto de \emph{HITS} se basa en la estrategia utilizada para calcular dichos ranking. En este caso utiliza el mismo enfoque de \emph{paseos aleatorios} del \emph{PageRank} (por eso se dice que es una combinación de ambos).

        \paragraph{}
        La generación de rankings a través de paseos aleatorios le otorgo una ventaja significativa a nivel de coste computacional respecto de \emph{HITS}, además de ofrecer resultados en tiempo de consulta, lo cual le diferencia de \emph{PageRank}. En el documento \emph{WTF: The Who to Follow Service at Twitter} \cite{gupta2013wtf} los autores describen cómo la red social \emph{Twitter} ha desarrollado una variación de este algoritmo para su sistema de recomendación de usuarios a los que comenzar a seguir.

      \subsection{SimRank}
      \label{sec:simrank}

        \paragraph{}
        El algoritmo \emph{SimRank} tiene un propósito distinto respecto de los descritos anteriormente. En este caso, en lugar de tratar de conocer el grado de importancia de un determinado vértice del grafo, lo que se pretende es obtener un ranking de similitud del resto de vértices del grafo respecto de uno concreto. En el trabajo \emph{SimRank: a measure of structural-context similarity} \cite{jeh2002simrank} de \emph{Jeh y Widom} se describe de manera el completa el modo de funcionamiento del algoritmo, así como la demostración acerca de la corrección del mismo. A continuación se realiza una breve descripción acerca de dicho modo de funcionamiento.

        \paragraph{}
        Al igual que el \emph{PageRank}, en este caso el algoritmo también se basa en el cálculo iterativo del ranking hasta llegar a un determinado índice de convergencia. La primera iteración se basa en el cálculo de un \emph{PageRank Personalizado} desde el vértice sobre el cual se pretende basar la comparación. Tras esta inicialización, el resto del algoritmo se basa en la repetición de la ecuación \eqref{eq:simrank_iteration} hasta llegar a un índice de convergencia.

        \begin{equation}
        \label{eq:simrank_iteration}
          Sim^{(k)}_{u_1}({u_2}) =
          \begin{cases}
            (1-c) * \frac{\sum_{\{(u_1,v_1),(u_2,v_2)\} \in E} Sim^{(k-1)}_{v_1}({v_2})}{d^-({u_1})*d^-({u_2})}, & \text{if} \ u_1 \neq u_2 \\
            1, & \text{if} \ u_1 = u_2
          \end{cases}
        \end{equation}

        \paragraph{}
        La ecuación \eqref{eq:simrank_iteration} calcula por tanto el grado de similitud del vértice $u_2$ respecto del vértice $u_1$ en la iteración $k$, lo cual se puede denotar como $Sim^{(k)}_{u_1}({u_2})$. Tal y como se ha indicado anteriormente, dicho valor se incrementa conforme los vértices $u_1$ y $u_2$ se relacionan con conjuntos similares de vértices desde el punto de vista de que estos sean los mismos. Esto no debe confundirse con otros problemas como el de \emph{Matchings} (referido a encontrar subestructuras con la misma forma dentro de un grafo).

        \paragraph{}
        Existen numerosas aplicaciones prácticas para este algoritmo como sistema de recomendaciones sobre conjuntos de datos con estructura de grafo. Algunos ejemplos donde podría ser utilizado es en problemas como la generación de anuncios de sitios web en un buscador, la generación de listas de reproducción de vídeo a partir de un determinado vídeo teniendo en cuenta la navegación que los usuarios han llevado a cabo previamente entre estos, o un sistema de recomendación de compras en una tienda virtual basándose en la misma idea.

    \section{Conclusiones}
    \label{sec:pagerank_conclusions}

      \paragraph{}
      Tal y como se ha visto a lo largo del capítulo, el cálculo de importancia sobre los vértices grafos es una tarea compleja, cuya dificultad se incrementa en gran medida cuando el tamaño del grafo es de tamaño masivo, lo cual dificulta la tarea de contenerlo de manera completa en memoria. Sin embargo, en este capítulo no se han discutido soluciones para dicho problema, sino que se ha realizado un estudio acerca del algoritmo \emph{PageRank}, el cual goza de gran popularidad gracias a motor de búsquedas \emph{Google}.

      \paragraph{}
      A partir de la descripción de este algoritmo se ha podido comprender la perspectiva algebraica para la resolución de un problema sobre grafos, que puede entenderse como uno problema de matrices. Además, se ha ilustrado la cercanía de este ranking respecto de las \emph{Cadenas de Markov} y los paseos aleatorios, que convergen hacia la \emph{distribución estacionaria}. Después se ha indicado cómo calcularla a partir de distintas estrategias (\emph{Algebraica}, \emph{Iterativa} o \emph{basada en paseos aleatorios})

      \paragraph{}
      Posteriormente también se habló del \emph{PageRank Personalizado} y la problemática referida a la cantidad de rankings que sería necesario mantener para llevar a cabo una solución exacta para dicho problema. Por último se ha hablado de estrategias similares para el ranking de vértices así como el algoritmo \emph{SimRank}, que indica el grado de similitud de vértices respecto de un determinado vértice.

      \paragraph{}
      Gracias a este algoritmo se puede obtener un indicador acerca de la importancia de los vértices de un determinado grafo, lo cual es una tarea interesante y aplicable sobre un gran conjunto de ámbitos para los cuales el problema puede ser modelizado como una red. La observación acerca de la importancia de los vértices desde el punto de vista estructural es un factor interesante que puede mejorar la toma de decisiones en un gran número de casos.

  \chapter{Implementación, Resultados y Trabajo Futuro}
  \label{chap:implementation}

    \section{Introducción}
    \label{sec:implementation_intro}

      \paragraph{}
      A lo largo de este documento se han descrito distintas ideas relacionadas con nuevas técnicas para tratar de hacer frente al problema de la complejidad derivada del tamaño de conjunto de datos de tamaño masivo, para el cual es necesario utilizar técnicas sofisticadas que agilicen dichos procedimientos. Dichos conceptos se han descrito desde una perspectiva teórica dejando de lado cuestiones de implementación u otros factores. Dicha abstracción ha permitido simplificar las descripciones teniendo en cuenta únicamente el enfoque algorítmico de las mismas.

      \paragraph{}
      Sin embargo, el enfoque que se seguirá en este capítulo pretende ser muy diferente, centrándose en los detalles de implementación y dejando de lado el contenido matemático. De esta manera, se pretende describir el código fuente desarrollado desde la perspectiva de su estructura y organización, ya que a pesar de basarse en una implementación que trata de ejemplificar conceptos descritos a lo largo del documento, se ha dedicado especial cuidado tratando de escribir código de calidad, mantenible y reutilizable.

      \paragraph{}
      Antes de profundizar en detalles relacionados con la implementación en si, es necesario realizar una explicación acerca de lo que se ha pretendido conseguir mediante el desarrollo de la misma, ya que debido al contexto en que se enmarca (\emph{Trabajo de Fin de Grado} de \emph{Ingeniería Informática}) y la metodología seguida para la realización  del mismo (\emph{Proyecto de Investigación}), esta implementación se encuentra en las primeras fases de su desarrollo, por lo cual aún no tiene el grado de madurez esperado para ser incluida en entornos de producción. A pesar de ello, se cree que la continuación en el desarrollo de la misma es una tarea interesante, que con las horas de trabajo necesarias, se podría convertir en una herramienta interesante frente a otras alternativas que existen actualmente.

      \paragraph{}
      Para entender lo que se ha pretendido conseguir con esta implementación, a continuación se ejemplifica un caso de una implementación similar que se ha llevado a cabo utilizando otras tecnologías en los últimos años. Dicha implementación (e ideas) se conoce como \emph{GraphX}, una biblioteca para el tratamiento de grafos masivos de manera distribuida presentada en \emph{2013} en el trabajo \emph{Graphx: A resilient distributed graph system on spark} \cite{xin2013graphx} desarrollado por \emph{Xin y otros}. Esta implementación se desarrolló inicialmente como un conjunto de utilidades y procedimientos sencillos para facilitar la representación de grafos y el desarrollo de algoritmos sobre estos.

      \paragraph{}
      \emph{GraphX} se ha desarrollado utilizando como base la plataforma de computación distribuida \emph{Spark} publicada en el trabajo \emph{Spark: Cluster computing with working sets} \cite{zaharia2010spark} de \emph{Zaharia y otros}. Esta plataforma se basa en el tratamiento de grandes conjuntos de datos mediante el procesamiento de los mismos en lotes, lo cual proporciona grandes mejoras respecto de otras soluciones como \emph{Hadoop}, presentado en el documento \emph{The hadoop distributed file system} \cite{shvachko2010hadoop} desarrollado por \emph{Shvachko y otros}.

      \paragraph{}
      Dichas plataformas tratan de abstraer la idea de procesamiento distribuido y hacerlo lo más transparente posible para el usuario, sin olvidar en ningún momento que los conjuntos de datos utilizados sobre los que se trabaja no se encuentran contenidos totalmente en una única máquina, lo cual implica distintas restricciones respecto de las estrategias de programación clásicas, como los procesos de acceso y escritura al sistema de almacenamiento. Sin embargo, en estos casos también existen soluciones que abstraen dichas tareas de almacenamiento distribuidas, algunas de ellas son \emph{Google File System} \cite{ghemawat2003google} o \emph{Hadoop File System} \cite{shvachko2010hadoop}.

      \paragraph{}
      Lo característico de \emph{GraphX} es que se ha desarrollado como una biblioteca para el tratamiento de grafos utilizando \emph{Spark} como plataforma base, pero tratando de mantener la independencia entre las mismas. Es decir, \emph{GraphX} ha sido desarrollado utilizando las utilidades que proporciona \emph{Spark}, pero en \emph{Spark} no existe ninguna dependencia hacia \emph{GraphX}. Por tanto, esto se puede entender como un sistema basado en capas, donde \emph{Spark} representa la capa inferior y \emph{GraphX} se coloca en una capa inmediatamente superior.

      \paragraph{}
      En este trabajo, se ha tratado de realizar una implementación semejante (a un nivel muy básico por las restricciones temporales en que se ha desarrollado), tratando de proporcionar igualmente una capa de abstracción que modeliza el concepto de grafo sobre otra plataforma de computación de alto rendimiento. En este caso se ha decidido utilizar la biblioteca de cálculo intensivo \emph{TensorFlow}, la cual se hizo pública en \emph{2016} en el trabajo \emph{Tensorflow: Large-scale machine learning on heterogeneous distributed systems} \cite{abadi2016tensorflow}, desarrollada por el departamente de investigación de \emph{Google} y actualmente publicada con licenciatura de código abierto.

      \paragraph{}
      \emph{TensorFlow} proporciona un \emph{framework} para la implementación de algoritmos cuyo funcionamiento se basa en el cálculo de operaciones sobre \emph{tensores} (una generalización del concepto de matriz). Se ha preferido posponer la descripción de esta plataforma hasta la sección \ref{sec:tensorflow}, ya que a continuación se describirá el conjunto de tecnologías utilizadas para la implementación realizada.

      \paragraph{}
      La motivación por la cual se ha decidido realizar la implementación de una biblioteca que simplifique el desarrollo de algoritmos sobre grafos utilizando una plataforma de cálculo matemático intensivo se debe a lo siguiente: una gran cantidad de analíticas sobre grafos pueden ser calculadas entendiendo dicho grafo como una estructura de datos matricial, a través de la \emph{matriz de adyacencia} (sección \ref{sec:adjacency_matrix}), u otras representaciones como la \emph{matriz laplaciana} (sección \ref{sec:laplacian_matrix}). Este marco conceptual conlleva el desarrollo de algoritmos con un alto grado de paralelización, que tal y como se verá posteriormente, satisface la plataforma \emph{TensorFlow}.

      \paragraph{}
      Sobre este contexto también se pueden desarrollar algoritmos de optimización, tales como planificación de rutas, recorridos de vehículos o cubrimiento de zonas mediante la modelización del grafo de manera conveniente y la utilización de distintas estrategias de programación lineal, lo cual se ha estudiado ampliamente en la literatura.

      \paragraph{}
      En este caso, la implementación realizada para este trabajo se encuentra en las primeras fases de su desarrollo. Por tanto, únicamente se ha basado en el conjunto de utilidades necesarias para llevar a cabo la implementación del \emph{Algoritmo PageRank} (capítulo \ref{chap:pagerank}), junto con un \emph{Sparsifier} (sección \ref{sec:sparsifiers}) que reduce el número de aristas del grafo para después comparar resultados a nivel de precisión.

      \paragraph{}
      Sin embargo, tal y como se indicará posteriormente, se pretende seguir trabajando en dicha biblioteca de grafos para ampliar su funcionalidad y desarrollar otras implementaciones que permitan obtener otras analíticas sobre el grafo.

      \paragraph{}
      El resto del capítulo se organiza de la siguiente manera: en la sección \ref{sec:implementation} se realiza una descripción acerca de las decisiones tomadas en la implementación de la biblioteca, indicando las tecnologías utilizadas (sección \ref{sec:used_technologies}), los servicios en que se ha apoyado el desarrollo (sección \ref{sec:used_services}) y el diseño que ha seguido dicha implementación (sección \ref{sec:implementation_design}). Posteriormente, se indican distintas vías a través de las cuales sería interesante seguir trabajando en la implementación en la sección \ref{sec:future_work} y, por último, se realiza una breve conclusión acerca del trabajo realizado en la sección \ref{sec:implementation_conclusions}

    \section{Implementación}
    \label{sec:implementation}

      \paragraph{}
      En esta sección se exponen distintas explicaciones acerca de la implementación realizada, la cual pretende comportarse como una biblioteca de utilidades que permita modelizar de manera sencilla el concepto de grafo, utilizando como base una plataforma de cálculo matemático intensivo. Dicha implementación se ha realizado prestando especial atención en la reducción de dependencias hacia el exterior, de tal manera que sea posible la distribución de la misma como un paquete compacto que integrar en otros sistemas de mayor envergadura. Por tanto, se ha utilizado el sistema de distribución de paquetes del lenguaje \emph{Python}, el cual simplifica dicha tarea. Sin embargo, antes de comenzar a describir distintos detalles acerca de las decisiones tomadas, a continuación se describen brevemente las tecnologías utilizadas, ya que son influyentes respecto de dichas decisiones.

      \subsection{Tecnologías Utilizadas}
      \label{sec:used_technologies}

        \paragraph{}
        La implementación se ha desarrollado utilizando el lenguaje \emph{Python}, junto con distintas bibliotecas que extienden su comportamiento y le otorgan una mayor funcionalidad. Además, se ha utilizado el sistema de control de versiones \emph{git}, que permite trabajar de manera ordenada en distintas partes del trabajo.

        \subsubsection{Python}
        \label{sec:python}

          \paragraph{}
          \emph{Python} se define como un lenguaje de propósito general sobre un paradigma imperativo pero con utilidades de programación funcional como funciones lambda o tratamiento de funciones como un valor más. Es orientado a objetos y no tipado, lo cual simplifica el trabajo a la hora de escribir código, pero limita la seguridad del mismo ante entradas incorrectas. Python es un lenguaje interpretado en tiempo de ejecución, por lo cual no es requiere de la utilización de un compilador. Internamente existen implementaciones de \emph{Python} en distintos lenguajes de programación compilados, sin embargo, en este caso el desarrollo se ha realizado sobre \emph{cpython}, que se basa en un intérprete desarrollado en el lenguaje \emph{C}.

          \paragraph{}
          En la revisión del estándar \emph{PEP-484} \url{https://www.python.org/dev/peps/pep-0484/} se añade un sistema de marcado de tipos para el lenguaje, el cual aún no está operativo en tiempo de ejecución (a través del intérprete), pero permite la comprobación estática del mismo. En la implementación realizado se ha utilizado este este sistema de comprobación de tipos, el cual se introdujo en \emph{Python 3.5}, por tanto está ha sido la versión escogida como mínima.

          \paragraph{}
          \emph{Python} implementa como estructura de datos indexadas \emph{listas enlazadas}, lo cual proporciona una gran versatilidad ya que permite tanto agregar como eliminar nuevos elementos de manera eficiente ($O(1)$). Sin embargo, el tiempo de acceso se ve altamente penalizado por dicha condición ($O(n)$). Por tanto, se han utilizado bibliotecas externas que mejoran dichos costes.

        \subsubsection{NumPy y Pandas}
        \label{sec:numpy_pandas}

          \paragraph{}
          Para solventar la problemática de la eficiencia en tiempo de acceso de las estructuras de datos indexadas en \emph{Python} existe una biblioteca que permite implementa dichas estructuras de datos de manera contigua, lo cual elimina el problema. Dicha biblioteca se conoce como \emph{NumPy} \cite{walt2011numpy}, la cual proporciona además un gran conjunto de operaciones matemáticas sobre esta estructura de datos. De esta manera se permite desarrollar algoritmos con una elevada carga matemática de manera muy eficiente y a la vez sencilla, al estilo de lenguajes como \emph{MatLab} o \emph{R}.

          \paragraph{}
          Para algunas partes del código implementado, se ha utilizado la biblioteca \emph{Pandas} \cite{mckinney2010data}. Esta biblioteca consiste en una extensión respecto de \emph{NumPy}, que permite ver la estructura de datos desde una perspectiva de conjunto de datos en lugar de estructura matemática, lo cual simplifica el trabajo para tareas como la lectura y escritura de conjuntos de datos en el espacio de almacenamiento.

        \subsubsection{TensorFlow}
        \label{sec:tensorflow}

          \paragraph{}
          \emph{TensorFlow} \cite{abadi2016tensorflow} es una biblioteca inicialmente desarrollada por \emph{Google}. La idea principal por la cual fue desarrollada es la simplificación de tareas para la implementación de algoritmos para aprendizaje automático y \emph{deep learning}, cuya carga computacional es altamente paralelizable y puede ser entendida como operaciones entre \emph{tensores}. Esta biblioteca proporciona interfaces para ser utilizada junto con los lenguajes \emph{Python}, \emph{C++}, \emph{Java} o \emph{Go}.

          \paragraph{}
          Un \emph{tensor} es una generalización del concepto de matriz, permitiendo que estas sean de cualquier número de dimensiones. Esto puede entenderse fácilmente diciendo que un tensor de grado 0 se corresponde con el concepto de escalar, uno de grado 1 puede ser visto como un vector, el grado 2 se corresponde con las matrices y así sucesivamente. A partir del conjunto de operaciones aplicable a dichos \emph{tensores} se crea un flujo de estas, que puede ser visto como un grafo de dependencias entre ellas. Por tanto, el la biblioteca se decidió llamar \emph{TensorFlow} (\emph{flujo de tensores}).

          \paragraph{}
          \emph{TensorFlow} proporciona por tanto, estructuras de datos para representar el concepto de \emph{tensor} así como un conjunto de operaciones básicas que aplicar entre ellos para así obtener nuevos \emph{tensores} con los resultados. La biblioteca se puede dividir en dos bloques bien diferenciados: el primero de ellos de bajo nivel y que se corresponde con operaciones matemáticas sencillas (suma, multiplicación, división, exponenciación, máximos, etc.) y otro segundo bloque construido a partir del primero al cual pertenece todo el conjunto de operaciones de alto nivel que permiten la implementación de algoritmos de aprendizaje automático fácilmente, tales como implementaciones del \emph{gradiente descendente} u otros conceptos semejantes. Independientemente de estos dos bloques, la biblioteca también proporciona otra serie de utilidades como capacidad para definir variables y constantes, guardar y recuperar modelos matemáticos y otras facilidades.

          \paragraph{}
          En esta implementación se han utilizado métodos relacionados con operaciones algebraicas, ya que se ha utilizado dicha librería centrándose únicamente en su perspectiva matemática y obviando por completo el bloque de aprendizaje automático, ya que no se corresponde con el tema de este trabajo ni con las ideas descritas en anteriores capítulos.

          \paragraph{}
          \emph{TensorFlow} proporciona distintas ventajas a nivel de rendimiento respecto de otras alternativas de computación numérica intensiva puesto que está construida como una interfaz de alto nivel que abstrae al usuario del sistema donde está siendo ejecutado el cómputo. Con esto nos estamos refiriendo a que la implementación realizada es ejecutable tanto en una \emph{CPU} clásica de un ordenador como en acelerados de computación externos tales como \emph{GPU's} \emph{CUDA} o las \emph{Tensor Processor Units} \cite{jouppi2017datacenter} diseñadas expecíficamente por \emph{Google} para ser utilizadas junto con esta librería.

          \paragraph{}
          Debido a estas ideas, el estilo de programación se divide en dos fases, una de definicion de tensores, operaciones y relaciones de dependencia entre ellas en la cual se construye el denominado flujo o grafo de operaciones, y una segunda fase correspondiente a la llamada para la ejecución de dichas operaciones. Esta idea cobra sentido debido tanto al elevado tamaño de los datos de entrada para las operaciones, así como el modelo de computación en aceleradores externos, que requiere de tareas de transferencia entre el sistema y dichas unidades. Al dividirse la definición de la ejecución, esto se puede optimizar de manera eficiente sobre lenguajes interpretados como \emph{Python}, que sino tendrían una alta penalización en temporal derivada del coste de transferencias.

        \subsubsection{pytest}
        \label{sec:tensorflow}

          \paragraph{}
          \emph{pytest} es una herramienta de generación de casos de prueba para el lenguaje \emph{Python}. Para ello se basa en la ejecución de distintas funciones definidas por el usuario, las cuales contienen un conjunto de asertos (palabra reservada \emph{assert} en \emph{Python}), los cuales deben superar satisfactoriamente para finalizar el test de manera satisfactoria. Para los casos de prueba, estos se han utilizado junto con los que proporciona \emph{NumPy} para comprobar la semejanza entre dichas estructuras de datos de manera eficiente.

        \subsubsection{sphinx}
        \label{sec:sphinx}

          \paragraph{}
          \emph{sphinx} consiste en una herramienta que permite extraer la documentación incorporada en el código a otras fuentes para facilitar su visualización, tales como un sitio web, documentos de \emph{PDF} y otras alternativas. Para ello se basa en la documentación interna del código. En el caso de \emph{Python}, esta documentación se denomina \emph{docstring} y permite añadir una breve explicación acerca de los bloques de código, así como las entradas y salidas de los métodos. En este caso se ha utilizado el estilo de documentación definido por \emph{Google} para tratar de asemejarse lo máximo posible a la documentación seguida por \emph{TensorFlow}

        \subsubsection{git}
        \label{sec:git}

          \paragraph{}
          \emph{git} se corresponde con un \emph{sistema de control de versiones} que permite el trabajo de manera colaborativa entre distintos usuarios, a través de distintas ramas de desarrollo, que después se combinan para llegar a un estado de desarrollo final. Además de esto, permite almacenar un historial de todos los cambios realizados, lo cual es de gran ayuda en puntos en los cuales es necesario entender la razón de cambios pasados así, como retroceder hasta un estado anterior si fuera conveniente. Otra de las ventajas de esta herramienta es la capacidad de sincronización entre distintos sistemas, así como la posibilidad de mantener una copia del repositorio en un servidor externo, lo cual previene de problemas relacionados con fallos del sistema local.

      \subsection{Servicios Utilizados}
      \label{sec:used_services}

        \paragraph{}
        Para el desarrollo de este trabajo se han utilizado distintos servicios de empresas externas que simplifican algunas de las tareas inherentes en la utilización de las tecnologías descritas anteriormente. Dichas tareas se pueden realizar de manera independiente a estos servicios, sin embargo, resulta interesante su utilización por el grado de mayor confiabilidad que proporcionan respecto del desarrollo completo utilizando únicamente una máquina local, que puede sufrir fallos perdiendo todo el trabajo. Todos los servicios externos utilizados han sido utilizados en su versión gratuita, que ofrece las funcionalidades suficientes para el correcto desarrollo del proyecto.

        \paragraph{GitHub}
        Proporciona un servidor \emph{git} externo sobre el cual se realizan copias del trabajo local de tal manera que se aumenta el grado de seguridad desde el punto de la aparición de pérdidas inesperadas. Además del servicio de control de versiones basado en \emph{git}, se proporcionan otra serie de funcionalidades como gestión de incidencias mediante el concepto de \emph{issue}, la fusión de ramas de manera segura a partir de \emph{pull request} o una gestión de proyectos mediante un tablero \emph{kanban} configurable según las necesidades especificas del mismo.

        \paragraph{Read the Docs}
        Ofrece un servicio de generación y publicación de sitios web basados en la documentación de proyectos sofware basada en \emph{sphinx} de manera sencilla, ya que únicamente requiere de la dirección de acceso al repositorio de trabajo \emph{git} junto con las opciones del entorno de desarrollo necesarias para utilizar \emph{sphinx} sobre dicho repositorio. Se cree conveniente la publicación de la documentación de código para que sea accesible fácilmente, por lo cual se decidió utilizar dicho servicio.

        \paragraph{Travis CI}
        Consiste en un entorno de realización de casos de prueba previamente configurados por el usuario a través de bibliotecas como \emph{pytest} u otras similares. El servicio ejecuta despliega un entorno de pruebas previamente configurado por el usuario para después ejecutar el conjunto de casos de prueba indicados. Este servicio indica tras cada cambio en el repositorio si las pruebas han sido superadas, o por contra han ocurrido fallos, indicando el punto de dichos fallos.

        \paragraph{Codecov}
        Es una utilidad interesante que determina el grado de cobertura de los casos de prueba sobre la implementación realizada, lo cual proporciona una buena estimación acerca del conjunto de lineas de código que están siendo verificadas por los tests. Sin embargo, a pesar de que esta utilidad proporciona una buena estimación, su resultado tan solo puede ser orientativo, ya que únicamente tiene en cuenta si existe un caso de prueba que analice una determinada línea de código. Esto no tiene en cuenta casos específicos como divisiónes entre cero, verificaciones de valores fuera de rango o casos similares.

        \paragraph{WakaTime}
        Es una utilidad destinada al seguimiento de horas de trabajo, que indica tanto la proporción de tiempo destinada a cada lenguaje como a cada proyecto. Esta se basa en la recolección de información sobre los editores de código. Por tanto, recoge el tiempo de trabajo destinado a dicha tarea. Esto es una métrica interesante, pero no tiene en cuenta la cantidad de tiempo destinado a tareas de aprendizaje e investigación leyendo artículos o leyendo documentación de bibliotecas utilizadas durante el desarrollo.

        \subsubsection{Astah}
        \label{sec:astah}

          \paragraph{}
          \emph{Astah} es un software para la generación de distintos diagramas relacionados con el diseño de software. A través de una interfaz de usuario permite el modelado de sistemas siguiendo el estándar \emph{UML}. Este software ha sido utilizado para la realización de diagramas de clases y componentes. Lo cual permite entender las relaciones entre las distintas clases implementadas de una manera rápida y simple.

        \paragraph{}
        Todos estos servicios se han utilizados de manera relacionada entre si, en algunos casos de manera directa, mientras que en otros de manera indirecta. El punto de conexión entre todos ellos es el sistema de control de versiones \emph{git}, que estos utilizan para relacionar la tarea que realizan con un determinado proyecto. El sistema de documentación de \emph{Read the Docs} realiza una nueva ejecución tras cada cambio en el repositorio \emph{git} de \emph{GitHub}. De la misma manera \emph{Travis CI} y \emph{Codecov} realizan sus tareas correspondientes. Por último, \emph{WakaTime} realiza un seguimiento constante sobre el tiempo dedicado al repositorio. De esta manera se obtiene un entorno de trabajo que permite un desarrollo ágil permitiendo al desarrollador centrarse únicamente en las tareas de investigación y desarrollo, para posteriormente validar los resultados de ejecución de estos servicios, lo cual reduce tiempo y costes.

      \subsection{Diseño de la implementación}
      \label{sec:implementation_design}

        \paragraph{}
        Una vez descritas las distintas dependencias externas utilizadas para la implementación realizada, ya se puede comenzar a describir la misma. Para ello, se han realizado distintos diagramas, entre los que se encuentran un \emph{Diagrama de Componentes}, \emph{Diagramas de Clases} que tratan de facilitar el entendimiento del código desde distintos puntos de vista y, por último, se han añadidos \emph{Diagramas de Operaciones} generados por \emph{TensorFlow}, que permiten entender en mayor medida el comportamiento de los algoritmos de una manera visual.

        \paragraph{}
        Sin embargo, antes de comenzar con la descripción a nivel de diseño de la implementación realiza, es interesante realizar una breve indicación acerca del sistema de distribución de paquetes utilizado por el lenguaje \emph{Python}, ya que se ha utilizado dicha estrategia para encapsular la implementación y facilitar su utilización en otros sistemas.

        \paragraph{}
        \emph{Python} proporciona un estándar sencillo de distribución de paquetes basados en módulos. Para utilizar dicha funcionalidad, es necesario incluir un fichero \texttt{/setup.py} en el directorio de nivel superior respecto del referido al módulo que se pretende empaquetar. En dicho fichero se anotan distintos valores acerca del módulo, entre las que se encuentran su localización, las dependencias sobre otros módulos externos, la versión mínima de \emph{Python} para su utilización, u otros meta-datos como el nombre del autor, el nombre del módulo, la versión o la licencia sobre la que se distribuye.

        \paragraph{}
        La implementación realizada se ha decido encapsular en un módulo al que se ha denominado \texttt{tf\_G}. Dicho módulo se ha nombrado de esta manera por la tecnología utilizada para su implementación, junto con su funcionalidad relacionada con \emph{Grafos}. Por tanto, se ha utilizado el acrónimo \texttt{tf}, utilizado como estándar al importar la biblioteca \emph{tensorflow} (\texttt{import tensorflow as tf}), al cual se le ha añadido la \texttt{G} para finalmente formar el nombre \texttt{tf\_G}.

        \paragraph{}
        En cuanto a la implementación, esta se ha incluido en el directorio \texttt{/src/}, que a su vez contiene otro directorio con el nombre del módulo (\texttt{tf\_G}). Además de la implementación en si, que se describirá a continuación, también se han incluido en el directorio \texttt{/tests/} distintos casos de prueba (a nivel muy básico y casi a modo demostrativo) para la comprobación acerca del correcto funcionamiento de la implementación. En el directorio \texttt{/examples/} se han incluido distintos scripts que permiten apreciar el funcionamiento de la implementación realizada de una manera práctica.

        \subsubsection{Diagrama de Componentes}
        \label{sec:component_diagram}

\begin{figure}[h!]
            \centering
            \includegraphics[width=\textwidth,height=0.9\textheight,keepaspectratio]{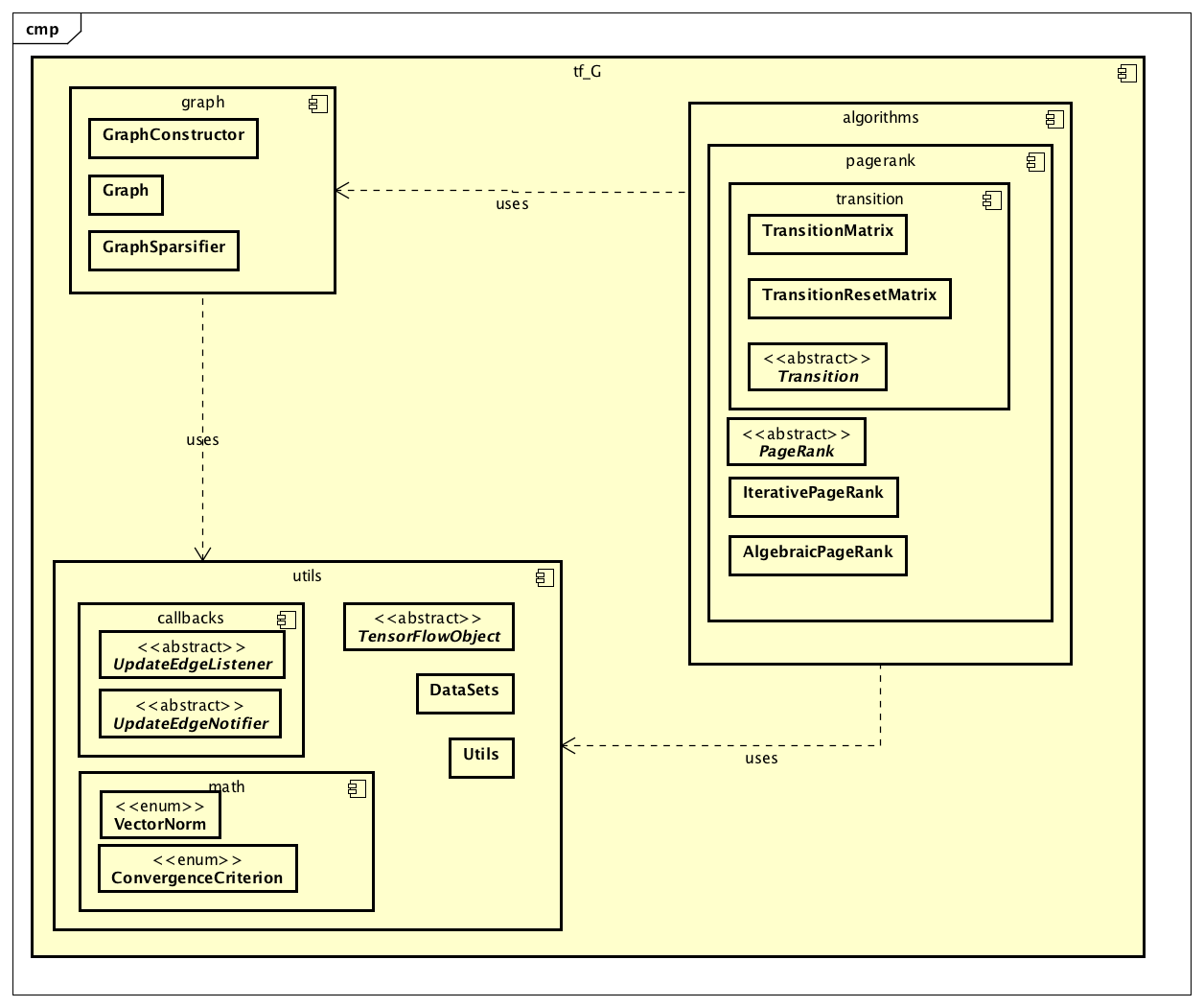}
            \caption{Diagrama de componentes referido al módulo \texttt{tf\_G}.}
            \label{img:component_diagram}
          \end{figure}

          \paragraph{}
          En la figura \ref{img:component_diagram} se muestra el diagrama de componentes seguido por el módulo \texttt{tf\_G}. Tal y como se puede apreciar, se ha decidido dividir la implementación en 3 paquetes denominados: \texttt{graph} (para la implementación de los grafos), \texttt{algorithms} (donde se encontrarán las implementaciones de los algoritmos para grafos) y \texttt{utils} (donde se contienen diferentes clases de carácter general necesarias para que el resto de paquetes funcionen de manera apropiada).

          \paragraph{}
          En cuanto al paquete \texttt{utils}, este contiene dos sub-paquetes, el referido a clases que ofrecen la funcionalidad de \texttt{callbacks} (esto se describirá a continuación junto con su respectivo diagrama de clases). Así como \texttt{math}, que contiene clases referidas a utilidades matemáticas (en este caso normas vectoriales y criterios de convergencia).

          \paragraph{}
          El paquete \texttt{algorithms} contiene las distintas implementaciones algorítmicas implementadas destinando un sub-paquete para cada una de ellas. En este caso, únicamente se ha implementado el algoritmo \emph{PageRank}, por lo que solo contiene un paquete, sin embargo, esto se espera extender en el futuro. El paquete \texttt{pagerank} contiene la implementación de una versión algebraica y otra iterativa, junto con sus respectivas matrices de transición contenidas en el sub-paquete \texttt{transition}. Posteriormente se describirá más en detalle la relación existente entre ellas en el respectivo diagrama de clases.

        \subsubsection{Diagrama de Clases}
        \label{sec:class_diagram}

          \paragraph{}
          Una vez descrito el diagrama de componentes de la implementación, lo siguiente es hablar del \emph{Diagrama de Clases} de la misma. Para ello, en primer lugar hay que tener en cuenta diversos factores, entre los que se encuentran las peculiaridades del lenguaje \emph{Python} respecto de las ideas de \emph{Orientación a Objetos}. En \emph{Python} no existen \emph{Clases Abstractas} ni \emph{Interfaces}. Sin embargo, este lenguaje permite emular su comportamiento mediante el concepto de \emph{Herencia Múltiple}, que permite que una clase tenga más de una clase padre.

          \paragraph{}
          A partir de dicho concepto se pueden describir los mismos conceptos que en otros lenguajes como \emph{Java} se harían utilizando las \emph{Clases Abstractas} e \emph{Interfaces}. Para ello, además de la \emph{Herencia Múltiple}, se ha utilizado la excepción \texttt{NotImplementedError} para notificar que no las clases descendientes no han implementado la funcionalidad requerida. Alternativamente, se podría haber utilizado la alternativa propuesta por el paquete \texttt{abc}, pero se ha preferido la otra opción.

          \paragraph{}
          Una vez descritas dichas caracterizaciones, se discutirá la implementación realizada. En este caso, tan solo se realizarán indicaciones desde el punto de vista de la estructura de clases, ya que si se desea conocer el comportamiento de cada método concreto se puede hacer uso de la documentación interna de cada clase visualizando el código fuente, o de manera online a través de \url{http://tf-g.readthedocs.io/en/latest/}.

\begin{figure}[h!]
            \centering
            \includegraphics[width=\textwidth,height=0.9\textheight,keepaspectratio]{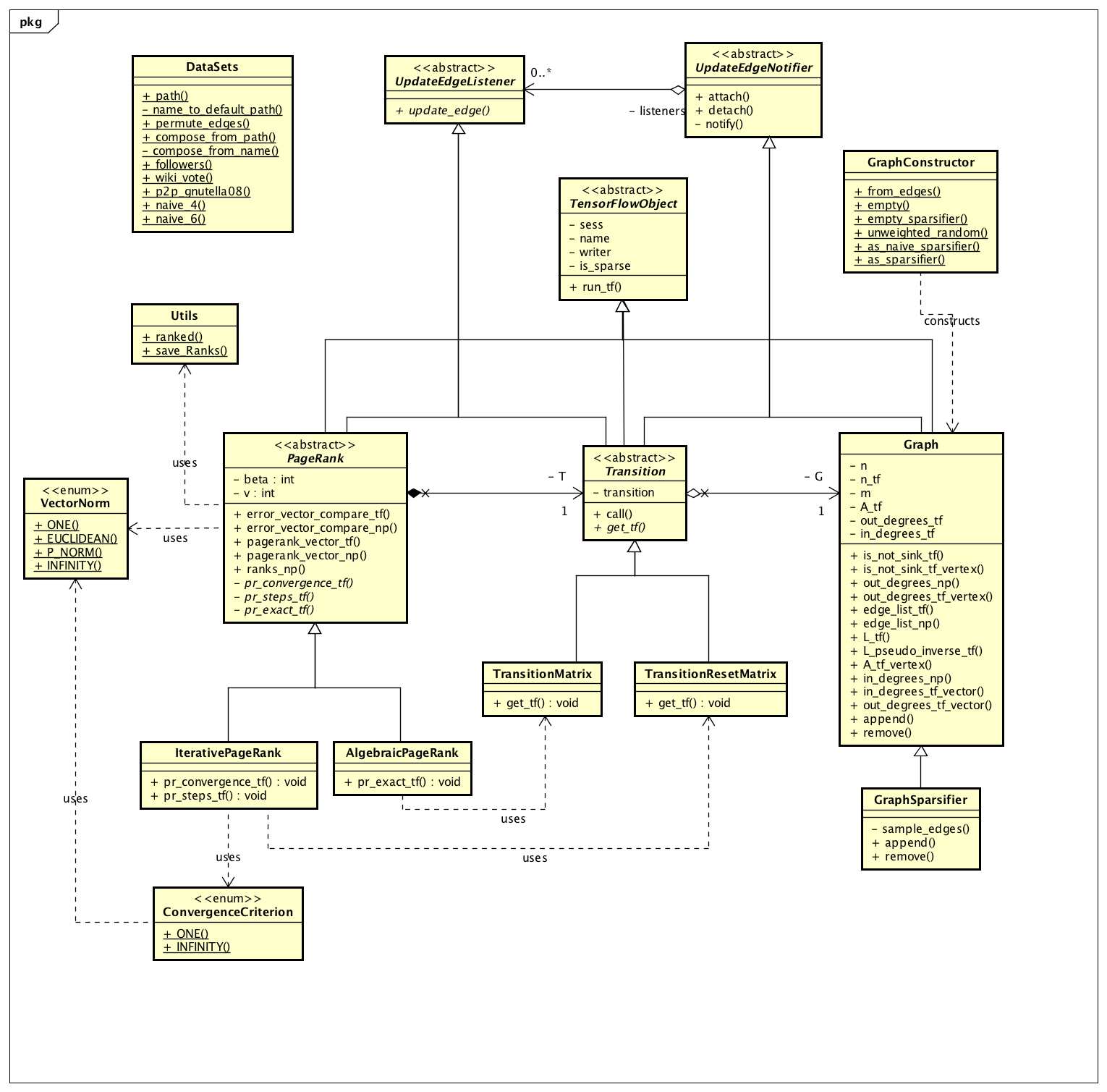}
            \caption{Diagrama de clases completo referido al módulo \texttt{tf\_G}.}
            \label{img:class_diagram}
          \end{figure}

          \paragraph{}
          El diagrama de clases completo se muestra en la figura \ref{img:class_diagram}. Tal y como se puede apreciar, a través de dicho dicho diagrama es complicado comprender el funcionamiento de la implementación, debido a las relaciones que se solapan unas a otras derivado de la herencia múltiple. Por tanto, se ha creído una solución más acertada dividir el diagrama en partes, para realizar la descripción desde distintos puntos de vista.

          \paragraph{}
          Sin embargo, en la figura \ref{img:class_diagram} se muestra una clase aislada. Esta clase se denomina \emph{DataSets} y no se relaciona de manera directa con el resto, sino que se suministra para facilitar las tareas de obtención de conjuntos de datos que representan las aristas de un grafo. Para ello proporciona funciones que permiten la generación de estructuras de datos que almacenan dichos aristas. Además, el módulo contiene distintos conjuntos de datos para poder realizar pruebas sobre la implementación. Por tanto, esta clase suministra los métodos para poder acceder a ellos.

          \begin{figure}[h!]
            \centering
            \includegraphics[width=\textwidth,height=0.9\textheight,keepaspectratio]{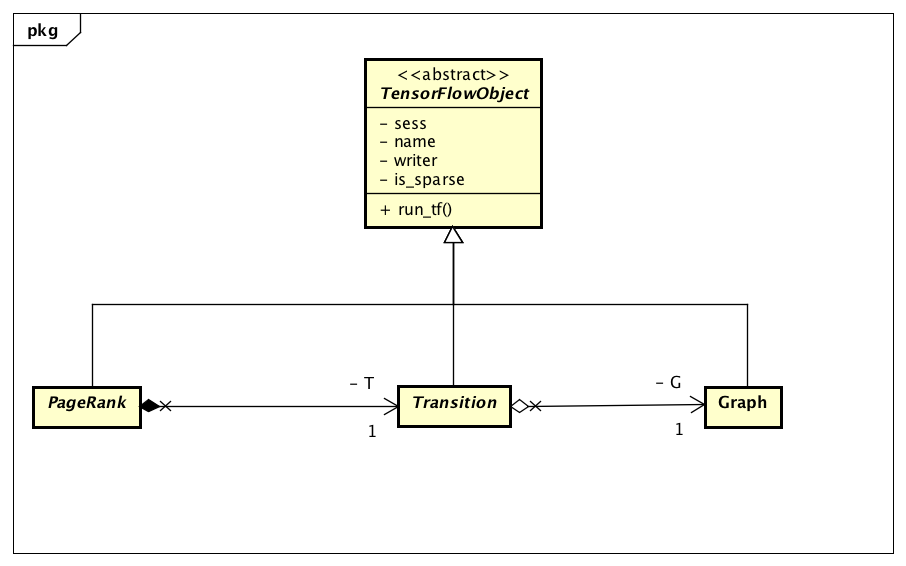}
            \caption{Diagrama de clases referido a las relaciones con la clase abstracta \texttt{TensorFlowObject} del módulo \texttt{tf\_G}.}
            \label{img:tensorflowobject_class_diagram}
          \end{figure}

          \paragraph{}
          En la figura \ref{img:tensorflowobject_class_diagram} se muestra una sub-conjunto de clases relacionadas entre sí a través de la herencia de la clase \texttt{TensorFlowObject}. Las relacionees entre ellas se discutirán posteriormente. Sin embargo, a continuación se expone la razón por la cual se ha implementado dicha clase abstracta. Puesto que la implementación se ha apoyado fuertemente en el uso de la biblioteca \emph{TensorFlow}, muchas de estas clases necesitaban atributos y operaciones comunes, como la necesidad de obtener el resultado de una determinada operación en \emph{TensorFlow} a través del método \texttt{run\_tf} o la de identificar a una variable por su nombre. Por tanto, a partir de \texttt{TensorFlowObject} se suministra dicha funcionalidad.

          \begin{figure}[h!]
            \centering
            \includegraphics[width=\textwidth,height=0.9\textheight,keepaspectratio]{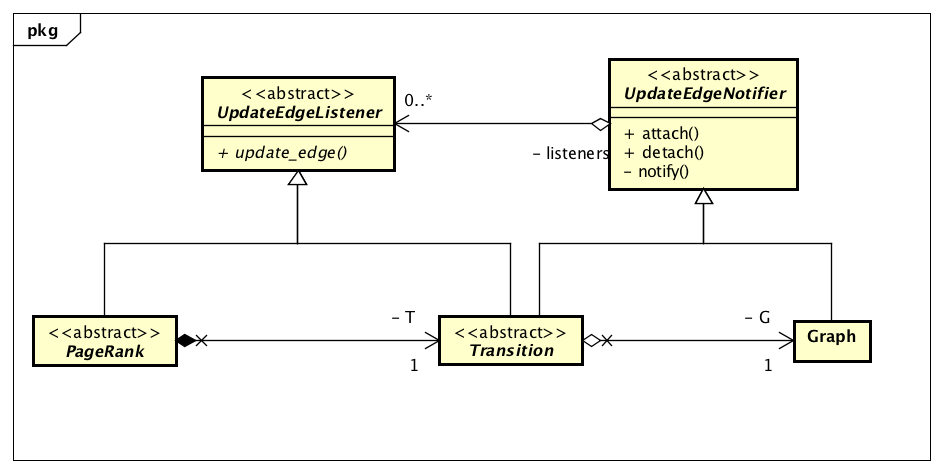}
            \caption{Diagrama de clases referido a las relaciones con las clases abstractas \texttt{UpdateEdgeListener} y \texttt{UpdateEdgeNotifier} del módulo \texttt{tf\_G}.}
            \label{img:update_edge_diagram}
          \end{figure}

          \paragraph{}
          El siguiente diagrama de clases incluido se muestra en la figura \ref{img:update_edge_diagram}. Dicho diagrama se corresponde con el conjunto de clases relacionadas con \texttt{UpdateEdgeListener} y \texttt{UpdateEdgeNotifier}. Estas clases son las contenidas en el paquete \texttt{utils.callbacks} y la funcionalidad que proporcionan es la de notificar y ser notificadas por otras clases, de algún cambio en el conjunto de aristas del grafo con el cual interaccionan.

          \paragraph{}
          Dicho comportamiento es muy similar al del patrón \emph{Observador}, sin embargo, no se puede denominar de dicha manera puesto que en el caso de las clases que descienden de \texttt{UpdateEdgeListener}, estas si que conocen al objeto observado. Por lo cual, en lugar de utilizar la denominación \emph{observer-observed} se ha preferido \emph{listener-notifier}. Para las clases descendientes se proporcionan los clásicos métodos \texttt{attach}, \texttt{detach} y \texttt{notify} en el caso de \texttt{UpdateEdgeNotifier} y \texttt{update\_edge} para \texttt{UpdateEdgeListener}.

          \paragraph{}
          Los conjuntos de clases que descienden de \texttt{UpdateEdgeListener} y \texttt{UpdateEdgeNotifier} tienen por tanto la capacidad de implementar sus algoritmos de manera dinámica. En este caso, puesto que tan solo se ha implementado el algoritmo \emph{PageRank}, esta es el único algoritmo que posee capacidad de dinamismo. A pesar de ello, el usuario del módulo puede implementar clases que desciendan de \texttt{UpdateEdgeListener} para así ser notificadas de cambios en el grafo al cual se refieran.

          \paragraph{}
          El resto de diagramas de clases que se han incluido se corresponden con implementaciones concretas que proporcionan funcionalidad directa al usuario. Estas se refieren a la representación de grafos sobre \emph{TensorFlow} (lo cual realiza por el paquete \texttt{graph}) y el cálculo del \emph{PageRank} (lo cual se lleva a cabo en el paquete \texttt{algorithms/pagerank}).

          \begin{figure}[h!]
            \centering
            \includegraphics[width=\textwidth,height=0.9\textheight,keepaspectratio]{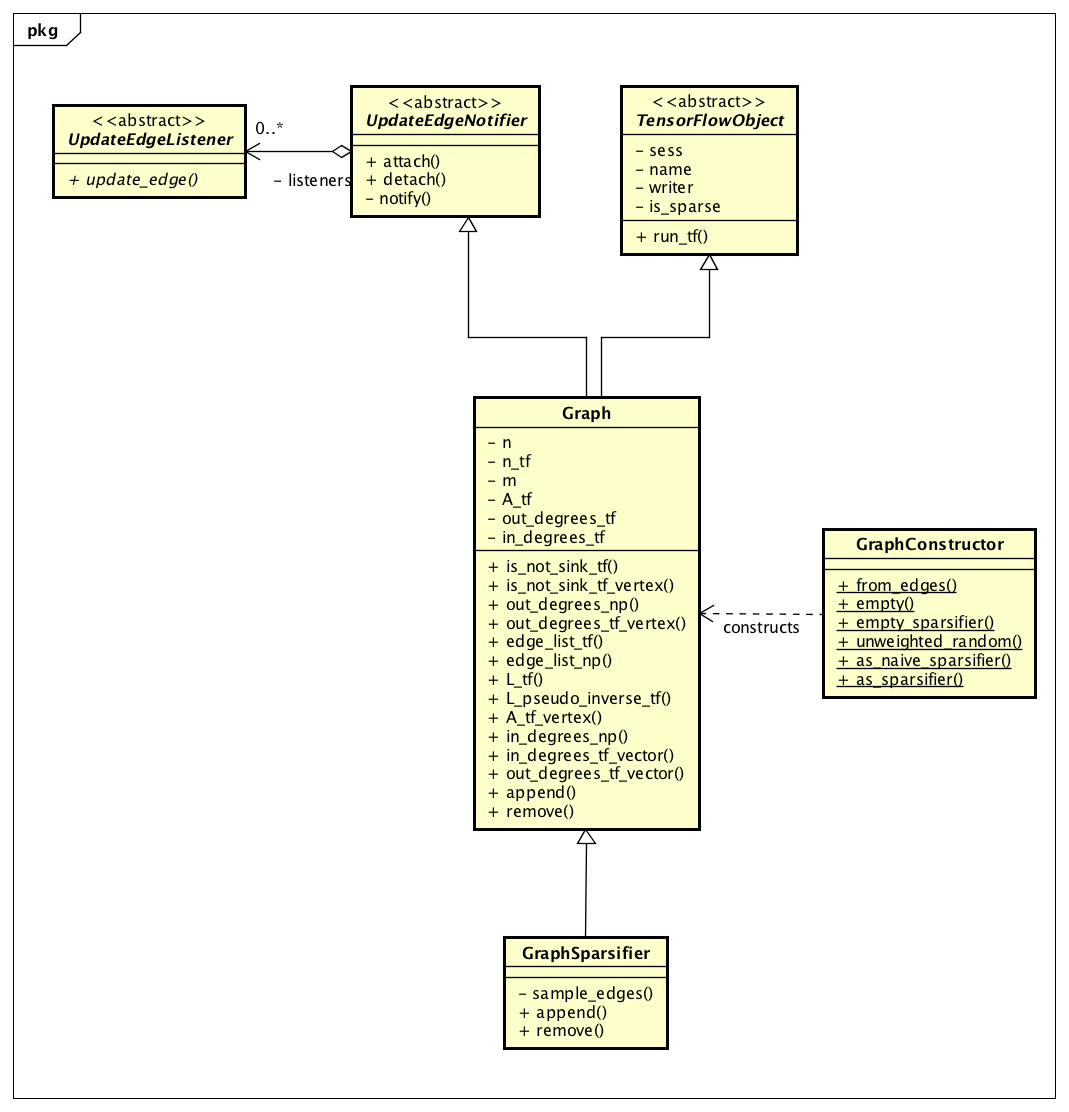}
            \caption{Diagrama de clases referido a las relaciones con la clase \texttt{Graph} del módulo \texttt{tf\_G}.}
            \label{img:graph_class_diagram}
          \end{figure}

          \paragraph{}
          La figura \ref{img:graph_class_diagram} muestra el conjunto de clases referidas a la representación de un grafo utilizando como base la biblioteca \emph{TensorFlow}. Dicha implementación se constituye únicamente por 3 clases: \texttt{Graph} (que representa un grafo), \texttt{GraphSparsifier} (que representa un \emph{pseudo-Sparsifier} de un grafo dado) y \texttt{GraphConstructor} (que proporciona distintas utilidades que permiten construir un grafo de manera simple).

          \paragraph{}
          Tal y como se indicó anteriormente, para entender la funcionalidad que provee cada método de la clase \texttt{Graph}, es más apropiado seguir la documentación interna de la clases o su versión publicada online. Debido a las restricciones de tiempo en la realización de la implementación, tan solo se han implementado aquellas funcionalidades necesarias para el desarrollo del algoritmo \emph{PageRank}.

          \paragraph{}
          En cuanto a la clase \texttt{GraphSparsifier}, esta se ha indicado anteriormente como \emph{pseudo-Sparsifier} puesto que no se ha realizado ninguna demostración acerca de la precisión de dicha implementación ni algoritmo. Por motivos derivados de las restricciones temporales no ha sido posible realizar una implementación apropiada de entre las discutidas en la sección \ref{sec:sparsifiers}. Sin embargo, se han seguido dichas ideas para la implementación del mismo, aunque tal y como se indica, \emph{este no ofrece ninguna garantía desde el punto de vista analítico}.

          \paragraph{}
          Por último, la clase \texttt{GraphConstructor} provee una serie de métodos que permiten la construcción de un grafo en sobre la implementación de la biblioteca de manera sencilla a través de una interfaz intuitiva. En el directorio \texttt{/examples/} se pueden visualizar distintos ejemplos que hacen uso de dicha clase.

          \begin{figure}[h!]
            \centering
            \includegraphics[width=\textwidth,height=0.9\textheight,keepaspectratio]{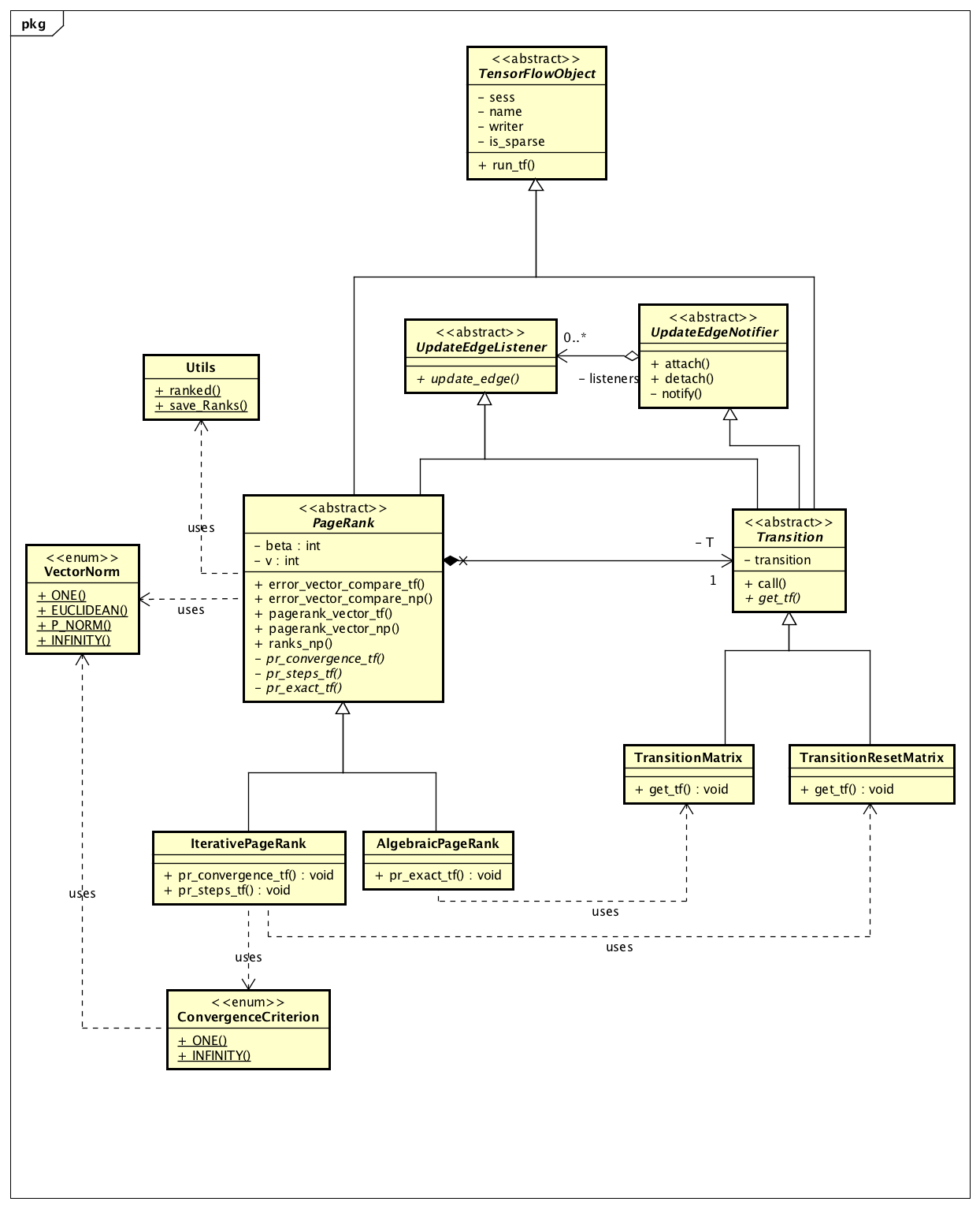}
            \caption{Diagrama de clases referido a las relaciones con la clase \texttt{PageRank} del módulo \texttt{tf\_G}.}
            \label{img:pagerank_diagram}
          \end{figure}

          \paragraph{}
          El último diagrama de clases que se incluye en esta implementación es el referido al algoritmo \emph{PageRank}. Dicho diagrama se corresponde con la figura \ref{img:pagerank_diagram}. Nótese que en dicho diagrama se ha obviado la relación entre la clase \texttt{Transition} y la clase \texttt{Graph}. La razón ha sido tratar de reducir la complejidad y mejorar el entendimiento del diagrama.

          \paragraph{}
          En cuanto a las herencias, en este caso tanto \texttt{PageRank} como \texttt{Transition} son descendientes de \texttt{TensorFlowObject}, que tal y como se indicó anteriormente ofrece la funcionalidad de ejecución de la estructura de datos \texttt{tf.Tensor} porporcionada por la biblioteca \emph{TensorFlow}. En cuanto a la notificación de modificaciones en las aristas del grafo, \texttt{PageRank} tiene la capacidad de ser notificada por dichos cambios, para recalcular el ranking PageRank conforme este se modifica tras los cambios en el conjunto de aristas del grafo.

          \paragraph{}
          Es necesario remarcar que \texttt{Transition} implementa tanto la funcionalidad de notificar como de ser notificado. Esto se debe a que cuando una arista es modificada, la matriz de transición debe ser notificada, y una vez hecho esto, se debe notificar al algoritmo PageRank para que recalcule el ranking a partir de ella. De ahí la necesidad de implementar las dos funcionalidades.

          \paragraph{}
          En cuanto a las clases descendientes de \texttt{PageRank} y \texttt{Transition}, estas implementan la funcionalidad del cálculo del ranking \emph{PageRank} de manera algebraica (\texttt{AlgebraicPageRank} junto con \texttt{TransitionMatrix}) y de manera iterativa (\texttt{IterativePageRank} junto con \texttt{TransitionResetMatrix}) respectivamente.

          \paragraph{}
          En cuanto a la clase \texttt{VectorNorm}, esta provee de distintas normas vectoriales utilizadas para realizar comparaciones entre diferentes rankings. Además, es utilizada para el cálculo del punto de convergencia en la clase \texttt{ConvergenceCriterion}, que utiliza la implementación iterativa del algoritmo \emph{PageRank}. Por último, la clase \texttt{Utils} proporciona la funcionalidad de generación del ranking de vértices a partir del valor del \emph{PageRank} obtenido, es decir, realiza una ordenación de los vértices.

          \paragraph{}
          Una vez descrita la implementación a partir de la estructura de sus clases y módulos, se ha decidido añadir diagramas referidos al conjunto de operaciones que se realizan para el cálculo del resultado en la siguiente sección.

        \subsubsection{Diagrama de Operaciones}
        \label{sec:operations_diagram}

          \paragraph{}
          La implementación realizada se basa en el desarrollo del algoritmo \emph{PageRank} sobre la biblioteca de cálculo matemático intensivo \emph{TensorFlow}. Sin embargo, los detalles de la implementación del algoritmo no son apreciables a través de los distintos diagramas ilustrados en anteriores secciones. Por dicha razón, se ha decidido incluir este apartado, en el cual se presentan los grafos de operaciones referidos a las implementaciones algebraica e iterativa del algoritmo.

          \paragraph{}
          Dichos árboles de operaciones requieren de la necesidad de estar familiarizado con la biblioteca utilizada para su apropiado entendimiento. Sin embargo, mediante su visualización rápida se puede apreciar una perspectiva de alto nivel acerca de las operaciones necesarias para el cálculo del ranking.

          \begin{figure}[h!]
            \centering
            \includegraphics[width=\textwidth,height=0.9\textheight,keepaspectratio]{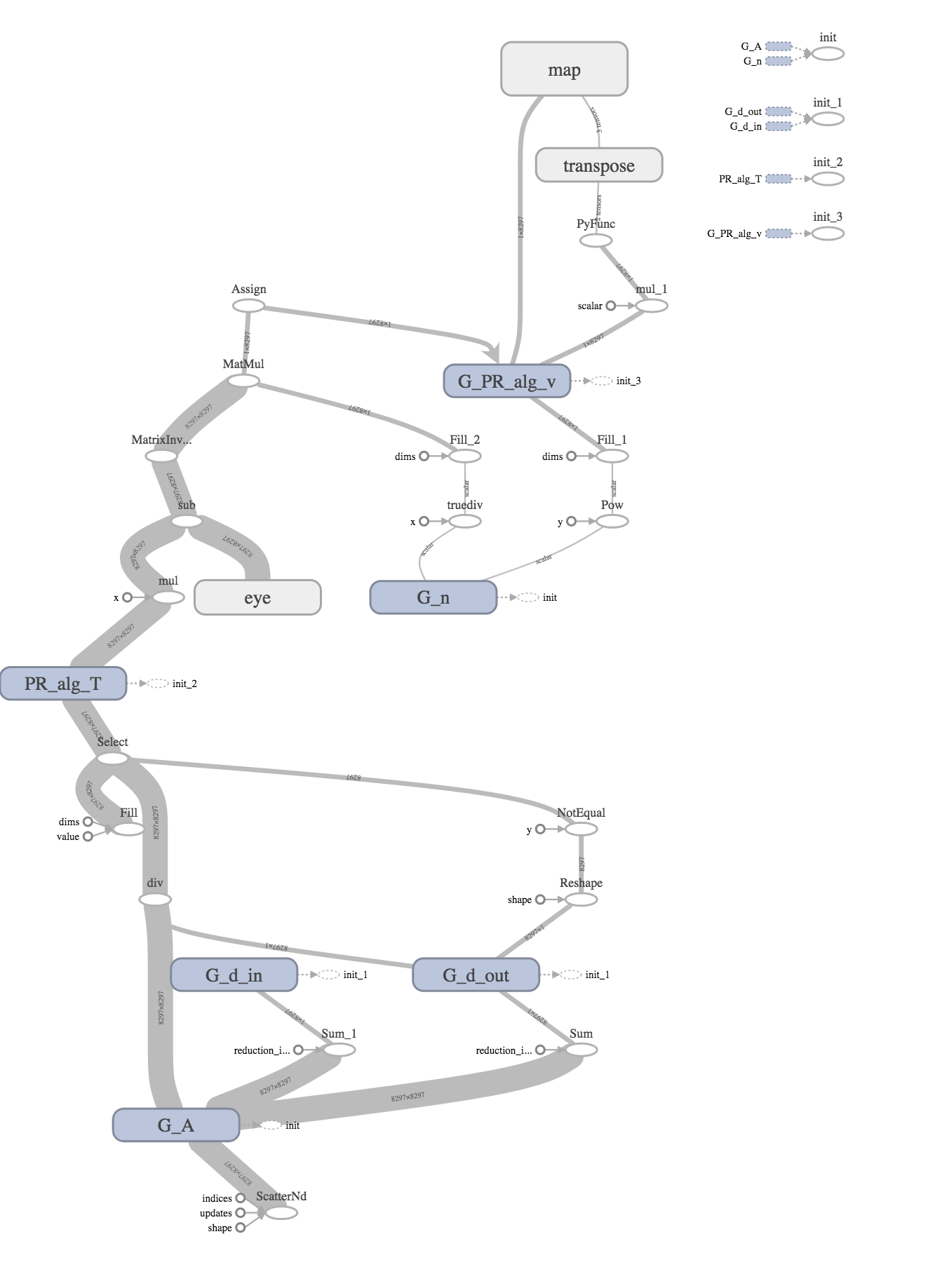}
            \caption{Diagrama de operaciones referido a la implementación algebraica del algoritmo \emph{PageRank} utilizando \emph{TensorFlow} del módulo \texttt{tf\_G}.}
            \label{img:pagerank_algebraic_diagram}
          \end{figure}

          \paragraph{}
          En la figura \ref{img:pagerank_algebraic_diagram}, se muestra el árbol de operaciones necesario para calcular el ranking \emph{PageRank} de manera algebraica. Dichas operaciones se corresponden con las previamente descritas en la sección \ref{sec:pagerank_algorithm_algebraic}, la cual se destinó íntegramente al estudio de dicho algoritmo.

          \begin{figure}[h!]
            \centering
            \includegraphics[width=\textwidth,height=0.9\textheight,keepaspectratio]{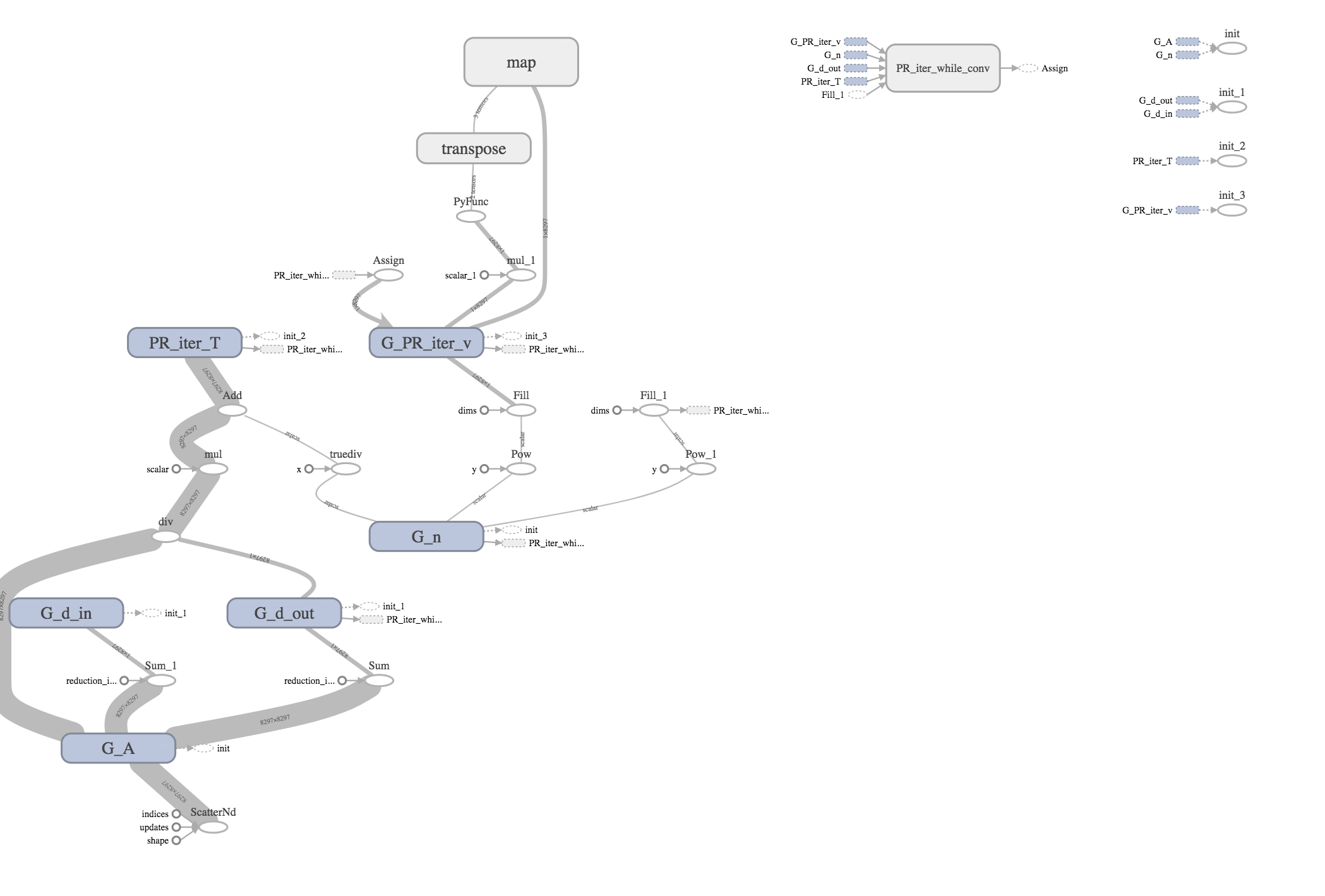}
            \caption{Diagrama de operaciones referido a la implementación iterativa del algoritmo \emph{PageRank} utilizando \emph{TensorFlow} del módulo \texttt{tf\_G}.}
            \label{img:pagerank_iterative_diagram}
          \end{figure}

          \paragraph{}
          En la figura \ref{img:pagerank_iterative_diagram} se muestra la versión iterativa implementada para el cálculo del \emph{PageRank}. En este caso, al igual que en el caso anterior, también se ha seguido el mismo algoritmo que el descrito en la sección \ref{sec:pagerank_algorithm_iterative}. Nótese por tanto, que en este caso es necesario el uso de un bucle que realiza las iteraciones, controlado por un determinado criterio de convergencia escogido de entre los implementados en la clase \texttt{ConvergenceCriterion}.

          \paragraph{}
          Nótese que estos diagramas representan una visión completamente distinta de la descrita en los diagramas de componentes o de clases, ya que estos ignoran por completo dicha organización, centrándose únicamente en el conjunto de operaciones necesarias para el cálculo del algoritmo.

      \paragraph{}
      En esta sección se han presentado una serie de detalles de implementación, comenzando por la descripción del conjunto de herramientas utilizadas para el desarrollo de la aplicación. Posteriormente se indicaron las dependencias con servicios externos que se han utilizado. El resto de la sección se ha basado en la descripción de la implementación a partir de diagramas de componentes, clases y operaciones. En este caso se ha prescindido del uso de diagramas de secuencia que indicasen el orden de llamadas entre métodos puesto que se ha creído que para una implementación de carácter algorítmico como la realizada, la inclusión del diagrama de operaciones generado por la biblioteca \emph{TensorFlow} podría aportar una información similar, que además muestra el conjunto de dependencias entre las variables utilizadas durante el cálculo.

    \section{Trabajo Futuro}
    \label{sec:future_work}

      \paragraph{}
      El trabajo realizado se puede enmarcar en una zona intermedia entre un trabajo de investigación y un trabajo de implementación práctica, puesto que una gran cantidad del mismo se ha destinado a adquirir nuevos conceptos relacionados con el ámbito del tratamiento de grandes conjuntos de datos. Esto se ve reflejado en los primeros capítulos del trabajo, marcadamente más teóricos que este último, en el cual se indican detalles y razonamientos acerca de las decisiones de implementación.

      \paragraph{}
      Debido al tiempo limitado en la realización del trabajo, así como el contexto académico junto con la compaginación de la realización al mismo tiempo que con otras asignaturas de la titulación, no se ha podido realizar de la manera tan extensa que se hubiera deseado. Además, se cree que el grado de extensión de los conceptos que se describen a lo largo del trabajo así como el proceso iterativo que se puede llevar a cabo para seguir trabajando en la implementación desarrollada lo hacen un trabajo continuable en el futuro.

      \paragraph{}
      Resultaría interesante seguir trabajando en el proyecto de una manera similar, siguiendo un equilibrio equitativo entre las horas de investigación en técnicas para la reducción de la complejidad de grafos de tamaño masivo, junto con la creación de un ecosistema de utilidades que permiten la implementación de estas técnicas así como nuevas métricas como el \emph{conteo de triángulos} u otros algoritmos, aprovechando la implementación inicialmente desarrollada.

      \paragraph{}
      Muchas de estas ideas podrían llegar resultados interesantes en el futuro mediante la realización de un estudio intensivo del problema así como una adecuada base matemática que facilite la comprensión de los trabajos desarrollados por los expertos en la materia, puesto que se cree que en el futuro será necesaria la utilización de técnicas como las estudiadas para hacer frente a los problemas que se están dando en la actualidad, tales como planificación de rutas u otros, que además, contiene una componente dinámica muy importante.

      \paragraph{}
      Al igual que en el caso de la investigación, se cree que continuar con el proceso de implementación de una biblioteca de utilidades que permita la implementación y desarrollo de soluciones basadas en grafos sobre una plataforma como \emph{TensorFlow} podría resultar muy interesante, al igual que sucedió con \emph{GraphX} y \emph{Spark}.

      \paragraph{}
      Por tanto, existen distintas lineas de trabajo sobre las cuales poder seguir en este área, todas ellas muy relacionadas entre si, y que podrían llegar a resultados satisfactorios con el correspondiente trabajo y dedicación.

    \section{Conclusiones}
    \label{sec:implementation_conclusions}

      \paragraph{}
      En esta sección se pretende realizar una descripción a nivel de resultados obtenidos tras realizar el \textbf{Trabajo de Fin de Grado} completo. Es decir, tanto la parte de investigación y estudio de \emph{Algoritmos para Big Data}, desde la modelización de \emph{Streaming} (capítulo \ref{chap:streaming}) y las \emph{Estrategias de Sumarización} (capítulo \ref{chap:summaries}) de información necesarias para agilizar la obtención de resultados sobre conjunto de datos masivos, como las aplicadas a \emph{Grafos} (capítulo \ref{chap:graphs}) y el estudio del algoritmo implementado \emph{PageRank} (capítulo \ref{chap:pagerank}).

      \paragraph{}
      Dichos estudios han permitido conocer de manera más profunda los factores que dificultan la tarea de diseño de algoritmos, cuya ejecución continúe siendo factible aún cuando el conjunto de datos crece de manera drástica y es necesario que las consultas sean realizadas en un periodo reducido de tiempo. Esto implica una tarea compleja, para la cual es necesario poseer una extensa base matemática que agilice las tareas de razonamiento y entendimiento acerca de la información encontrada.

      \paragraph{}
      Sin embargo, a pesar de la dificultad propiciada por dicha carga matemática, se ha tratado de hacer frente al trabajo de investigación mediante la lectura de una gran cantidad de artículos científicos en las cuales aparecen explicaciones e ideas muy acertadas. Algo que ha servido para introducirse y conocer cómo es el proceso de investigación, en numerosas ocasiones frustrante y complejo, pero que también ofrece un alto índice de satisfacción personal cuando se consiguen los propósitos alcanzados.

      \paragraph{}
      Desde la perspectiva de los resultados obtenidos a nivel de investigación, estos no han sido satisfactorios desde el punto de vista del descubrimiento o análisis de una nueva técnica aplicable sobre el área de investigación que se ha estudiado a lo largo del trabajo. Sin embargo, esto no se entiende como algo totalmente negativo puesto que a través de este proceso se han aprendido una gran cantidad de nuevos conceptos más amplios respecto de los prefijados para un \emph{Graduado en Ingeniería Informática}, los cuales se han creído interesantes y útiles en el futuro.

      \paragraph{}
      La introducción en el ámbito de la investigación, el aprendizaje de distintas estrategias para leer artículos (ya que requieren de práctica y metodología en comparación con otro tipo de literatura), y la gestión del tiempo en dichas tareas han sido un conjunto de conocimientos transversales, que además del aprendizaje inherente relacionado con el \emph{Big Data} se han creído extremadamente útiles para el futuro. Además, el estudio en profundidad de la problemática mediante artículos donde los autores originales de los trabajos exponen sus ideas se ha creído muy enriquecedor para las tareas posteriores de implementación.

      \paragraph{}
      Respecto de los resultados obtenidos desde el punto de vista de la implementación, se ha creído interesante el aprendizaje completo en cuanto al despliegue de un entorno de desarrollo que realice pruebas automatizadas, genere documentación y mantenga un seguimiento acerca de las horas de trabajo, así como la encapsulación de la implementación en un paquete auto-contenido de \emph{Python}, distribuible e instalable en otros sistemas es una tarea interesante, que hasta el momento no se había llevado a cabo debido a la rama escogida en los dos últimos años del grado (centrada mayoritariamente en aspectos algorítmicos y computacionales).

      \paragraph{}
      La implementación realizada ha servido para entender en profundidad el algoritmo \emph{PageRank}, así como las ideas subyacentes que permiten entender su procedencia así como su convergencia hacia un resultado satisfactorio. En cuanto a la implementación de los \emph{Sparsifiers}, partir de estos se ha conseguido entender en mayor medida las ventajas y dificultades derivadas de la adicción de un grado de indeterminismo sobre un \emph{Grafo}, tratando de perturbar al mínimo posible la estructura del mismo, lo cual es una tarea compleja, pero que se ha creído muy interesante para el futuro.

      \paragraph{}
      El \emph{Trabajo de Fin de Grado} ha servido para comprender el grado de dificultad que representa la realización de un proyecto de mayor envergadura que una práctica de una asignatura del grado, con una fecha prefijada y un periodo relativamente largo de tiempo para su ejecución. En este trabajo se ha tenido muy en cuenta la necesidad de organización y constancia personal diaria para poder asimilar la gran cantidad de conceptos estudiados así como la complejidad de las tecnologías utilizadas para la implementación, que muchas no habían sido utilizadas anteriormente. Esto ha proporcionado un grado de experiencia que se cree favorable y necesario para la finalización de los estudios por un graduado en \emph{Ingeniería Informática}. Sin embargo, este trabajo debe marcar un punto de comienzo sobre el cual mejorar en futuras ocasiones a partir de la experiencia, ya se los errores cometidos durante el desarrollo del mismo no se deben entender como algo negativo, sino como factores que no se deben repetir en futuros proyectos similares.

  \appendix

  \chapter{Metodología de Trabajo}
  \label{chap:methodology}

    \paragraph{}
    La metodología seguida para la realización de un proyecto de gran envergadura para un estudiante de grado, tal como es el trabajo de fin de grado requiere de una definición adecuada de la misma. De esta manera, se puede clarificar el camino a seguir para la elaboración del mismo.

    \paragraph{}
    Tal y como se puede apreciar en el documento \emph{Portal of research methods and methodologies for research projects and degree projects} \cite{haakansson2013portal}, el cual se ha utilizado como base de partida para conocer las posibles metodologías a seguir para un proyecto de investigación, esta decisión debe ser tomada al comienzo del mismo. De esta manera, se simplifica en gran medida la búsqueda de objetivos a logra durante el trabajo. En dicho documento \cite{haakansson2013portal} se realiza una diferenciación entre estrategias de investigación y métodos de investigación para después describir cada una de ellas. Posteriormente en este apéndice se indicará cómo se ha desarrollado este trabajo siguiendo dichas diferenciaciones

    \paragraph{}
    Un problema común que surge durante el desarrollo de los primeros proyectos de investigación es la dificultad por concretar el tema de estudio en un ámbito específico y analizable apropiadamente. Existen distintas razones por las cuales puede suceder dicho problema, sin embargo, la razón más destacada es la elevada interrelación entre las distintas ramas del conocimiento, que muchas veces complica la labor de \say{aislar} un pequeño sub-conjunto de ideas sobre las cuales realizar una profundización más extensa.

    \paragraph{}
    Tal y como se ha indicado anteriormente, es necesario indicar tanto las estrategias de investigación como lo métodos de investigación seguidos durante este trabajo. Sin embargo, lo primero es indicar a qué se refieren cada una de ellas. Cuando se habla de \emph{Estrategia de Investigación} nos estamos refiriendo al propósito final que se pretende conseguir con la realización del proyecto, es decir, es algo similar a los objetivos del mismo. En cambio, cuando se habla de \emph{Métodos de Investigación} nos estamos refiriendo al conjunto de \say{herramientas} conceptuales que se utilizan para conseguir llegar al propósito del trabajo.

    \paragraph{}
    La estrategia de investigación seguida a lo largo del desarrollo de este proyecto ha sido un estudio (\emph{survey}) realizado con la finalidad de tratar de comprender mejor el amplísimo área de investigación relacionado con el \emph{Big Data}, a partir del cual se ha ido profundizando en un ámbito más concreto: el estudio de grafos de tamaño masivo y la implementación del \emph{PageRank}, que en conjunto han otorgado una visión detallada acerca de dichas áreas desde una perspectiva clara. En cuanto al método seguido, lo primero fue prefijar el tema del \emph{Big Data} por el tutor del proyecto. A partir de este punto se puede diferenciar el método en dos partes, la primera de ellas correspondiente a un periodo de \emph{investigación descriptiva}, basada en obtener una visión global sobre las áreas de investigación del \emph{Big Data}, por tanto, esta fase fue guiada por las competencias de distintos cursos impartidos en universidades de a lo largo del mundo, centrados en la materia. Tras haber conseguido un nivel de comprensión adecuado del mismo, se modificó la metodología a seguir, por la que en \cite{haakansson2013portal} denominan \emph{investigación fundamental} (\emph{Fundamental Research}). Esta se caracteriza por la focalización del trabajo en un ámbito concreto a partir de la curiosidad personal, que en este caso, poco a poco fue acercandose hacia el estudio de grafos.

    \paragraph{}
    Puesto que el trabajo de fin de grado se refiere a una titulación de \emph{Ingeniería Informática}, se creyó interesante que este no se basara únicamente en tareas de investigación, sino que también contuviera una pequeña parte de implementación de código fuente. Para dicha labor, se escogió el algoritmo \emph{PageRank} puesto que a partir de su estudio se abarcan una gran cantidad de conceptos relacionados con el resto de los estudiados.

    \paragraph{}
    Muchas de estas decisiones fueron tomadas conforme avanzaba el proyecto, lo cual presenta distintas ventajas e inconvenientes. La ventaja más notoria se corresponde con el grado de libertad que se ha tenido durante todo el proceso, lo cual ha permitido centrarse en aquellas partes más motivadoras. Sin embargo, esto también genera una serie de desventajas, entre las que se encuentra la dificultad al llegar a puntos a partir de los cuales no saber hacia qué dirección continuar. Sin embargo, esta desventaja también representa un aprendizaje, que ayudará en futuras ocasiones a hacer frente a problemas semejantes con un grado de presión mucho menor debido a la experiencia adquirida en este caso.

    \paragraph{}
    Muchos proyectos de ingeniería informática se realizan siguiendo distintas metodologías de gestión de proyectos, tales como \emph{metodologías en cascada} o las denominadas \emph{ágiles} como \emph{SCRUM}. En las primeras semanas de la realización de este proyecto, se pretendió seguir una metodología basada en \emph{SCRUM}, tratando de realizar una serie de tareas divididas en bloques de dos semanas. Sin embargo, dicho enfoque se avandonó rápidamente por la naturaleza inherente de investigación seguida para este proyecto. La razón se debe a que es muy complicado compaginar las tareas de investigación, las cuales se basan en la adquisición de conocimiento con un determinado concepto concreto y un periodo de tiempo acotado.

    \paragraph{}
    Las razones que han llevado a pensar esto están relacionadas con el proceso de investigación basado en la lectura de artículos de carácter científico, los cuales se relacionan fuertemente unos con otros. Dicho suceso conlleva la necesidad de tener que leer un grupo de artículos relacionados con el que se pretende comprender mediante el conjunto de citaciones que asumen el conocimiento de los términos Extraídos de otro artículo. Estas razones dificultan la tarea de estimación temporal necesaria para entender la idea descrita en un artículo, que muchas veces se reduce a unos pocos trabajos, que además han sido comprendidos previamente, mientras que en otras ocasiones es necesario adquirir una gran cantidad de nuevos conceptos.

    \paragraph{}
    A este factor se le añade otra dificultad derivada del mismo, en un gran número de ocasiones dichas dificultades no se conocen hasta que no se ha profundizado en el entendimiento del trabajo, por lo que no se pueden estimar a priori. Sin embargo, son de gran ayuda los estudios (\emph{surveys}) realizados sobre temas específicos, que permiten obtener una visión panorámica acerca del tema mediante la lectura de un único trabajo, que después puede ser ampliada por el lector mediante la lectura de las referencias contenidas en el estudio.

    \paragraph{}
    A partir de todos estos factores se ha permitido conocer en mayor detalle cómo es el proceso de investigación, así como los retos metodológicos que surgen durante la realización de dichos proyectos. Ha habido muchos puntos que se podrían haber realizado de una manera más apropiada, con la consiguiente reducción de tiempo en dichas tareas. Sin embargo, se cree que todas estas complicaciones han permitido adquirir una experiencia que en futuras ocasiones agilizará el proceso y permitirá evitar dichos errores.

  \chapter{¿Cómo ha sido generado este documento?}
  \label{chap:how_it_was_build}

    \paragraph{}
    En este apéndice se describen tanto la estructura como las tecnologías utilizadas para redactar este documento. El estilo visual que se ha aplicado al documento se ha tratado de almoldar lo máximo posible a las especificaciones suministradas en la \emph{guía docente} de la asignatura \emph{Trabajo de Fin de Grado} \cite{uva:tfg-teaching-guide}.

    \paragraph{}
    Este documento ha sido redactado utilizando la herramienta de generación de documentos \LaTeX \cite{tool:latex}, en concreto se ha utilizado la distribución para sistemas \emph{OS X} denominada \emph{MacTeX} \cite{tool:mactex} desarrollada por la organización \emph{\TeX \ User Group}. Mediante esta estrategia todas las labores de compilación y generación de documentos \emph{PDF} (tal y como se especifica en la guía docente) se realizan de manera local. Se ha preferido esta alternativa frente a otras como la utilización de plataformas online de redacción de documentos \LaTeX \ como \emph{ShareLateX} \cite{tool:sharelatex} u \emph{Overleaf} \cite{tool:overleaf} por razones de flexibilidad permitiendo trabajar en lugares en que la conexión a internet no esté disponible. Sin embargo, dichos servicios ofrecen son una buena alternativa para redactar documentos sin tener que preocuparse por todos aquellos aspectos referidos con la instalación de la distribución u otros aspectos como un editor de texto. Además garantizan un alto grado de confiabilidad respecto de pérdidas inesperadas.

    \paragraph{}
    Junto con la distribución \LaTeX \ se han utilizado una gran cantidad de paquetes que extienden y simplifican el proceso de redactar documentos. Sin embargo, debido al tamaño de la lista de paquetes, esta será obviada en este apartado, pero puede ser consultada visualizando el correspondiente fichero \texttt{thestyle.sty} del documento.

    \paragraph{}
    Puesto que la alternativa escogida ha sido la de generar el documento mediante herramientas locales es necesario utilizar un editor de texto así como un visualizador de resultados. En este caso se ha utilizado \emph{Atom} \cite{tool:atom}, un editor de texto de propósito general que destaca sobre el resto por ser desarrollado mediante licencia de software libre (\emph{MIT License}) y estar mantenido por una amplia comunidad de desarrolladores además de una extensa cantidad de paquetes con los cuales se puede extender su funcionalidad. En este caso, para adaptar el comportamiento de \emph{Atom} a las necesidades de escritura de texto con latex se han utilizados los siguientes paquetes: \emph{latex} \cite{tool:atom-latex}, \emph{language-latex} \cite{tool:atom-language-latex}, \emph{pdf-view} \cite{tool:atom-pdf-view} encargados de añadir la capacidad de compilar ficheros latex, añadir la sintaxis y permitir visualizar los resultados respectivamente.

    \paragraph{}
    Puesto que el \emph{Trabajo de Fin de Grado} se refiere a algo que requiere de un periodo de tiempo de elaboración largo, que además sufrirá una gran cantidad de cambios, se ha creído conveniente la utilización de una herramienta de control de versiones que permita realizar un seguimiento de los cambios de manera organizada. Para ello se ha utilizado la tecnología \emph{Git} \cite{tool:git} desarrollada originalmente por \emph{Linus Torvalds}. En este caso en lugar de confiar en el entorno local u otro servidor propio se ha preferido utilizar la plataforma \emph{GitHub} \cite{tool:github}, la cual ofrece un alto grado de confiabilidad respecto de posibles perdidas además de alojar un gran número de proyectos de software libre. A pesar de ofrecer licencias para estudiantes que permiten mantener el repositorio oculto al público, no se ha creído necesario en este caso, por lo cual se puede acceder al través de la siguiente url: \url{https://github.com/garciparedes/tf_G}

    \paragraph{}
    Una vez descritas las distintas tecnologías y herramientas utilizadas para la elaboración de este trabajo, lo siguiente es hablar sobre la organización de ficheros. Todos los ficheros utilizados para este documento (obviando las referencias bibliográficas) han sido incluidos en el repositorio indicado anteriormente.

    \paragraph{}
    Para el documento, principal alojado en el directorio \texttt{/document/} se ha seguido una estructura modular, dividiendo los capítulos, apéndices y partes destacadas como portada, bibliografía o prefacio entre otros en distintos ficheros, lo cual permite un acceso sencillo a los mismos. Los apéndices y capítulos se han añadido en los subdirectorios separados.  Para la labor de combinar el conjunto de ficheros en un único documento se ha utilizado el paquete \emph{subfiles}. El fichero raiz a partir del cual se compila el documento es \texttt{document.tex}. La importación de los distintos paquetes así como la adaptación del estulo del documento a los requisitos impuestos se ha realizado en \texttt{thestyle.sty} mientras que el conjunto de variables necesarias como el nombre de los autores, del trabajo, etc. se han incluido en \texttt{thevars.sty}.

    \paragraph{}
    En cuanto al documento de resumen, en el cual se presenta una vista panorámica acerca de las distintas disciplinas de estudio relacionadas con el \emph{Big Data} se ha preferido mantener un único fichero debido a la corta longitud del mismo. Este se encuentra en el directorio \texttt{/summary/}.

    \paragraph{}
    Por último se ha decidido añadir otro directorio denominado \texttt{/notes/} en el cual se han añadido distintas ideas de manera informal, así como enlaces a distintos cursos, árticulos y sitios web en que se ha basado la base bibliográfica del trabajo. En la figura \ref{fig:repository-tree} se muestra la estructura del repositorio en forma de árbol.

    \begin{figure}
      \centering
\begin{Verbatim}
.
+-- document
|   +-- bib
|   |   +-- article.bib
|   |   +-- ...
|   +-- img
|   |   +-- logo_uva.eps
|   |   +-- ...
|   +-- tex
|   |   +-- appendices
|   |   |   +-- how_it_was_build.tex
|   |   |   +-- ...
|   |   +-- chapters
|   |   |   +-- introduction.tex
|   |   |   +-- ...
|   |   +-- bibliography.tex
|   |   +-- ...
|   +-- document.tex
|   +-- ...
+-- notes
|   +-- readme.md
|   +-- ...
+-- summary
|   +-- summary.tex
|   +-- ...
+-- README.md
+-- ...
\end{Verbatim}
      \caption{Árbol de directorios del repositorio}
      \label{fig:repository-tree}
    \end{figure}

  \chapter{Guía de Usuario}
  \label{chap:user_guide}

    \paragraph{}
    En esta sección se describe el proceso de instalacción y uso de la implementación realizada. Para ello existen distintas alternativas, entre las que se encuentran la instalación utilizando el comando \texttt{python} u otras basadas en gestores de modulos como \texttt{easy\_install} o \texttt{pip}. En este caso se realiza una descripción para instalar el proyecto usando el gestor \texttt{pip}.

    \paragraph{}
    Antes de nada es necesario tener instalado en el sistema el lenguaje de programación \texttt{Python}, en su versión \textbf{3.5} o superior, junto con su correspondiente versión de \texttt{pip}. Para el proceso de instalacción del modulo se puede recurrir al repositorio alojado en \emph{GitHub} o instalarse directamente a través de la copia local.

    \paragraph{}
    El comando a ejecutar para instalar la implementación en el sistema a partir del repositorio de \emph{Github} (nótese que en este caso es necesario tener instalada la utilidad \texttt{git}) se muestra a continuación:
    \begin{verbatim}
    $ pip install git+https://github.com/garciparedes/tf_G.git
    \end{verbatim}

    \paragraph{}
    En el caso de preferir instalar la copia local del repositorio, tan solo es necesario ejecutar la siguiente orden:

    \begin{verbatim}
    $ pip install .
    \end{verbatim}

    \paragraph{}
    Una vez completado el proceso de instalación con éxito, ya se está en condiciones suficientes como para utilizar la implementación realizada. Para ello, esta se puede importar en ficheros que contengan código fuente \texttt{python} que se ejecute sobre intérpretes cuya versión sea \textbf{3.5} o superior.

    \paragraph{}
    También se puede utilizar la implementación sobre un intérprete ejecutandose sobre la línea de comandos del sistema mediante comandos como \texttt{python3} o la versión extendida \texttt{ipython3}. Una vez en el intérprete se puede importar el módulo simplemente con ejecutar:

    \begin{verbatim}
    >>> import tf_G
    \end{verbatim}

    \paragraph{}
    Una vez ejecutada dicha sentencia se tiene acceso al ecosistema de clases descrito en la documentación. La cual se encuentra contenida en el propio código a través del estándar \texttt{docstring}. Dicha documentación también es accesible en forma de sitio web a través de la siguiente url: \url{http://tf-g.readthedocs.io/en/latest/}.

    \paragraph{}
    En el caso de poseer una copia local del repositorio, también es posible realizar una ejecución del conjunto de pruebas de test que confirman que la corrección del código. Para ello es necesario poseer la utilidad \texttt{pytest}. Únicamente con la ejecución de dicha prueba sobre el fichero raiz del repositorio, se realizarán todas las pruebas unitarias contenidas en el repositorio. Para ello se debe ejecutar la siguiente sentencia:

    \begin{verbatim}
    $ pip install -e .
    $ pytest
    \end{verbatim}

    \paragraph{}
    En la copia local, además de incluirse los distintos casos de prueba a partir de los cuales se comprueba el correcto funcionamiento del código, se incluyen una serie de ejemplos. Dichos ejemplos consisten en un conjunto de pequeños scripts que realizan distintas llamadas a al módulo desarrollado para después imprimir los resultados en pantalla. Estos son accesibles a través del directorio \texttt{/examples/}.

    \paragraph{}
    Tal y como se puede apreciar mediante este manual de instalación y uso, gracias al sistema de gestión de módulos de \emph{Python}, las tareas de distribución de los mismos, así como la gestión de dependencias se simplifican drásticamente, límitandose únicamente al comando de instalacción, junto con la correspondiente importación necesaria para su uso.

  \bibliographystyle{alpha}
  \addcontentsline{toc}{chapter}{\protect\numberline{}Bibliografía}
  \bibliography{article,book,examples,inproceedings,lecture,misc,tools,uva}

\end{document}